\documentclass[%
 reprint,
 amsmath,amssymb,
 aps,
]{revtex4-2}

\usepackage{graphicx}
\usepackage{dcolumn}
\usepackage{bm}
\usepackage{booktabs} 
\newcommand{\ra}[1]{\renewcommand{\arraystretch}{#1}}
\newcolumntype{L}[1]{>{\raggedright\arraybackslash\setlength{\baselineskip}{2pt}}p{#1}}

\newcount\colveccount
\newcommand*\colvec[1]{
        \global\colveccount#1
        \begin{pmatrix}
        \colvecnext
}
\def\colvecnext#1{
        #1
        \global\advance\colveccount-1
        \ifnum\colveccount>0
                \\
                \expandafter\colvecnext
        \else
                \end{pmatrix}
        \fi
}

\usepackage{color}

\begin{document}

\preprint{APS/123-QED}

\title{Macroscopic approximation methods for the analysis of adaptive networked agent-based models: The example of a two-sector investment model}

\author{Jakob J. Kolb}
 \email{kolb@pik-potsdam.de}
 \affiliation{%
 FutureLab on Game Theory and Networks of Interacting Agents, \\ Potsdam Institute for Climate Impact Research, Potsdam, Germany}
 \affiliation{%
 Department of Physics, Humboldt University Berlin, Berlin, Germany
}%
 
\author{Finn M\"{u}ller-Hansen}%
\affiliation{%
 Mercator Research Institute on Global Commons and Climate Change, Berlin, Germany
}
\affiliation{
Potsdam Institute for Climate Impact Research, Potsdam, Germany
}%

\author{J\"{u}rgen Kurths}
\affiliation{
Potsdam Institute for Climate Impact Research, Potsdam, Germany
}%
\affiliation{
 Department of Physics, Humboldt University Berlin, Berlin, Germany
}%
\author{Jobst Heitzig}
 \affiliation{%
 FutureLab on Game Theory and Networks of Interacting Agents, \\ Potsdam Institute for Climate Impact Research, Potsdam, Germany}

\date{\today}

\begin{abstract}

In this paper, we propose a statistical aggregation method for agent-based models with heterogeneous agents that interact both locally on a complex adaptive network and globally on a market. The method combines three approaches from statistical physics: (a) moment closure, (b) pair approximation of adaptive network processes, and (c) thermodynamic limit of the resulting stochastic process.
As an example of use, we develop a stochastic agent-based model with heterogeneous households that invest in either a fossil-fuel or renewables-based sector while allocating labor on a competitive market. Using the adaptive voter model, the model describes agents as social learners that interact on a dynamic network.
We apply the approximation methods to derive a set of ordinary differential equations that approximate the macro-dynamics of the model. 
A comparison of the reduced analytical model with numerical simulations shows that the approximation fits well for a wide range of parameters.

The method makes it possible to use analytical tools to better understand the dynamical properties of models with heterogeneous agents on adaptive networks.
We showcase this with a bifurcation analysis that identifies parameter ranges with multi-stabilities.
The method can thus help to explain emergent phenomena from network interactions and make them mathematically traceable.
\end{abstract}

\maketitle

\section{Introduction}
\label{sec:intro}

Agent-based modeling is a computational approach to simulate systems composed of a large number of similar sub-units with many applications in ecology \citep{Grimm2005}, business \citep{Bonabeau2002}, sociology \citep{Macy2002} and economics \citep{Tesfatsion2006, Hamill2016}.
ABMs are used to study aggregate phenomena emerging from local interactions \citep{Epstein1999}.
These interactions can be structured by spatial embedding of agents or by social networks \citep{Gross2008,Holme2006a,Bargigli2014, Granovetter2005}.
In economics, ABMs have been used to study for example business cycles \citep{DelliGatti2008}, market power \citep{Tesfatsion2006} and trade \citep{Hamill2016}.

ABMs are a promising alternative to dynamic stochastic general equilibrium (DSGE) modeling, the current workhorse of theoretical macroeconomics. 
DSGE models usually build on the representative agent approach, i.e., they represent all individuals of one type such as firms or consumers by one representative decision maker.

The representative agent approach implies that theoretical macroeconomics reduces macroeconomic phenomena to assumptions about a few different representative agents, leaving out many explanatory mechanisms for fluctuations in aggregate variables based on intra-group interaction and heterogeneity \footnote{Approaches to represent heterogeneous agents in DSGE models have been used to counter this criticism and add more realism regarding the distribution of agent attributes \citep[see for example the review by][]{Heathcote2009}.
Particularly, because the representative agent approach cannot account for interactions within a heterogeneous group, models using this approach do not allow for the representation of emergent phenomena \citet{Kirman1992}.
But their solution require complex numerical methods and cannot integrate local interactions between agents.}.
Furthermore, DSGE model often assume rational expectations, i.e., agents know the constraints and dynamics of the entire economy, which has been criticised as philosophically unsound and empirically unjustified \citep{Kirman2014}.
But, due to these assumptions, most DSGEs allow for a thorough analytical analysis.

ABMs allow implementing various individual decision models that are behaviorally more realistic than full economic rationality.
Agents are often assumed to be boundedly rational and adapt their expectations, which is compatible with the Lucas critique \citep{Evans2006}.
In ABMs, fluctuations in aggregate variables do not only arise from exogenous shocks as in DSGE models but primarily from irregularities in local interactions.
Therefore, they offer an avenue for explaining various emergent phenomena \footnote{We use here a weak notion of emergence, which allows explaining macro-phenomena on the basis of micro-interactions of the systems constituents that differ from the explained macro-phenomena. This is opposed to strong emergence, that embraces the irreducibility of macro-phenomena to lower-level dynamics. For a discussion see \citet{Bedau1997}.} studied in empirical macroeconomics.

On the other hand, ABMs are often very detailed so that an analytic treatment is unfeasible. 
Therefore, in ABMs, the difficulties arising from the aggregation of heterogeneous and interacting agents are usually solved computationally.
Because the model mechanisms are difficult to trace in the `black box' of a computational model, the results of ABMs are often difficult to interpret and cannot provide mathematically sound proofs of relationships between model variables. Results may therefore be difficult to generalize \citep{Leombruni2005}.
There has been some progress in the standardization of model descriptions for ABMs \citep{Grimm2006}, but the lack of standardization, e.g. of decision rules, makes the models difficult to compare \citep[][p. 239]{Hamill2016}. Even though there are various techniques available for comprehensive model analysis \citep{Lee2015}, a systematic model exploration is uncommon and mostly limited to sensitivity analysis with respect to crucial parameters.

Methods from theoretical physics have been applied successfully to various problems in economics for many years \citep{Mantegna1999}. Here, aggregation methods from statistical physics can bridge the gap between analytic macroeconomic models such as DSGE approaches and agent-based computational models \citep[for a review of physics methods in social modeling, see refs.][]{castellano2009statistical, Martino2006}. In contrast to macroeconomic models, these approaches account for local interactions and use aggregation techniques to derive macro-dynamics, providing a true microfoundation of the resulting macromodel.
These kinds of approximation methods have found much interest in the fields of financial economics, behavioral finance and evolutionary game theory recently and have produced interesting and promising results, e.g. to explain macroeconomic fluctuations \citep[e.g.][]{Acemoglu2015} and understand propagation of financial shocks and the resulting systemic risk \citep[e.g.][]{DiGuilmi2012}.

Many authors use mean field approximations to aggregate interactions between heterogeneous agents, e.g. making use of stochastic differential equations, Master or Fokker-Planck equations \citep{Aoki1998, Aoki2007, DelliGatti2000, Gualdi2015, Gualdi2017, DiGuilmi2008, Chiarella2011a, Landini2014, Bouchaud2013, Fiaschi2010}.
Such approaches assume that each agent pair interacts with the same probability.
But many social and economic interactions are structured and the structure can be described by complex networks \citep{Friedkin2011}. To also capture the dynamics arising from structured interactions, so-called moment closure methods take the micro-structure of networks into account when deriving macroscopic quantities \citep[e.g.][]{Alfarano2008a, Lux2016}. Thereby, they are able to show that often the network structure, whether fixed or evolving, has a crucial influence on the dynamics not only quantitatively but also qualitatively in enriching the stability landscape and introducing additional (meta-) stable dynamical regimes, e.g. due to effects related to clustering and community structure.

Yet, most of the literature regards either the network between agents or the states of agents as static, implicitly assuming different time scales for dynamics of and processes on the network.
However, recent literature on opinion formation processes and the spreading of social norms in the field of computational social sciences suggests that both happen on a comparable timescale and can therefore not be treated separately \citep{Gross2008, gross2009adaptive}.
For such adaptive networks \citep{Gross2008}, moment closure techniques have been introduced in the physics literature to aggregate the feedback between complex adaptive network dynamics and dynamics of single node states \citep{Do2009, Demirel2014, Wiedermann2015, Min2017}.
Here, we introduce these techniques to economic modeling and combine them with approaches from macroeconomics where interactions also happen globally via aggregated variables.

The technical challenges of analytic approximation methods for agent-based model has so far hampered their wide-spread use in economics. But they have a huge potential in providing profound insights into dynamical properties of economic systems: First, they help increasing performance of computer simulations, making calculation of single model runs much faster and therefore allowing for a wider range of bifurcation and parameter analyses. Second, in contrast to stochastic simulations, they make formal proofs of relations between macroscopic variables possible. Third, they allow the derivation of analytical expressions of relations between model variables from the dynamic equations, which is not possible from single simulation runs. This paper makes a step forward in showcasing how such methods can be used to combine interactions on complex adaptive networks with macroeconomic modeling. It is therefore a contribution to integrate non-standard behavioral assumptions into macroeconomic models.

The agent-based model we introduce as an illustration of these methods is designed to investigate low-carbon transitions in an economy in the context climate economics and features both local interactions on a network and system-level interaction through markets.
We use an adaptive network approach for our model to demonstrate how the individual approximation techniques mentioned above may be combined. In our model, the network of interactions between agents as well as the spreading of strategies between agents on this interaction network happen on a comparable timescale.
In particular, we combine the different approximation techniques mentioned above, namely moment closure, pair approximation, and large system limit approximations to derive an aggregate description for the dynamics of our model
\citep[for an overview of the different techniques, see][]{Kuehn2016}.
The model consists of heterogeneous households that interact and learn from neighbors on a social network and a two-sector productive economy.
The households differ in their investment strategy: they invest their savings either in the ``dirty'' or the ``clean'' sector, each representing a separate capital market through which the agents interact.
Agents imitate the investment strategy of acquaintances that are better off with a higher probability.
To the best of our knowledge this is the first study that applies such a combination of approximation methods on a model that combines structured local with global interactions of heterogeneous agents in a socioeconomic setting.
By successfully applying approximation techniques for adaptive networks to our model, we demonstrate that they are useful for investigating economic relationships within considerably complex models.
Even though our reference application is an economic one, this approximation method can also be used to describe similarly structured models in other fields of research such as social ecology, neuroscience or computational social science.

In the remainder of the paper, we first describe the details of the model (Sec.~\ref{sec:Model_Description}). Then, we derive an aggregate description of the model by applying three approximation techniques, moment closure, pair approximation, and large system limit (Sec.~\ref{sec:Approximation}). We discuss commonalities and differences between computer simulations and the approximation approach. Before concluding, we illustrate how the derived macro-approximation can be used in a bifurcation analysis to better understand the qualitative properties of the non-linear model (Sec.~\ref{sec:bifurcation-analysis}).

\section{Model Description}
\label{sec:Model_Description}

To illustrate the use of the methods that we put forward, we develop a model of a stylized economy that captures the shift from a fossil-fuel-based to a renewable energy-based sector.
Decarbonization pathways consistent with the Paris agreement require a rapid shift of investments away from fossil fuel exploration and extraction to the development and deployment of renewable energies \citep{IPCC2014}.
However, the implementation of climate policies is uncertain and expectations cannot be based on self-consistent beliefs about the future. 
In conventional macroeconomic models such shifts can only occur due to price signals either from improvements in green technology, increasing scarcity of fossil reserves, or carbon pricing.
While price signals are certainly important, movements advocating for the divestment from fossil fuels point to the role of social norms and practices regarding investment decision to initiate and accelerate the energy transition \citep{Ans2013}.
To better understand such culturally driven situations of socioeconomic change, it is important to develop models that can incorporate endogenous preferences \citep{Mattauch2016, Mattauch2018} and aspects of bounded rationality \citep{Gsottbauer2011} such as imperfect foresight and information as well as learning.

Our model is designed to incorporate social dynamics that influence investment decisions \citep{Hong2009, Williams2007}. In the context of climate economics and policy, the literature on social influence and norms has pointed out that such mechanism are a leverage point to induce rapid change in socioeconomic systems \citep{Griskevicius2008, Masson2014, Stern2011, Rabinovich2011, Nyborg2016}.
The model focuses on two important mechanisms: First, investment strategies are spread on a network, which can be understood as a social learning process \citep{bandura1977} influenced by social norms \citep{Friedkin2001}. Secondly, the network adapts endogenously based on simple rules that model homophily \citep{Centola2007HomophilyGroups, Kimura2008}.
In the following, we explain the different parts of our two-sector model in detail. Table~\ref{tab:variable_list} provides an overview of the variables used for its formal description and Table~\ref{tab:Parameter_list} a list of parameters.

\begin{figure}[t]
  \centering
  \includegraphics[width=.5 \textwidth]{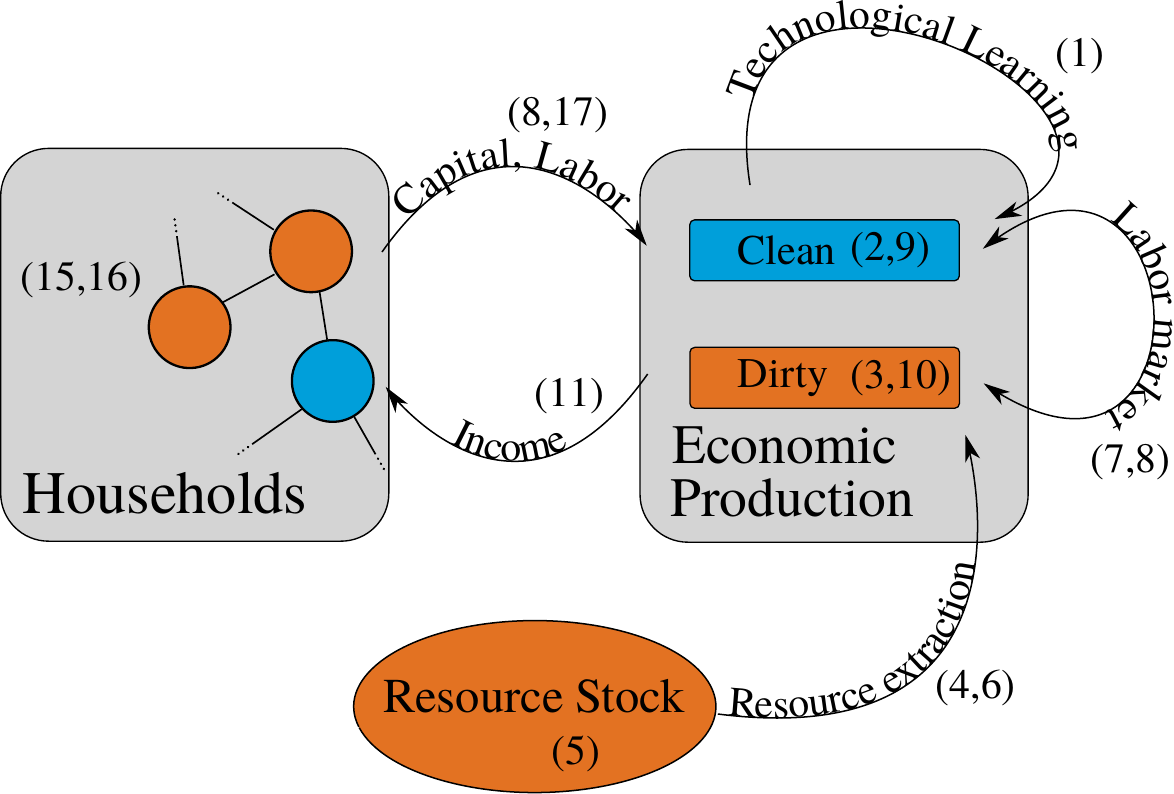}
  \caption{Schematic figure of the model consisting of two production sectors of which one depends on an exhaustible fossil resource stock as well as a set of heterogeneous households that interact on an adaptive complex network and use social learning to decide upon which of two production sectors to invest in. Boxes and bubbles denote modeled entities, arrows denote interactions. Numbers in brackets refer to equations that describe the specific part of the model.}
  \label{fig:model_scheme}
\end{figure}

\subsection{Economic Production}
\label{sec:model_description}

Our model as outlined in Fig.~\ref{fig:model_scheme} consists of two sectors for production and a set of heterogeneous households that interact via a complex adaptive social network. The two production sectors employ different technologies. The production technology in one sector depends on the input of an exhaustible (fossil) energy resource $R$ that is used up in the process whereas the technology in the other sector does not. We call them the \textit{dirty} and the \textit{clean} sector accordingly. We assume that physical capital is technology-specific and can not be reallocated between the two sectors.
Therefore, the heterogeneous households in the model provide different types of capital $K_j$ as well as labor $L$ to the sectors.
We assume that the technology in the dirty sector is fully developed and adequately described in terms of a fixed technological factor subsumed in the constant $b_d$, the so-called total factor productivity.
For fossil fuels, price elasticities of demand, i.e., changes in demand in response to increasing or decreasing prices, are low in real economies \cite{IMF2011, Hosslinger2017, Labandeira2017}, even with the choice between alternative technologies factored in. We approximate this by assuming that the fossil resource cannot be substituted by other production factors (capital, labor) in the dirty sector. This is in line with critique of commonly assumed substitutability of natural resources in some widely used production functions in neoclassical models \citep{Daly1997,georgescu1975energy,georgescu1979comments, Ayres2007, Ayres2013}.
However, we acknowledge that a shift in the output of economic production from manufacturing to services can lead to substitution of resources by capital and labor \citep{Mulder2012} and argue that our model pictures this in a shift of economic production from the dirty to the clean sector, which is described in the following.

The clean sector represents a circular economy in which the output of final goods depends on the machinery, knowledge and effort used in its production and is not limited by resource scarcity on the timescale under consideration. The technology $C$ used in the clean sector is assumed to be still in development and is therefore explicitly modeled.
Following \cite{argote1990learning}, we model technological progress as learning by doing according to Wright's law \citep{wright1936factors, Nagy2013}.  We assume that $C$ is proportional to cumulative production but also depreciates with a constant rate $\chi$. Depreciation can be regarded as a human capital effect that leads to knowledge depreciation over time as in \cite{Kahouli-Brahmi2008}. This is also in line with the empirically observed decrease in learning rates for maturing technologies \cite{argote1990learning}
\begin{equation}
	\dot{C} = Y_c - \chi C.
	\label{eq:learning_by_doing}
\end{equation}

Capital, labor and technology/knowledge are assumed to be mutual substitutes. To satisfy these requirements, we use the following production functions:
\begin{align}
	Y_c &= b_c C^{\gamma} L_c^{\alpha_c}K_c^{\beta_c}, \label{eq:clean_production} \\
	Y_d &= {\rm min}\left( b_d L_d^{\alpha_d}K_d^{\beta_d}, e R \right), \label{eq:dirty_production}
\end{align}
Subscripts $c$ and $d$ denote the clean and dirty sector respectively, $L_c$ and $L_d$ are labor in the two sectors, $\alpha$ and $\beta$ are elasticities of the respective input factors, $b_c$ and $b_d$ are the total factor productivities and $K_c$ and $K_d$ are the capital stocks for the respective sector. Measuring unit production cost in the number of working hours as in the original study by \cite{wright1936factors}, $\gamma$ is equivalent the elasticity of learning by doing in the clean sector as outlined in \cite{Kahouli-Brahmi2008}.

We assume an efficient usage of resources in the dirty sector, such that
\begin{equation}
    b_d L_d^{\alpha_d}K_d^{\beta_d} = e R
    \label{eq:efficient_dirty_resources}
\end{equation}
where $1/e$ is the resource intensity of the sector, i.e., the amount of fossil resource needed for one unit of final product. The usage of the fossil resource $R$ depletes a geological resource stock $G$ with the initial stock $G(t=0) = G_0$:
\begin{equation}
    \dot{G} = -R. 
    \label{eq:resource_depletion}
\end{equation} 
In line with the assumptions common in the literature \citep{Dasgupta1974, Perman2003}, the cost of the fossil resource extraction and provision $c_R$ depends on the resource flow $R$ and the remaining fossil resource stock $G$ such that $\partial c_R / \partial R >0$ and $\partial c_R / \partial G < 0$. We chose the specific form to be
\begin{equation}
	c_R = b_R R^{\rho}\left( \frac{G_0}{G} \right)^{\mu}; \quad \rho \geq 1, \quad \mu > 0,
	\label{eq:resource_cost}
\end{equation}
such that at some point $\partial Y_d / \partial R < \partial c_R / \partial R$ to take into account that some part of the resource is not economic, i.e., its marginal cost exceeds its marginal productivity.
We assume perfect labor mobility and competition for labor between the two sectors. This leads to an equilibrium wage $w$ that equals the marginal return for labor, i.e., the production increase from an additional unit of labor:
\begin{equation}
	w = \frac{\partial Y_c}{\partial L_c} = \frac{\partial Y_d}{\partial L_d} - \frac{\partial c_R}{\partial L_d}
	\label{eq:equilibrium_wage}
\end{equation}
with the sum of labor in both sectors equal to a constant total amount of labor:
\begin{equation}
	L_c + L_d = L.
	\label{eq:population}
\end{equation}
As discussed before, we assume physical capital to be specific to the technology employed such that it can only be used in the sector that it has been invested in originally. This means that there are separate capital markets for the two sectors. We assume these capital markets to be fully competitive resulting in capital rents equal to marginal productivity, after accounting for energy costs:
\begin{align}
	r_c &= \frac{\partial Y_c}{\partial K_c} \label{eq:clean_capital_rent}\\
	r_d &= \frac{\partial Y_d}{\partial K_d} - \frac{\partial c_R}{\partial K_d} \label{eq:dirty_capital_rent}
\end{align}

\subsection{Adaptive Network Model for Investment Decision Making}
\label{sec:investment_decision_making_descr.}

We model households as boundedly rational decision makers \citep{simon1972theories, simon1982models, gigerenzer2002bounded}:
Households take their investment decisions, i.e., whether to invest their savings in the clean or the dirty sector, not by forming rational expectations \citep{Evans2006, Kirman2014} but by engaging in social learning \citep{bandura1977} to obtain successful strategies \citep{Traulsen2010} with reasonable effort.
The outcomes of social learning crucially depend on the structural properties of the complex network of social ties amongst the households \citep{Barkoczi2016}.
The strong and still increasing polarization of some societies on climate change issues suggests that social dynamics reinforce opposed positions in the population \citep{Fisher2013, Farrell2016a, Dunlap2016, McCright2011, Hart2012, Williams2015}. In static network models, such effects cannot be represented.
Therefore, we model the adaptive formation of the social network endogenously.
A well established principle for the emergence of structured ties in social networks is homophily, i.e., the tendency that similar individuals get linked \citep{McPherson2007, Centola2007HomophilyGroups, Centola2011}.
The following model specification uses social learning in combination with endogenous network formation based on homophily to model the investment decisions of the households.

We model $N$ heterogeneous households denoted with the index $i$ as owners of one unit of labor $L^{(i)} = L/N$ and capital $K_c^{(i)}$ and $K_d^{(i)}$ in the clean and dirty economic sector respectively.
Households generate an income $I^{(i)}$ from their labor and capital income which they use for consumption $F^{(i)}$ and savings $S^{(i)}$. The rate at which households save their income is assumed to be fixed and is given by the savings rate $s$:
\begin{align}
	I^{(i)} &= w L^{(i)} + r_c K_c^{(i)} + r_d K_d^{(i)}, \label{eq:household_income} \\
	F^{(i)} &= (1-s) I^{(i)}, \label{eq:consumption} \\
	S^{(i)} &= s I^{(i)}. \label{eq:savings}
\end{align}
A binary decision parameter $o_i \in [c,d]$ denotes the sector in which the households decide to invest. As motivated above, we model decision making that is driven by two processes: social learning via the imitation of successful strategies and homophily towards individuals exhibiting the same behavior. \\
We describe households as the nodes in a graph of acquaintance relations that change according to the following rules. 
\begin{enumerate}
  \item Households get active at a constant rate $1/\tau$. \label{r1}
        \item When a household $i$ becomes active, it interacts with one of its acquaintances $j$ chosen uniformly at random. 
        \item If they follow the same strategy, i.e., they invest in the same sector, nothing happens. 
        \item If they follow a different strategy, i.e., they invest in different sectors, one of two actions can happen:
        \begin{enumerate}
                \item Homophilic network adaptation: with probability $\varphi$, the households end their relation and household $i$ connects to another household $k$, that follows the same strategy. 
                \item Imitation: with probability $1-\varphi$, household $i$ engages in social learning, i.e., it imitates the strategy of household $j$ with a probability $p_{ji}$ that increases with their difference in income. \label{rn}
        \end{enumerate}
\end{enumerate}
We follow previous results on human strategy updating in repeated interactions from \cite{Traulsen2010}, when we assume the imitation probability as a monotonously increasing sigmoidal function of the relative difference in consumption between both households:
\begin{equation}
	p_{ji} =  \left(1 + \exp \left(- \frac{a(F^{(i)} - F^{(j)})}{F^{(i)} + F^{(j)}} \right) \right)^{-1}.
    \label{eq:imitation_probability}
\end{equation}
As opposed to the absolute difference in the original study by \cite{Traulsen2010}, the probability in our model depends on relative differences. 
We set $a = 8$ to conform to their empirical evidence. This dependence on relative differences in per household quantities is crucial for our method as we will discuss later at the end of Sec. \ref{sec:large_system_limit}.
We model strategy exploration as a fraction $\varepsilon$ of events that are random, e.g., rewiring to a random other household or randomly investing in one of the two sectors.
Given the savings decisions of the individual households, and assuming equal capital depreciation rates $\kappa$ in both sectors, the time development of their capital holdings is given by

\begin{align}
	\dot{K}_c^{(i)} =& \delta_{o_ic} s \left( r_c K_c^{(i)} + r_d K_d^{(i)} + w L_i \right) - \kappa K_c^{(i)}, \label{eq:clean_investment}\\
	\dot{K}_d^{(i)} =& \delta_{o_id} s \left( r_c K_c^{(i)} + r_d K_d^{(i)} + w L_i \right) - \kappa K_d^{(i)}, \label{eq:dirty_investment}
\end{align}
where $\delta_{ij}$ is the Kronecker Delta. The total capital stocks in the two sectors are made up of the sum of the individual capital stocks
\begin{equation}
K_j = \sum_i^N K_j^{(i)} = N k_j,
\end{equation}
where $k_j$ is the average per household capital stock of a given capital type.

We acknowledge the fact that different model specifications are possible and interesting.
For instance, we only consider fixed savings rates and the decision between two capital assets and leave the analysis of the interesting possible effects of households setting their savings rates individually to another study \citep{Asano2019}.
However, we want to point out that the approximation methods that we develop in the following are highly useful to gain insights from different but similar models that rely on complex adaptive interaction networks.

\begin{table}
	\centering \ra{1.3}
	\begin{tabular}{lL{5.5cm}}
	    \toprule
	    Symbol & Variable description \\
    	\midrule
		$Y_c$, $Y_d$ & clean and dirty production (flows) \\
		$L_c$, $L_d$ & labor employed in the clean and dirty sector \\
		$K_c$, $K_d$ & physical capital stocks of the clean and dirty sector \\
		$C$ & clean technology \\
		$R$ & fossil resource use (flow) \\
		$G$ & fossil resource stock \\
		$c_R$ & resource extraction cost \\
		$w$ & equilibrium wage \\
		$r_c$, $r_d$ & equilibrium capital rents in the clean and dirty sector \\
		$I^{(i)}$, $F^{(i)}$, $S^{(i)}$ & income, consumption expenses and savings of individual $i$  (flows)\\
		$K_c^{(i)}$, $K_d^{(i)}$ & individual capital stocks in the clean and dirty sector \\
        \bottomrule
	\end{tabular}
	\caption{List of variables in the agent-based model. All variables are without units of measurement.}
	\label{tab:variable_list}
\end{table}

\begin{table}[h]
    \centering \ra{1.3}
	\begin{tabular}{llL{5.5cm}}
	    \toprule
	    Symbol & Value & Parameter description \\
    	\midrule
    	$N$  & 200  & Number of households \\
    	$M$  & 2000  & Number of network linkes between the households \\
		$b_c$ & 1. & Total factor productivity in the clean sector \\
		$b_d$ & 4. & Total factor productivity in the dirty sector \\
		$b_R$ & .1 & Initial resource extraction cost \\
		$e$   & 1 & Resource conversion efficiency \\
		$\kappa$   & 0.06 & Capital depreciation rate \\
		$\chi$      & 0.1 & Knowledge depreciation rate \\
		$\gamma$	   & 0.1 & Elasticity of knowledge in the clean sector \\
		$\alpha_c$ & 0.5 & Elasticity of labor in the clean sector \\
		$\alpha_d$ & 0.5 & Elasticity of labor in the dirty sector \\
		$\beta_c$ & 0.5 & Elasticity of capital in the clean sector \\
		$\beta_d$ & 0.5 & Elasticity of capital in the dirty sector \\
		$\varphi$ & 0.5 & Fraction of rewiring events in opinion formation \\
		$1/\tau$ & 1. & Rate of opinion formation events \\ 
		$\varepsilon$ & 0.05 & Fraction of noise events in opinion formation \\ 
        $G_0$ & 1000000 & Initial resource stock \\
        $L$ & 100 & Total labor \\
        $s$ & 0.25 & Savings rate \\
        $\rho$ & 1  & exponent for resource flow in extraction cost \\
        $\mu$  & 2  & exponent for resource stock in extraction cost \\
        \bottomrule
	\end{tabular}
	\caption{List of model parameters with their default values. Note that the parameter values are set to mirror plausible values observed in real-world economies but are not the result of a detailed model estimation procedure.}
	\label{tab:Parameter_list}
\end{table}

\subsection{Numerical Modelling and Results}
\label{sec:numerical_results}
With the model specifications from Sec.~\ref{sec:Model_Description}, the parametrization in Tab.~\ref{tab:Parameter_list} and appropriate initial conditions for the dynamic variables, the model can be simulated numerically.
For this, we implemented the dynamics in the multi-purpose programming language Python. The implementation of the ABM as well as the numerical analysis using the approximation methods described in the following are available on the github software versioning service in \cite{kolb2018}.
In the following, we discuss the resulting aggregate dynamics.

\begin{figure*}[ht]
  \centering\includegraphics[width=.8\linewidth]{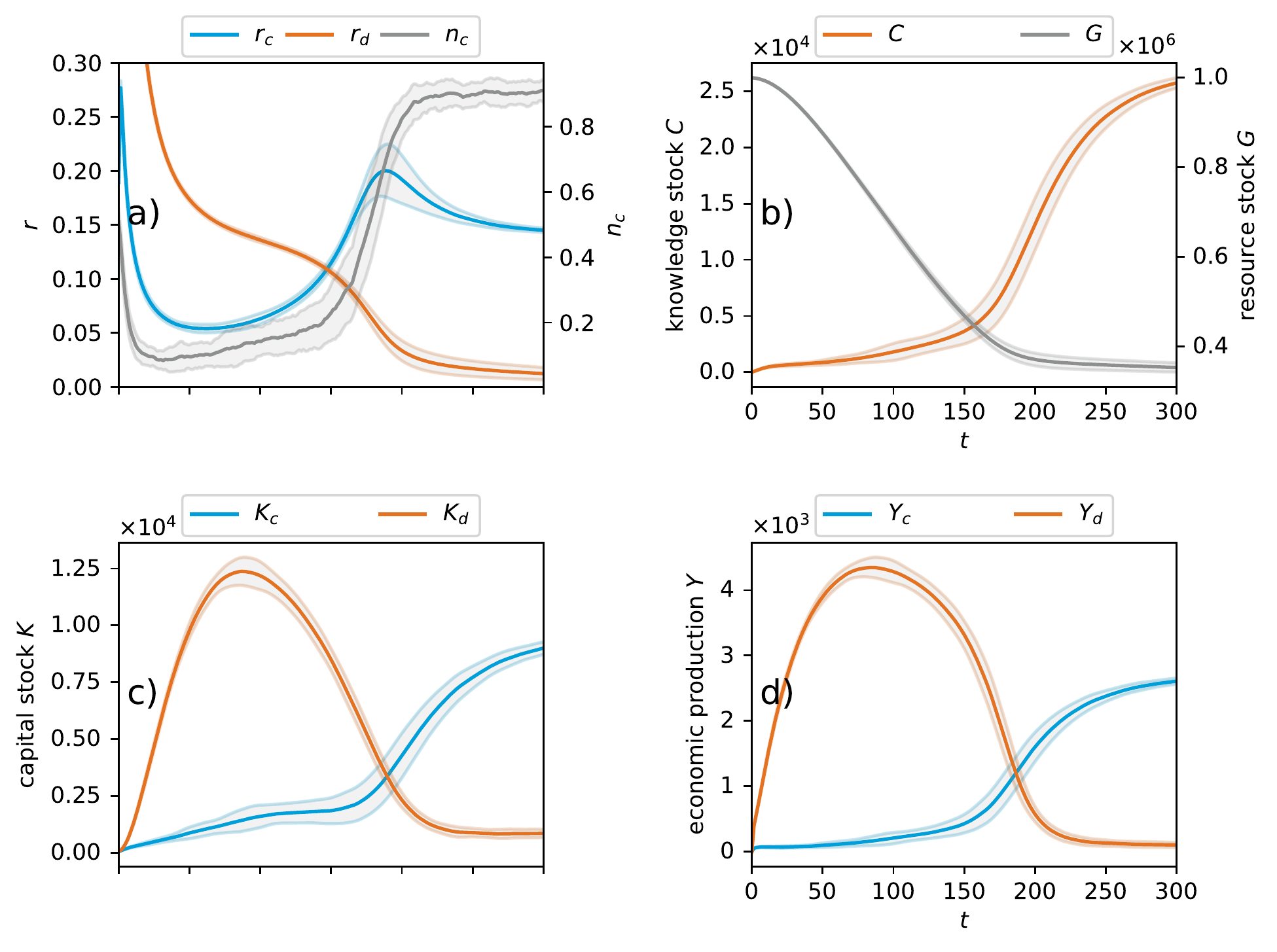}
  \caption{\textbf{Example trajectory of the ABM.} Solid lines show mean results from 100 runs of the model. Grey areas around solid lines show their standard deviation. The panels show capital rents in the clean and dirty sector $r_c$ and $r_d$ as well as the fraction of households investing in the clean sector $n_c$ in panel (a), knowledge and resource stock $C$ and $G$ in panel (b), output of clean and dirty sector $Y_c$ and $Y_d$ in panel (c) and capital stocks $K_c$ and $K_d$ in the clean and dirty sector in (d).
Initial conditions are $G=G_0$, $C=1$, $K_j^{(i)}=1$ for the economic subsystem. For the investment decision process, the initial opinions of the $N=200$ households are drawn from a uniform distribution. Their initial acquaintance structure is an Erd\H{o}s-Renyi random graph with mean degree $k=10$.}
\label{fig:example_trajectory}
\end{figure*}

Figure \ref{fig:example_trajectory} displays an exemplary average evolution of our model calculated as the mean of 100 simulation runs.
The simulation starts with initial conditions of abundant fossil resources $G$ and low clean technology knowledge stock $C$ (panel b) as well as equally low capital stocks in the clean and dirty sector $K_c$ and $K_d$ (panel c). As we show later (see Sec.~\ref{sec:bifurcation-analysis}), the rest of the initial configuration of the model is rather irrelevant for the selected parameter values listed in Tab.~\ref{tab:Parameter_list}, since there is only one stable dynamical equilibrium as long as resource extraction costs are negligibly low.
The high initial capital rents $r_c$ and $r_d$ are a direct result of our model assumptions and initial conditions, more precisely, the assumption that capital rent equals marginal productivity in Eq.~\ref{eq:clean_capital_rent} and \ref{eq:dirty_capital_rent} and that of decreasing marginal productivity due to our choice of $\beta_i$ in combination with the initial condition of low capital and a fixed labor supply. Also as a direct consequence of these assumptions, the capital rents $r_c$ and $r_d$ decrease over time as the capital stock is built up.
Initially (from $t=0$ to $t=100$), as a result of our choice of total factor productivities $b_i$ and due to low fossil resource extraction costs, capital productivity (and therefore capital rent $r$) is higher in the dirty sector than the clean sector (see panel a). 
Consequently, the majority of households invest in the dirty sector which leads to a high capital stock $K_d$ (panel c) and high production output $Y_d$ (panel d) in this sector.

Regarding the capital rents, we would expect the system to move towards a dynamic equilibrium in which the capital rent is equal in both sectors, i.e., $r_d = r_c$, if everything else remained constant. However, we find that there is a persisting difference between $r_c$ and $r_d$ between $t=50$ and $t=100$.
This difference can be explained by the exploration of investment strategies even if they perform worse, which brings the shares of clean and dirty investors closer together. In terms of the depicted variables this means that it brings $n_c$ closer to $0.5$. 

For $t>100$ the depletion of the fossil resource leads to significantly increasing resource extraction costs. Consequently, the marginal productivity of dirty capital $K_d$ decreases and so does $r_d$, leading to a peak in accumulation of capital in the dirty sector around $t=100$ (panel c).
Once the relative return on capital in the clean sector increases, households start to adopt a clean investment strategy visible in an increase in $n_c$ in panel a.
When the fossil resource stock reaches its economically exploitable share at around $t=200$, the overall productivity in the dirty sector reaches zero, leading to full employment of all available labor in the clean sector.
This drives demand for capital in the clean sector up, accelerating the change from dirty to clean investment.
As all households except for the share caused by exploration are investing in the clean sector, the system reaches an equilibrium with high capital in the clean sector and low capital in the dirty sector.

Notably, we find an increasing variance in the fraction of households investing in the clean sector before and around the transition, which means that due to the stochasticity of the social learning process the transition happens earlier for some simulation runs than for others. Nevertheless, we find that the inertia of the model resulting from the large accumulated stock of capital that is specific to the dirty sector eventually leads to an almost entire depletion of the fossil resource.

The adaptation dynamics in our model can lead to a fragmentation of the network with stark economic consequences. As results in Appendix~\ref{app:rewiring} show, an increased rewiring rate $\varphi$ in the network adaptation process leads to a strongly delayed shift of investment from one sector to the other during the transition, even though the incentive in terms of an increased return $r_c$ for the investment in this sector is high. This fragmentation is equivalent with a strong decline in the fraction of active edges in the network, e.g. the fraction of edges that connect households investing in different sectors of the economy.
This finding is consistent with a major result of adaptive network modeling studies that show that adaptation will lead to fragmentation of a network at high rewiring rates $\varphi$ \citet{Gross2006,Do2009,Bohme2011,Min2017}. Such network properties emerging from adaptation dynamics have been studied for example in the context of opinion dynamics, epidemics and social-ecological systems \citep{Gross2008,Rogers2012,Rogers2013,Wiedermann2015}. 
 One could suspect that the slow-down in the transition from one sector to the other results from the decreased rate of imitation events as their frequency scales with $1-\varphi$. However, the results in Appendix~\ref{app:fully_connected} show that this effect is particular to the adaptive network model and cannot be reproduced in a well-mixed system simply by adjusting for the reduced frequency of imitation events. Appendices~\ref{app:rewiring} and \ref{app:fully_connected} discuss further differences between the full model and special cases without adaptation as well as well-mixed interaction. 

\section{Approximate Analytical Solution}
\label{sec:Approximation}

Structurally, the model described in Section \ref{sec:Model_Description} consists of a set of coupled ordinary differential equations \eqref{eq:learning_by_doing}, \eqref{eq:resource_depletion}, \eqref{eq:clean_investment} and \eqref{eq:dirty_investment} with algebraic constraints \eqref{eq:efficient_dirty_resources}, \eqref{eq:equilibrium_wage}, \eqref{eq:population}, \eqref{eq:clean_capital_rent} and \eqref{eq:dirty_capital_rent} for the economic production process and a stochastic adaptive network process for the social learning component that is described by the rules \ref{r1} to \ref{rn} in Section \ref{sec:investment_decision_making_descr.}. The state space of this combined process consists of two degrees of freedom of the knowledge stock and the geological resource stock as well as $2N$ degrees of freedom for the capital holdings of the set of all individual households plus the configuration space of the adaptive network process of the social learning component. We denote the variables of this process by capital letters ($C, G, K_j^{(i)}\dots$).
To find an analytic description of the model in terms of a low dimensional system of ordinary differential equations, we approximate it via a Pair Based Proxy (PBP) process, a stochastic process in terms of aggregated quantities, thereby drastically reducing the dimensionality of the state space. We denote the variables of this process with capital letter with bars ($\bar{X}$, $\bar{Y}$, $\bar{Z}$, $\bar{K}_l^{(k)}\dots$).

The derivation of this approximate process is done in three steps: First, we solve the algebraic constraints to the economic production process given by market clearing in the labor market and efficient production in the dirty sector -- loosely following \cite{Nitzbon2017}. Second we use a pair approximation to describe the complex adaptive network process of social learning in terms of aggregated variables, similar to \cite{Rogers2012}. Third, we use a moment-closure method to approximate higher moments of the distribution of the capital holdings of the heterogeneous households by quantities related to the first moments of their distribution.

Finally, we take the limit of infinitely many households (large system- or thermodynamic limit) to obtain a deterministic description of the system.

\subsection{Algebraic Constraints}

To calculate labor $L_c$ and $L_d$ as well as wages in the two sectors, we use equations \eqref{eq:resource_cost} and \eqref{eq:equilibrium_wage} and for simplicity assume $\rho=1$ and $\mu=2$. We also assume equal labor elasticities in both sectors $\alpha_d = \alpha_c = \alpha$ resulting in
\begin{align}
	w &= \frac{\partial Y_d}{\partial L_d} - \frac{\partial c_R}{\partial L_d} \nonumber \\
	&= \frac{\partial Y_d}{\partial L_d} - \frac{\partial c_R}{\partial R} \frac{\partial R}{\partial L_d} \nonumber = \frac{\partial Y_d}{\partial L_d} - \frac{\partial c_R}{\partial R} \frac{\partial}{\partial L_d} \frac{Y_d}{e} \nonumber \\
	&= \frac{\partial Y_d}{\partial L_d} - b_R\frac{G_0^2}{G^2} \frac{\partial}{\partial L_d} \frac{Y_d}{e} = b_d \alpha L_d^{\alpha-1} K_d^{\beta_d}\left( 1-\frac{b_R}{e}\frac{G_0^2}{G^2} \right)
	\label{eq:dirty_wages}
\end{align}
for the dirty sector and
\begin{equation}
	w = b_c \alpha L_c^{\alpha-1} K_c^{\beta_c} C^{\gamma}
	\label{eq:clean_wages}
\end{equation}
for the clean sector. Combining these results via equation \eqref{eq:population} and substituting
\begin{align}
	&X_c = (b_c K_c^{\beta_c}C^{\gamma})^{\frac{1}{1-\alpha}}, \qquad X_d = (b_d K_d^{\beta_d})^{\frac{1}{1-\alpha}}, \nonumber \\
	&X_R = \left( 1 - \frac{b_R}{e}\frac{G_0^2}{G^2} \right)^{\frac{1}{1-\alpha}}
	\label{eq:substitutions}
\end{align}
and solving for $w$ yields:
\begin{equation}
	w = \alpha L^{\alpha-1}\left( X_c + X_d X_R \right)^{1-\alpha}.
	\label{eq:wage_result}
\end{equation}
Plugging \eqref{eq:wage_result} into equations \eqref{eq:dirty_wages} and \eqref{eq:clean_wages} results in 
\begin{align}
	L_c &= L \frac{X_c}{X_c + X_d X_R}, \label{eq:clean_labor} \\
	L_d &= L \frac{X_d X_R}{X_c + X_d X_R} \label{eq:dirty_labor}
\end{align}
for labor in the two sectors, and plugging this into \eqref{eq:efficient_dirty_resources} leads to
\begin{equation}
	R = \frac{b_d}{e}K_d^{\beta_d}L^{\alpha}\left( \frac{X_d X_R}{X_c + X_d X_R} \right)^{\alpha}
	\label{eq:R_result}
\end{equation}
for the use of the fossil resource. Using the results for $L_c$ and $L_d$ together with equations \eqref{eq:clean_capital_rent} and \eqref{eq:dirty_capital_rent}, the return rates on capital result in
\begin{align}
	r_c &= \frac{\beta_c}{K_c}X_c L^{\alpha}\left( X_c + X_d X_R \right)^{-\alpha}, \label{eq:r_c_result}\\
	r_d &= \frac{\beta_d}{K_d}\left(X_d X_R\right) L^{\alpha}\left( X_c + X_d X_R \right)^{-\alpha}. \label{eq:r_d_result}
\end{align}

It is also worth noting that if we assume constant returns to scale with respect to capital and labor, e.g.,

\begin{equation}
	\beta_c = \beta_d = 1-\alpha,
	\label{eq:elasticities_restriction}
\end{equation}
(even though it is not necessary for our method) this yields zero profits in both sectors:
\begin{align}
	Y_c &= w L_c + r_c K_c, \nonumber \\
	Y_d &= w L_d + r_d K_d + c_R. \nonumber
\end{align}

To sum up, we solved the algebraic constraints to the ordinary differential equations describing the economic production process resulting in the following equations:
\begin{subequations}
\begin{align}
	X_c =& (b_c K_c^{\beta_c}C^{\gamma})^{\frac{1}{1-\alpha}}, \quad X_d = (b_d K_d^{\beta_d})^{\frac{1}{1-\alpha}}, \nonumber \\ 
	X_R =& \left( 1 - \frac{b_R}{e}\frac{G_0^2}{G^2} \right)^{\frac{1}{1-\alpha}}, \\
	w =& \alpha L^{\alpha-1}\left( X_c + X_d X_R \right)^{1-\alpha}, \label{eq:equilibrium_wage_solution}\\
	r_c =& \frac{\beta_c}{K_c}X_c L^{\alpha}\left( X_c + X_d X_R \right)^{-\alpha}, \\
	r_d =& \frac{\beta_d}{K_d}X_d X_R L^{\alpha}\left( X_c + X_d X_R \right)^{-\alpha}, \\
	R =& \frac{b_d}{e}K_d^{\beta_d}L^{\alpha}\left( \frac{X_d X_R}{X_c + X_d X_R} \right)^{\alpha}, \\
	\dot{G} =& - R \label{eq:sum_up_resource_depletion}, \\
	\dot{K}_c^{(i)} =& s \delta_{o_i,c} (r_c K_c^{(i)} + r_d K_d^{(i)} + w L^{(i)}) \nonumber \\
	&- \kappa K_c^{(i)}, \label{eq:sum_up_clean_capital_accumulation} \\
	\dot{K}_d^{(i)} = &s \delta_{o_i,d} (r_c K_c^{(i)} + r_d K_d^{(i)} + w L^{(i)}) \nonumber \\
	&- \kappa K_d^{(i)}, \label{eq:sum_up_dirty_capital_accumulation} \\
    \dot{C} = &Y_c- \chi C.\label{eq:sum_up_learning}
\end{align}
\end{subequations}

\subsection{Pair Approximation}
\label{sec:pair_approximation}
To derive a macroscopic approximation of the social learning process described by rules \ref{r1} to \ref{rn} in Sec. \ref{sec:investment_decision_making_descr.}, we make use of a Pair based proxy (PBP) process that is derived via pair approximation from the adaptive network process. This proxy process is not equivalent but sufficiently close to the microscopic process approximating it in terms of aggregated quantities by making certain assumptions about the properties of their microscopic structure. The aggregated quantities of interest are: the number of households investing in clean capital $N^{(c)}$, the number of households investing in dirty capital $N^{(d)}$, the number of links between agents of the same group $[cc]$ and $[dd]$ as well as between the two groups $[cd]$. Since the total number of households $N$ and links $M$ are fixed, these five variables reduce to three degrees of freedom, which we parameterize as follows:

\begin{equation}
	\bar{X} = N^{(c)} - N^{(d)}, \quad \bar{Y} = [cc] - [dd], \quad \bar{Z} = [cd].
	\label{eq:opinion_formation_macro_variables}
\end{equation}

These three degrees of freedom span the reduced state space of the social process $\mathbf{\bar{S}} = (\bar{X}, \bar{Y}, \bar{Z})^T$. The investment decision making process can then be described in terms of jump lengths $\Delta \mathbf{\bar{S}}_j$ and jump rates $W(\mathbf{\bar{S}},\mathbf{\bar{S}} + \Delta \mathbf{\bar{S}}_j)$ in this state space for the different events $j$ in the set $\Omega$ of all possible events.
Their derivation is illustrated by the example of a clean household imitating a dirty household: The approximate rate of this event is given by
\begin{equation}
	W_{c \rightarrow d} = \frac{N}{\tau} (1-\varepsilon) (1 - \varphi) \frac{N^{(c)}}{N}\frac{[cd]}{[cd] + 2 [cc]}p_{cd}.
	\label{eq:cdswitchingprob}
\end{equation}
In some more detail this results from
\begin{itemize}
	\item $N/\tau$ the rate of social update events, i.e., the rate of events per household times the number of households,
	\item $(1-\varepsilon)$ the probability of the event not being a noise event,
	\item $(1-\varphi)$ the probability of imitation events (versus network adaptation events),
	\item $N^{(c)}/N$ the probability of each active household to invest in clean capital,
	\item $[cd]/(2[cc] + [cd])$ the approximate probability of interaction with a household investing in dirty capital. Here, we approximate the distribution of dirty neighbors among clean households with its first moment i.e., we act as if links between clean and dirty households were evenly distributed among all households. 
	\item $p_{cd}$ is the expected value of the probability of each active household imitating its randomly chosen neighbor depending on the difference in consumption between households investing in clean and dirty capital as given in equation \eqref{eq:imitation_probability}. The expression is derived in detail as part of the moment closure in subsection \ref{moment_closure}.
\end{itemize}
The corresponding change in the state space variables is a little more tricky. Since the event is a clean household imitating a dirty household, we already know about one of the neighbors of the household. As laid out in detail by e.g.~\citep{Do2009}, the state of the remaining neighbors in the full model is determined by the frequency of higher order network motifs, e.g., [dcd] and [dcc]. The frequency of these higher order motifs is approximated by the expected value of the states of additional neighbors as follows: Summing over the excess degree of the node $q^{c}$ by drawing $k^{c} - 1$ times from the distribution of neighbors which is, as before, approximated by an even distribution of edges between same and different households among all households. Again, this approximates the respective full distributions with their first moments. If one wanted to include higher-order effects in the network dynamics, one could follow one of the various ways laid out by, e.g., \cite{Demirel2014}. Thus the probability for a neighbor to be dirty, $p^{(d)}$, or clean, $p^{(c)}$, reads:
\begin{equation}
	p^{(c)} = \frac{2 [cc]}{2[cc] + [cd]}; \qquad p^{(d)} = \frac{[cd]}{2[cc] + [cd]}.
\label{eq:neighbordist}
\end{equation}
This results in an expected number of $n^{(c)}$ additional clean neighbors and $n^{(d)}$ additional dirty neighbors:
\begin{equation}
	n^{(c)} = (1-1/k^{(c)})\frac{2[cc]}{N^{(c)}}, \quad n^{(d)} = (1-1/k^{(c)})\frac{[cd]}{N^{(c)}},
	\label{eq:additional_neighbors}
\end{equation}
where $k^{(c)}$ is the mean degree, e.g., the mean number of neighbors of a clean household in the network. 
With the results from \eqref{eq:additional_neighbors} the changes in the expected values of the state space variables can be approximated as follows:
\begin{align}
	\Delta N^{(c)} &= -1, \nonumber \\
	\Delta N^{(d)} &= 1, \nonumber \\
	\Delta [cc] & \approx \left( 1 - \frac{1}{k^{(c)}} \right)\frac{2[cc]}{N^{(c)}}, \nonumber \\
	\Delta [dd] & \approx \left( 1 - \frac{1}{k^{(c)}} \right)\frac{[cd]}{N^{(c)}}, \nonumber \\
	\Delta [cd] & \approx -1 + \left( 1 - \frac{1}{k^{(c)}} \right)\frac{2[cc] - [cd]}{N^{(c)}}, \nonumber
\end{align}
and, summing up, the change in the state vector is approximately given by:
\begin{equation}
	\Delta \mathbf{\bar{S}}_{c \rightarrow d} \approx \colvec{3}{-2}{-k^{(c)}}{-1 +  \left( 1 - \frac{1}{k^{(c)}} \right)\frac{2[cc] - [cd]}{N^{(c)}} }.
	\label{cdstatespacechange}
\end{equation}

In terms of the jump lengths $\Delta \mathbf{\bar{S}}$ and the rates $W$, the dynamics of the PBP can be written as a master equation for the probability distribution $P$ on the state space of $\mathbf{\bar{S}}$:

\begin{align}
	\frac{{\partial} P(\mathbf{\bar{S}}, t)}{\partial t} = \sum_{j \in \Omega} &P(\mathbf{\bar{S}} - \Delta \mathbf{\bar{S}}_j, t) W(\mathbf{\bar{S}} - \Delta \mathbf{\bar{S}}_j,\mathbf{\bar{S}}) \nonumber \\
	&- P(\mathbf{\bar{S}}, t) W(\mathbf{\bar{S}},\mathbf{\bar{S}} + \Delta \mathbf{\bar{S}}_j). \label{eq:PBP}
\end{align}

\subsection{Moment Closure}
\label{moment_closure}

To describe the capital structure in the model that consists of $2N$ equations of type \eqref{eq:clean_investment} and \eqref{eq:dirty_investment}, we use the cohort of $N^{(c)}$ households investing in clean and the cohort of $N^{(d)}$ households investing in dirty capital and look at the aggregates of their respective capital holdings:
\begin{align}
  \bar{K}_l^{(k)} = \sum_{i}^{N} \delta_{o_ik} K_l^{(i)}.
	\label{eq:moments_definition}
\end{align}
Here, the upper index in $\bar{K}_l^{(k)}$ indicates the shared investment decision of the cohort of households as opposed to the index of the individual household before. The lower index still denotes the capital type. $\delta_{o_ik}$ is the Kronecker Delta.

Later, we use the fact that in the limit of $N \rightarrow \infty$ these aggregates should converge to their expected values, e.g., the first moments of their distribution with probability one.
The time derivative of the aggregates defined in \eqref{eq:moments_definition} is given by the deterministic process of capital accumulation \eqref{eq:sum_up_clean_capital_accumulation} and \eqref{eq:sum_up_dirty_capital_accumulation} as well as terms resulting from the stochastic process of agents switching their saving decisions. 
\begin{equation}
      \begin{aligned}
          \dot{\bar{K}}_c^{(c)} =&  \\
          \dot{\bar{K}}_d^{(c)} =&  \\
          \dot{\bar{K}}_c^{(d)} =&  \\
          \dot{\bar{K}}_d^{(d)} =& 
      \end{aligned}
  \underbrace{ 
      \begin{aligned}
      &(sr_c - \alpha)\bar{K}_c^{(c)} + s r_d \bar{K}_d^{(c)} + s w \bar{L} \\
      &- \alpha\bar{K}_d^{(c)} \\
      &- \alpha\bar{K}_c^{(d)} \\
      &sr_c \bar{K}_c^{(d)} + (s r_d - \alpha)\bar{K}_d^{(d)} + s w \bar{L}
      \end{aligned}
  }_{\textstyle D^{(i)}_{l} } ~ + \rm{switching~terms.} \label{eq:sterm0}
\end{equation}
The switching terms for $\bar{K}_c^{(c)}$ result from agents changing their saving decision, thereby moving their capital endowments from the aggregate capital of the cohort of clean investors to the aggregate of the cohort of dirty investors and vice versa. We assume that each household switching to the other cohort is endowed with the mean capital of the cohort and that their capital endowment is independent of the probability of switching such that we can describe the switching terms as a product of both factors. Then, we can write down the changes in capital stocks explicitly including the switching terms as a simple stochastic differential equation:
\begin{equation}
	{\rm d}\bar{K}_{l}^{(k)} = D^{(k)}_{l} {\rm d}t + \underbrace{\frac{\bar{K}_l^{(j)}}{N^{(j)}} {\rm d} N^{j \rightarrow k} -  \frac{\bar{K}_l^{(k)}}{N^{(k)}} {\rm d} N^{k \rightarrow j}, }_{\text{switching terms}}
	\label{eq:aggregated_capital_time_derivative}
\end{equation}
where the first term of the right-hand side refers to the change in aggregates without switching, as given by the equations of capital accumulation \eqref{eq:sterm0} and the following terms denote the influx and outflux of capital from the aggregate due to households changing their savings decisions.
${\rm d} N^{j \rightarrow k}$ denotes the stochastic process of households switching from one opinion to another according to the rules outlined in \ref{sec:investment_decision_making_descr.}. In line with the pair approximation described in \ref{sec:pair_approximation} we approximate them as
\begin{equation}
{\rm d} N^{j \rightarrow k} = \sum_{l \in \Omega_{j \rightarrow k}}W_l {\rm d}t
\end{equation}
where $\Omega_{j \rightarrow k}$ denotes the set of all events that result in a household changing from cohort $j$ to cohort $k$ and $W_l$ is the rate of the respective event analogously to \eqref{eq:cdswitchingprob}.

The imitation probability $p_{cd}$ in Eq.~\eqref{eq:cdswitchingprob} is approximated as the expected value of a linearized version of Eq.~\eqref{eq:imitation_probability} when drawing a pair of neighboring households $i$, $j$ as specified. More precisely, we perform a Taylor expansion of Eq.~\eqref{eq:imitation_probability} in terms of the consumption of the two interacting households $F^{(c)}$ and $F^{(d)}$ around some fixed values $F^{(c)*}$ and $F^{(d)*}$ up to linear order. To maintain the symmetry of the imitation probabilities with respect to the household incomes, we change variables to $\Delta F = F^{(c)} - F^{(d)}$ and $F = F^{(c)} + F^{(d)}$ and expand around $\Delta F = 0, F = F_0$, where $F_0$ is yet to be fixed to a value. In linear order this results in:
\begin{align}
	p_{cd} &= \frac{1}{2} - \frac{a}{4 F_0} \Delta F, \label{eq:approx_p_cd}\\
	p_{dc} &= \frac{1}{2} + \frac{a}{4 F_0} \Delta F. \label{eq:approx_p_dc}
\end{align}

To make the approximation work in the biggest part of the systems state space, we set the reference point $F_0$ to be the middle of the sum of the estimated upper and lower bounds for the attainable income of households investing in the clean, resp. dirty sector. The minimum attainable income is assumed to be zero. The maximum attainable income for a household investing in the clean sector is assumed to be reached in equilibrium given all other households also invest in the clean sector, e.g., we calculate $F^{(c)*}$ as half of an average household income at the steady state of $\dot{K}_c = s b_c L^\alpha K_c^{\beta_c} C^\gamma - \delta K_c$ and $\dot{C} = b_c L^\alpha K_c^{\beta_c} C^\gamma - \delta C$:
\begin{equation}
	C^* = \left( \frac{b_c L^\alpha s^{\beta_c}}{\delta}\right)^{\frac{1}{1-\beta_c-\gamma}}, \quad K_c^* = \left( \frac{b_c L^\alpha s^{1-\gamma}}{\delta}\right)^{\frac{1}{1-\beta_c-\gamma}}.
	\label{eq:clean_steady_state}
\end{equation}
Equivalently, we calculate $F^{(d)*}$ as half of an average household income at the steady state of $ \dot{K}_d = s \left(1 - \frac{b_R}{e} \right) b_d K_d^{\beta_d} P^{\alpha} - \delta K_d $:
\begin{equation}
	K_d^* = \left( \frac{s b_d L^\alpha}{\delta} \left(1 - \frac{b_R}{e} \right)\right)^{\left(\frac{1}{1 - \beta_d} \right)}.
	\label{eq:dirty_steady_state}
\end{equation}
With these results, using the fact that we set $\beta_c = \beta_d = \alpha = 1/2$, the reference point $F_0$ is
\begin{align}
	F_0 &= \frac{1}{2}\left(F^{(c)*} + F^{(d)*}  \right) \nonumber \\
	&= \frac{1-s}{2N}\left(r_c^* K_c^* + w L + r_d^* K_d^* + w L\right) \label{eq:inc_2}\\
	&= \frac{1-s}{2N}\left( \left( \frac{s b_c L^{\alpha}}{\delta^{\beta_c + \gamma}} \right)^{\frac{1}{1-\beta_c - \gamma}} + \frac{s}{\delta}\left( \left( 1 - \frac{b_R}{e} \right) b_d L^{\alpha} \right)^2 \right),
\end{align}
where $r_c^*$ and $r_d^*$ in \eqref{eq:inc_2} are the capital return rates \eqref{eq:clean_capital_rent} and \eqref{eq:dirty_capital_rent} in the respective equilibria \eqref{eq:clean_steady_state} and \eqref{eq:dirty_steady_state}.

Given this linear approximation of the imitation probabilities, we approximate the consumption $F_c$ and $F_d$ of the randomly selected households $i$ and $j$ as the household consumption of the average household investing in clean and dirty capital using the aggregated variables as introduced in \eqref{eq:moments_definition}. In the large system limit, this is equivalent to taking the expected value over all households in the respective cohorts:

\begin{align}
	p_{cd} = \frac{1}{2} - \frac{a}{4 F_0} &\left(r_c\left( \bar{K}_c^{(c)} - \bar{K}_c^{(d)} \right) \right. \nonumber \\ 
	& \left. + r_d\left( \bar{K}_d^{(c)} - \bar{K}_d^{(d)} \right) + w\frac{L}{N}\left( N^{(c)} - N^{(d)} \right) \right), \label{eq:approx_p_cd_final}\\
	p_{dc} = \frac{1}{2} + \frac{a}{4 F_0} &\left(r_c\left( \bar{K}_c^{(c)} - \bar{K}_c^{(d)} \right) \right. \nonumber \\ 
	& \left. + r_d\left( \bar{K}_d^{(c)} - \bar{K}_d^{(d)} \right) + w\frac{L}{N}\left( N^{(c)} - N^{(d)} \right) \right).  \label{eq:approx_p_dc_final}
\end{align}
With this approximation, we have now reached an approximate description of the microscopic dynamics in terms of stochastic differential equations for the aggregate variables.
\subsection{Large System Limit}
\label{sec:large_system_limit}
The description of the model in terms of equations \eqref{eq:sum_up_resource_depletion}, \eqref{eq:sum_up_learning} \eqref{eq:PBP} and \eqref{eq:sterm0} poses a significant reduction of complexity, yet it is still a description in terms of a stochastic process rather than in terms of ordinary differential equations, as typically used in macroeconomic models. To further reduce it to ordinary differential equations, we do an expansion in terms of system size, which in our case is given by the number of households $N$.
Therefore, following \citet[p. 244]{VanKampen1992}, we introduce the rescaled variables
\begin{equation}
	x = \frac{X}{N}, \quad y = \frac{Y}{M}, \quad z = \frac{Z}{M}, \quad k = \frac{2M}{N}.
	\label{eq:rescalled_pbp_variables}
\end{equation}
and expand the master equation \eqref{eq:PBP} that describes the social learning process in terms of a small parameter $N^{-1}$. In the leading order, the time development of the rescaled state vector $\mathbf{s} = (x, y, z)$ is given by 
\begin{equation}
	\frac{\rm{d}}{\rm{d}t}\mathbf{s} = \alpha_{1,0}(\mathbf{s}),
	\label{macroscopic_equation}
\end{equation}
where $\alpha_{1,0}$ is the first jump moment of $W$. In terms of the rescaled variables $\mathbf{s}$, $\alpha_{1,0}$ is given by
\begin{equation}
	\alpha_{1,0}(\mathbf{s}) = \int \Delta \mathbf{s} W(s, \Delta \mathbf{s}) {\rm d} \Delta \mathbf{s},
	\label{eq:jump_moment}
\end{equation}
which in the case of discrete jumps in state space simplifies to
\begin{equation}
	\frac{\rm{d}}{\rm{d}t}\mathbf{s} = \sum_{j \in \Omega}  \Delta \mathbf{s_j} W_j,
	\label{eq:lsl_transitions}
\end{equation}
where $\Omega$ is the set of all possible (discrete) events in the opinion formation process.

As for the economic processes, we keep the aggregated quantities $(\bar{K}_i^j, C, G)$ fixed and formally go to a continuum of infinitesimally small households. As people and also households for that matter are finite entities, a continuum of households makes no sense. But practically, this can be understood as an interpretation of the heterogeneous households as a weighted sample of a very large population of heterogeneous individuals and increasing the sample size up until the point where a continuum of households is a sufficiently good approximation of reality in terms of the model. 
The only element in the approximation of the economic model that depends on per household quantities is the imitation probability \eqref{eq:imitation_probability} or rather its approximation \eqref{eq:approx_p_cd} and \eqref{eq:approx_p_dc}. Since we have chosen this to depend on relative differences in income, their dependence on the number of households $N$ cancels out and the limit of $N \rightarrow \infty$ becomes trivial resulting in the following deterministic approximation for the capital endowments in sector $l$ of households investing in sector $k$ described in Eq.~\eqref{eq:aggregated_capital_time_derivative}:

\begin{equation}
  \dot{\bar{K}}_l^{(k)} = D_l^{(k)} + \frac{\bar{K}_l^{(j)}}{N^{(j)}}\sum_{l \in \Omega_{j \rightarrow k}}W_l - \frac{\bar{K}_l^{(k)}}{N^{(k)}}\sum_{l \in \Omega_{k \rightarrow j}}W_l,
  \label{eq:lsl_capital}
\end{equation}
where $D_l^{(k)}$ are the capital accumulation terms as given in \eqref{eq:sterm0} and $\Omega_{l \rightarrow k}$ is the set of all opinion formation events, where a household changes its opinion from $l$ to $k$.

Together with equations \eqref{eq:sum_up_resource_depletion} and \eqref{eq:sum_up_learning} the sets of equations specified by \eqref{eq:lsl_transitions} and \eqref{eq:lsl_capital} fully describe the approximate dynamics of the original model as specified in Section \ref{sec:Model_Description}. The full set of equations is given in Appendix \ref{si:odes}.

Our approximation reduces the full model to a set of first order differential equations with nine degrees of freedom. For comparison, the full model has $2N + 2$ degrees of freedom in the economic system plus the configuration space of the social network component. The right-hand sides of the set of differential equations are continuously differentiable and depend on 12 parameters for the economic system and two parameters for the social network process. The state space of the system is bounded between $-1$ and $1$ in $x$ and $y$ and between $0$ and $1$ in $z$ as well as by $0$ from below in the variables of the economic system $\bar{K}_l^{(k)}$, $G$ and $C$. As the equations are bulky, it is recommended to use a computer algebra system to work with them.

The freedom to chose equations for economic production that are not scale-invariant critically depends on the assumption that household interaction only depends on relative differences. For individual interaction that depends on absolute differences, one can show that the large system limit only works if the system is scale-invariant in terms of aggregated quantities. Nevertheless, it would be possible to relax both of these assumptions and to work with the PBP process with the results explicitly depending on the number of households, which in return could lead to interesting finite size effects.

\begin{table}
	\centering \ra{1.3}
	\begin{tabular}{lL{5.5cm}}
	    \toprule
	    Symbol & Variable description \\
    	\midrule
		$[cc]$, $[dd]$, $[cd]$ & number of links between clean and dirty households \\
		$p_{kl}$ & probability that household investing in sector $l \in \{c, d\}$ imitates household investing in sector $k \in \{c, d\}$ \\
		$W$ & jump rates for the stochastic process approximating the network dynamics \\
		$\bar{\mathbf{S}}$ & jump lengths for the network dynamics \\
		$\bar{K}_l^{(k)}$ & aggregate capital investments in sector $l \in \{c, d\}$ of all households in category $k \in \{c, d\}$ \\
        \bottomrule
	\end{tabular}
	\caption{List of variables used in the macroscopic approximation}
	\label{tab:approximation_variable_list}
\end{table}

\subsection{Results of the Model Approximation}

\begin{figure*}[ht!]
\centering\includegraphics[width=.8\linewidth]{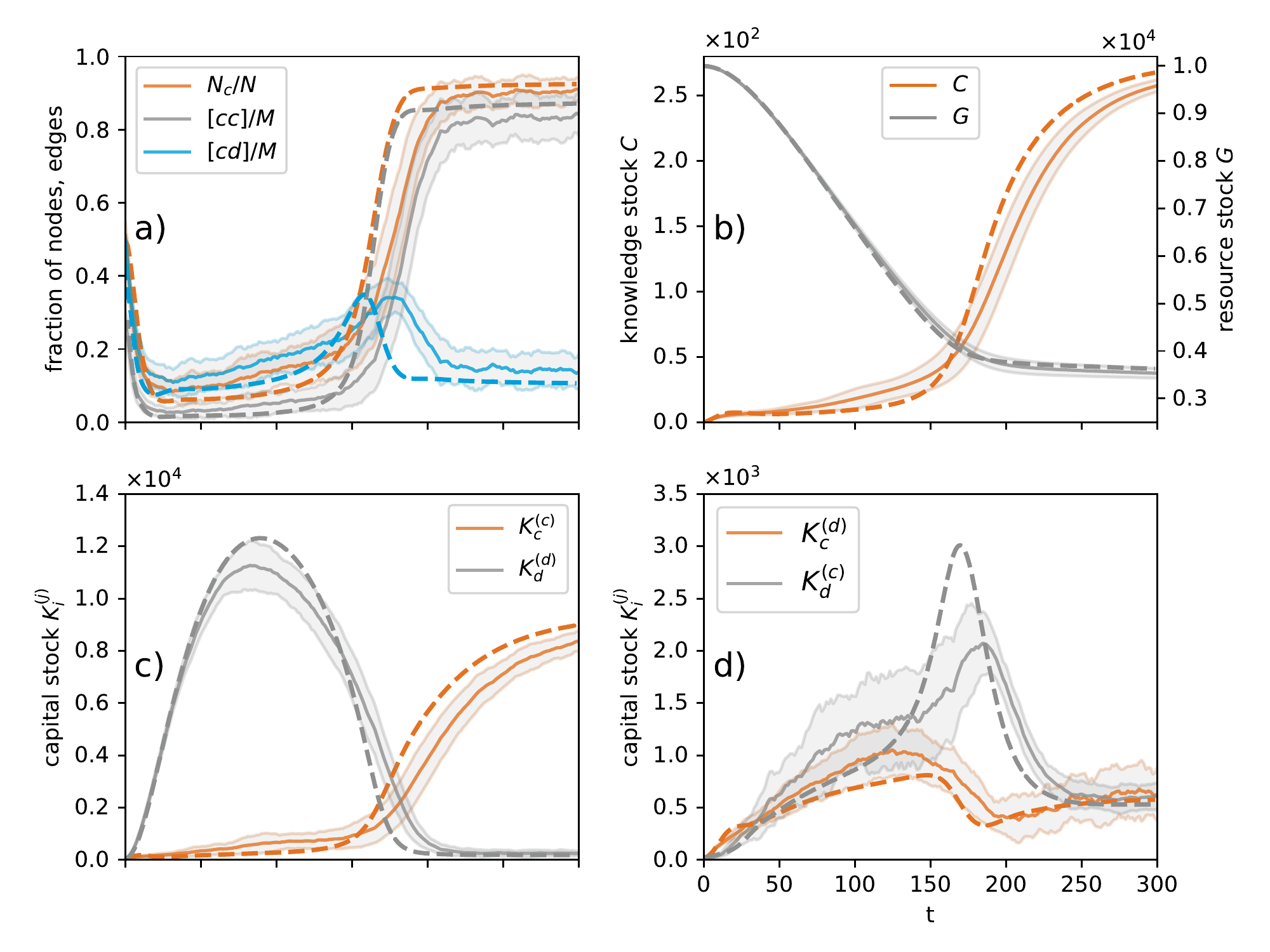}
\caption{\textbf{Trajectories of dynamic variables from the macro approximation and from measurement in ABM simulations.} The results from ABM simulations (solid lines) are obtained as an ensemble average from 50 runs with standard errors indicated by gray areas. Initial conditions are given by equal shares of the $N=200$ households investing in both sectors and equal endowments in both sectors for all households. The initial acquaintance network amongst the households is an Erd\H{o}s-Renyi random graph with mean degree $k=10$. Other initial conditions are $C_0=0.5$ and $G_0=5 \times 10^5$. All other parameter are given in table \ref{tab:Parameter_list}. The results from the macro approximation (dashed lines of the same colors) are obtained by integration of the ODEs that are obtained from the large system limit with fixed per household quantities. The initial conditions are drawn from the same distribution as previously for the ABM simulations e.g. $N_c$, $[cc]$ and $[cd]$ are calculated from an Erd\H{o}s-Renyi random graph with mean degree $k=10$.
}
\label{fig:comparison2}
\end{figure*}

The results in  Fig.~\ref{fig:comparison2} are to some extent complementary to the results in Fig.~\ref{fig:example_trajectory} that we discussed in Sec.~\ref{sec:numerical_results}. Fig.~\ref{fig:comparison2}d shows capital in both sectors belonging to households that actually invest in these sectors, which is almost equivalent to the variables in fig.~\ref{fig:example_trajectory}d as it makes up almost the entirety of these capital stocks. This can be seen in Fig.~\ref{fig:comparison2}c: It shows capital of households in the sector that they do not currently invest in, which is approximately an order of magnitude smaller (note the different scale of the vertical axis in the figure).

A comparison of the results of the approximation (dashed lines) with those of the numerical simulation of the ABM (solid lines) in Fig.~\ref{fig:comparison2} shows that the approximation exhibits the same qualitative features, such as trends, timing and order of magnitude of the displayed variables, as the microscopic model.

Particularly, these results show that for the given parameter values the macroscopic approximation is capable of reproducing very closely the quasi-equilibrium states before and after the transition from the dirty to the clean sector, as it lies within the standard error of the ensemble of ABM runs. Also, the approximation is reasonably capable to reproduce the timing of and the transient states during the transition. This is somewhat surprising since in other works, macro-approximations were less well able to get the timing of transition right.

In the following, we discuss the existing differences between the results of the approximated model and the numerical simulation results. 

For instance, we find that the approximation estimates the transition from investment in the dirty sector to investment in the clean sector a bit too early (best visible in panel a). The reason for this might be the slight underestimation of the share of clean investing households, leading to a slight overestimation of the share of dirty capital in the system which is also visible in panel \ref{fig:comparison2}c.

We find a second obvious discrepancy between the micro-model and the approximation in the overestimation of dirty capital of clean investors ($K_d^{(c)}$) (panel d) during the transition phase between $t\approx 150$ and $t \approx 200$. This can be explained by the inequality in capital holdings amongst households. In the approximation, all households investing in dirty or clean capital are assumed to have the same income respectively. Therefore, the probability to change their investment behavior will change for all of them at once during the transition phase leading to a rapid shift of dirty investors changing to invest in clean capital but taking their dirty capital endowments with them (hence the sharp peak in dirty capital of clean investors during the transition phase, see Fig.~\ref{fig:comparison2}d dashed grey line). 

Also, in the micro-model, households changing from a dirty to a clean investment strategy take their -- presumably high -- endowments in dirty capital with them. Therefore, the endowments in dirty capital of households investing in the clean sector are relatively wide-spread (see grey area around solid orange line in Fig.~\ref{fig:comparison2}d. 
This has effects on the estimated timing of the transition, too. In the micro-model, income of households is heterogeneous. Therefore, for each of them the probability to change their investment behavior changes at different points in time, i.e., poorer households are likely to switch earlier during the transition than richer households. Together this leads to a slower, more spread-out transition dynamic the micro-model resulting in a flatter peak in the dirty capital endowments of clean-investing households.

Another effect at play during the transition is related to the assumptions in equations \ref{eq:neighbordist} and \ref{eq:additional_neighbors}. Namely, that all households that invest in the same type of capital have the same distribution of clean and dirty neighbors.

In the reality of the micro-model, however, these assumptions that are essential to the pair approximation may well be wrong -- especially so during a rapid transition. For example, a household that has only recently changed its state has a neighborhood that is atypical for its group and adapts only slowly. Consequently, when many changes in the state of the system happen in a short time, a significant proportion of the population is not well described by the assumed approximate distribution.

A number of these effects that lead to discrepancies between the micro-model and the approximation can be mitigated by higher-order moment closure for the distribution of heterogeneous agent-properties or higher-order motif approximation of the network dynamic.

For instance, a higher-order moment closure approximation that tracks the variance and skewness of the distribution of capital endowments can also account for the likelihood of capital endowments of agents that switch their investment decision to be biased. This would presumably mitigate the overestimation of dirty capital of clean investors ($K_d^{(c)}$) during the transition as well as the underestimation of ($K_d^{(c)}$) before the transition and therefore also estimate the timing of the transition even more precisely. 

Similarly, a higher-order motif approximation of the network dynamic can describe the heterogeneity in the local distribution of opinions in the neighborhood of individual agents and correct for the effects of this especially during periods of transient non-equilibrium dynamics in the approximated model.\\

In the previous section we derived a set of ordinary differential equations describing the stochastic dynamics of an agent-based model in terms of aggregated variables in the large system limit. We intend this derivation to be a prototypical example for a macroeconomic model with true microfoundations based on heterogeneous agents, given their microscopic interactions are of similar complexity. As such, it might also serve as a starting point for the application and development of similar models for other kinds of social dynamics. For example, an extension to continuous opinions requiring a Fokker-Planck-type description would follow naturally and would grant compatibility to a large body of models for social influence \citep[see ref.][pp. 988 f.]{Mueller-Hansen2017}.

\section{Bifurcation Analysis}
\label{sec:bifurcation-analysis}

\begin{figure*}[ht!]
  \centering\includegraphics[width=.8\linewidth]{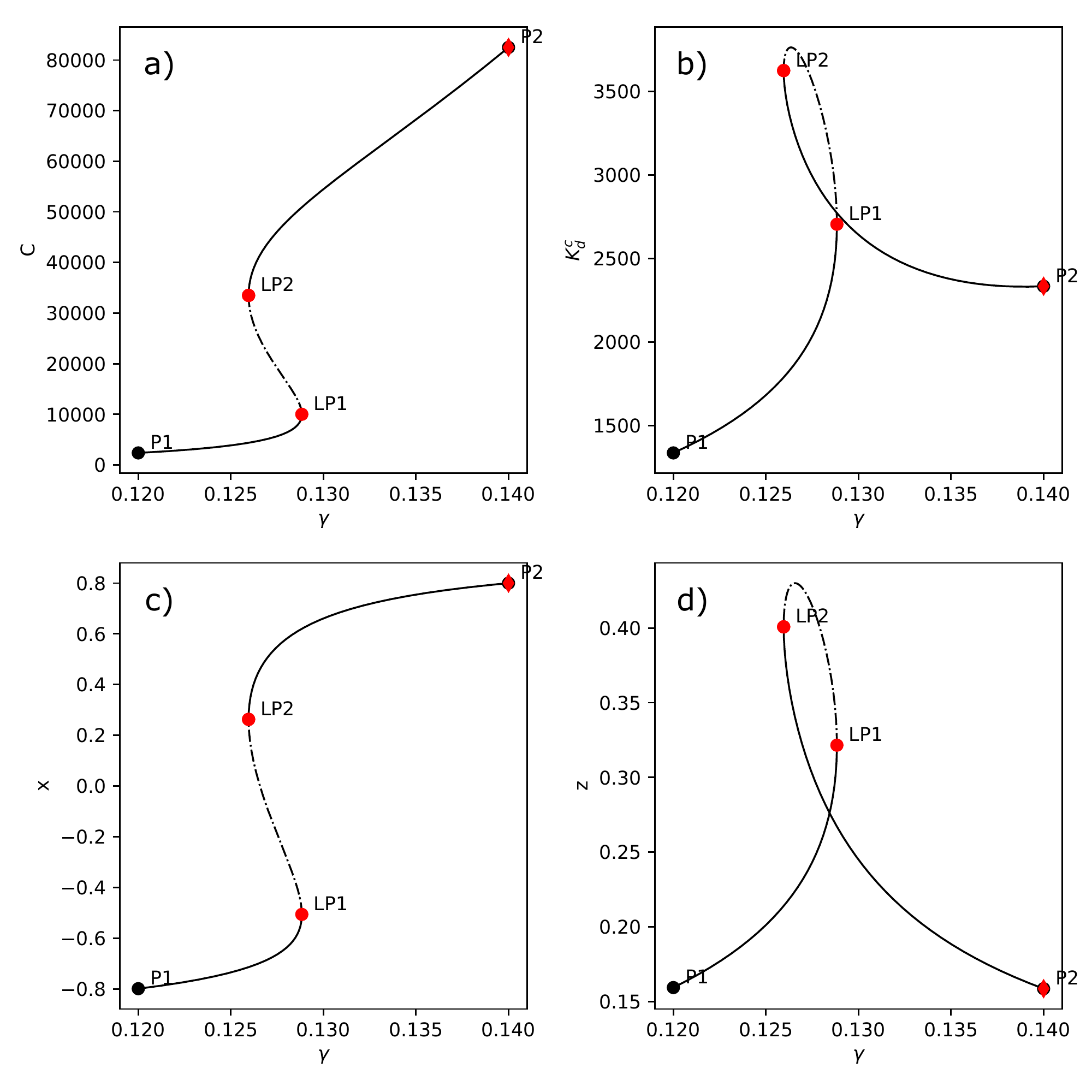}
\caption{\textbf{Bifurcation diagram:} Continuation of the stationary solution of the macroscopic approximation without resource depletion, i.e., with $\dot{G} = 0$ instead of the rate $R$ as given by Eq.~\eqref{eq:sum_up_resource_depletion}. Bifurcation parameter is $\gamma$, the elasticity of knowledge in the clean sector that also reflects the elasticity of learning by doing of the respective technology. The points labeled P1 and P2 are the beginning and end points of the continuation line, the points labeled LP1 and LP2 are the bifurcation points of two fold bifurcations. The stable unstable manifold is indicated by a dotted line, the stable manifold is indicated by solid line. Note that the intersections of the curves in the two right panels do not actually mean that the stationary manifold is not a bijective function of the bifurcation parameter $\gamma$ but rather a result of the projection of the multidimensional manifold onto the two-dimensional space.\label{fig:bifurcation_analysis}}
\end{figure*}

The description of the model as a system of ordinary differential equations allows for the analytical analysis of emergent model properties such as multi-stability, tipping and phase transitions. 
As a proof of concept application we subsequently show the results of a bifurcation analysis.

\subsection{Methods}
Bifurcation theory is the analysis of qualitative changes of dynamical systems under parameter variation, for example between a regime with a unique equilibrium (fixed point) and a multi-stable regime.
The parameter value at which a qualitative change, for example in the stability of an equilibrium, occurs is called a critical value or bifurcation point. Bifurcations are classified according to the changes in dynamical properties of the system \cite{Strogatz1994,Kuznetsov1998}.
Analytical methods have limited scope to identify bifurcation points in non-linear systems. Methods like numerical continuation can handle complex systems of ordinary differential equations like the one derived in Sec.~\ref{sec:Approximation} \citep{Allgower2003}.
Consequently, we use numerical continuation from PyDSTool \citep{pydstool,10.1371/journal.pcbi.1002628}, a Python package for dynamical systems modeling and analysis \footnote{PyDSTool is building on the AUTO-07p continuation library \citep{Doedel07auto-07p:continuation}.}.

A common bifurcation type that appears in our model is the fold bifurcation that is also known as saddle-node bifurcation. This type is a local bifurcation in which a stable fixed point collides with an unstable one and both disappear. 

Varying two bifurcation parameters at the same time can result in even richer qualitative changes of the dynamics. A prevalent example for such a bifurcation is the cusp geometry \citep[][p.\.397]{Kuznetsov1998}. A change of the second bifurcation parameter in this geometry beyond a certain value results in the so-called cusp catastrophe: the multi-stability of the system disappears for all values of the first bifurcation parameter. As we will show in the following, the macro-approximation of our model indeed exhibits a cusp bifurcation. 

\subsection{Discussion of Results}

\begin{figure}[ht!]
\centering\includegraphics[width=\linewidth]{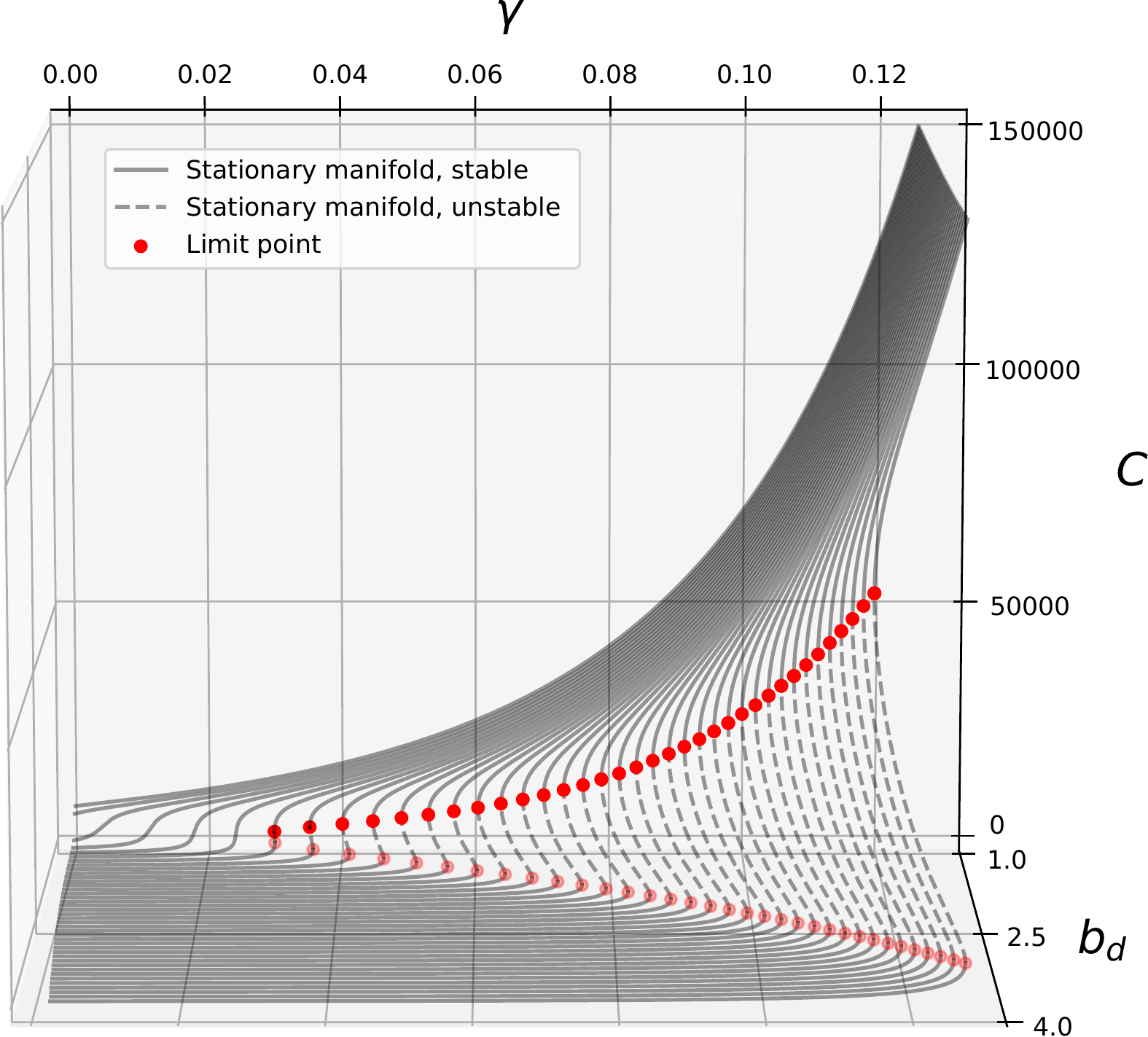}
\caption{\textbf{Cusp Bifurcation diagram:}
Stationary manifold from Fig.~\ref{fig:bifurcation_analysis} panel a for different values of the total factor productivity on the dirty sector $b_d$. Red dots indicate the limit points of the one dimensional fold bifurcation separating the stable and the unstable parts of the stationary manifold indicated by a solid and a dashed line respectively. For a critical value of $b_d \approx 1.4$ and $\gamma \approx 0.03034$ the two limit points converge and annihilate each other. This codimension-two bifurcation with bifurcation parameters $\gamma$ and $b_d$ is called a cusp catastrophe. In our two-sector economic model, this results in a lock in effect in the dirty sector, i.e., below this point, there is a smooth transition of production from the dirty to the clean sector and above this point production in the dirty sector is continued even though production in the clean sector would be more efficient. \label{fig:cusp}}
\end{figure}

A considerable advantage of the description of our model in terms of ordinary differential equations \eqref{eq:sum_up_resource_depletion}, \eqref{eq:sum_up_learning} \eqref{eq:lsl_transitions} and \eqref{eq:lsl_capital} over agent-based modeling is the fact that it allows for the usage of established tools for bifurcation analysis.
As a proof of concept, we show some results in Fig.~\ref{fig:bifurcation_analysis}.
Here, we analyze the possible steady states of the system with abundant fossil resources, e.g., the possible equilibrium states of the model in the regime before the fossil resource becomes scarce and acts as an external driver on the system pushing it towards clean investment.
Therefore, we set the resource depletion to zero, i.e., we keep the resource stock in Eq.~\eqref{eq:sum_up_resource_depletion} constant $G(t) \equiv G_0$ such that the resource usage cost in Eq.~\eqref{eq:resource_cost} still depends on resource use $R$ but is not increased by deceasing resource stock $G$. Thereby, we eliminate the rising resource extraction cost as the constraint in \eqref{eq:equilibrium_wage} and \eqref{eq:dirty_capital_rent} that eventually halts production in the dirty sector. 
We chose the learning rate $\gamma$ as bifurcation parameter as we expect it to yield interesting results.
Generally, in nonlinear dynamical systems, exponential factors are expected to have a strong influence on dynamical properties. Therefore, changing these factors is expected to lead to bifurcation behavior.
Consequently, in Fig.~\ref{fig:bifurcation_analysis} panel a and c we see that for certain learning rates $\gamma$ the macroscopic approximation exhibits a bistable regime limited by two fold bifurcations with bifurcation points indicated by LP1 and LP2.
In this regime both low investment in the clean sector together with high investment in the dirty sector and low knowledge as well as high investment in the clean sector together with low investment in the dirty sector and high knowledge are stable states of the economic system. This means that in this region economic outcomes are highly path dependent. Starting with slightly different knowledge about clean technologies may lead to widely differing adoption levels of the technology in the long run.

Figure \ref{fig:cusp} shows an example of how this bifurcation structure of the dynamical system depends on other parameters. Varying the total factor productivity in the dirty sector, $b_d$, the system undergoes a cusp bifurcation. Above a certain value of $b_d$ the system exhibits bi-stability whereas below this value it does not.

Clearly, this choice of bifurcation parameters is only one of many and other choices may very well lead to interesting results. However, we had to limit ourselves to this proof of concept study since an extensive analysis of all possible combinations would be well beyond the scope of this paper.

Multi-stability of the economy would mean that policies could make use of inherent dynamical properties of the system to reach a desired state or bring the system onto a desired pathway. For example, policy measures such as regulation or taxes can help driving the system into another basin of attraction, i.e., a region of the phase-space in which trajectories approach another equilibrium in the long term. To do so, the system has to cross a separatrix, the boundary between two basins of attraction.
After this boundary is crossed, the policy measure can be discontinued, the system's dynamics guarantee that it reaches the new equilibrium. 
Figure \ref{fig:cusp} shows that such an intervention could be complemented by an additional policy measure, lowering the total factor productivity in the dirty sector, effectively reducing the distance of the stable manifold from the separatrix and thereby presumably making the first measure less costly.
Another possibility to take advantage of the system's inherent dynamical structure is to use its hysteresis, i.e., to find policy measures that change the first bifurcation parameter $\gamma$ across a bifurcation point or to change the second bifurcation parameter $b_d$ to move the bifurcation point past the current state of the system (or a combination of both) after which the system would fall to the other branch of the stable manifold. Afterwards, the policy can be discontinued and the system would remain in its new state.
For such considerations, tools from dynamical systems theory and topology can be used to classify the phase-space of the system into regions with respect to the reachability of a desirable state \citep{Heitzig2016,Nitzbon2017}. This allows designing temporary policies that leverage the multi-stability of the socio-economic system.

\section{Conclusion}

This paper combines a set of methods to overcome shortcomings of current approaches to base macroeconomic models on microfoundations.
While representative agent approaches are unable to capture dynamics that emerge from structured and local interactions of multiple heterogeneous agents, computational agent-based approaches have the disadvantage that they make tractable model analysis difficult and computationally challenging.
We demonstrated that a combination of approximation techniques allows finding a macro description of a multi-agent system in which heterogeneous agents interact locally on a complex adaptive network as well as via aggregated quantities. 
In contrast to previous analytic work, where the network structure was either static \cite{Lux2016}, restricted to star-like clusters \cite{DiGuilmi2012} or approximated by a mean-field interaction approach and hence neglected \cite{Aoki1998, Aoki2007, Alfarano2008a, DiGuilmi2008, Chiarella2011a}, we explicitly treat the structure of the adaptive complex interaction network with appropriate approximation methods.

We develop a stylized two-sector investment model, in which investment decisions are driven by a social imitation process, to showcase the three approximations:
First, a pair approximation of networked interactions takes into account the heterogeneity in interaction patterns.
Second, a moment closure approximation makes it possible to deal with heterogeneous attributes that characterize the agents.
Third, the large-system limit abstracts from effects due to finite population size.
It is only possible to take this limit if the model has at least one of the following properties: (i) individual interaction depend only on relative rather than absolute quantities such that the size of households can be decreased while taking the number of households to infinity or (ii) the economic production functions exhibit constant returns to scale such that they scale linearly with the number of households $N$.
The resulting set of ordinary differential equations captures the effect of local interactions at the system level while still allowing for analytical tractability.

A comparison between a computational version of the ABM and the macro-description reveals that the approximation works well for parameter values distinct from special cases even if only accounting for first moments. Taking more moments into account would increase accuracy but comes at the cost of higher dimensionality and complexity of the macroscopic dynamical system.

Our model shows that social imitation dynamics add inertia to the investment decisions in the system that cannot be captured by a representative agent approach.
The imitation process results in social learning such that agents tend to direct their investments into the more profitable sector over time.
Because of this, the shift of investments from the dirty (fossil) to the clean (renewable) sector is driven only by economic factors, namely increasing exploration and extraction costs for the fossil energy resource.
Thus, we conclude that neutral imitation of better-performing peers is not a feasible mechanism to initiate a bottom-up transformation of the economy. Directed imitation, for example driven by changes in social norms, and supporting policies that make dirty production less profitable are needed to initiate a transformation towards a sustainable economy in the absence of fossil resource shortage.

Finding a system of ordinary differential equations to approximate ABMs is useful because it makes the analysis of the dynamical properties of the model much easier. One promising application here is bifurcation theory, as illustrated in Sec.~\ref{sec:bifurcation-analysis}.
Furthermore, it opens the possibility to mathematically prove model properties such as the dependency between different parameters and variables in the model.

In the context of climate economics and policy, the proposed techniques are especially important because they allow investigating the interplay of learning agents adapting to new policies and effects of shifts in values and preferences. The resulting changes in individual behavior and their impact on macroeconomic dynamics can be studied in a comprehensive modeling framework. 
Large shifts in investments that are required to reach the goals of the Paris agreement are likely to profit from both, policies that rely on price signals, as well as policies that target individual norm change, interaction and behavior not unlike those researched in, e.g., the public health context \cite{Zhang2016, Zhang2015, Centola2011}. The presented techniques can help to better understand how such behavioral interventions would impact the macro-level dynamics of the economic system.

On this regard, there are several promising avenues to develop the model and approximation techniques further: For example, instead of binary opinions, the social interaction model can use continuous variables to represent gradual opinions, drawing on a variety of models of social influence \citep[see ref.][pp. 988 f.]{Mueller-Hansen2017}.
An approximation of the agent ensemble would then need a Fokker-Planck-type description rather than a master equation.

Our model could be extended to explicitly include policy instruments such as a carbon tax and explore its impact on the investment decisions of the heterogeneous agent population.
Another promising modification could include consumption decisions into our two-sector model. Consumption decisions are strongly influenced by social norms and interactions \citep{Peattie2010}. Their inclusion could inform the discussion about green consumption as a potential mechanism for a bottom-up transformation towards a more sustainable economy.

Finally, the techniques proposed in this paper could be used to approximate other systems that interact both locally on a network and in an aggregate way on the system level, for example social-ecological systems or neural networks.

\section*{Acknowledgements}
The authors declare that they do not have any conflict of interest. Jakob J. Kolb acknowledges funding by the Foundation of German Industries. Finn M\"{u}ller-Hansen acknowledges funding by the DFG (IRTG 1740/TRP 2011/50151-0).
Jobst Heitzig acknowledges funding by the Leibniz Assiciation (project DominoES).
Numerical analyses were performed on the high-performance computer system of the Potsdam-Institute for Climate Impact Research, supported by the European Regional Development Fund, BMBF, and the Land Brandenburg.

\appendix
\begin{widetext}

\section{Comparing adaptive with fully connected network}
\label{app:fully_connected}

\begin{figure*}[ht]
    \centering
    \includegraphics[width=.8 \textwidth]{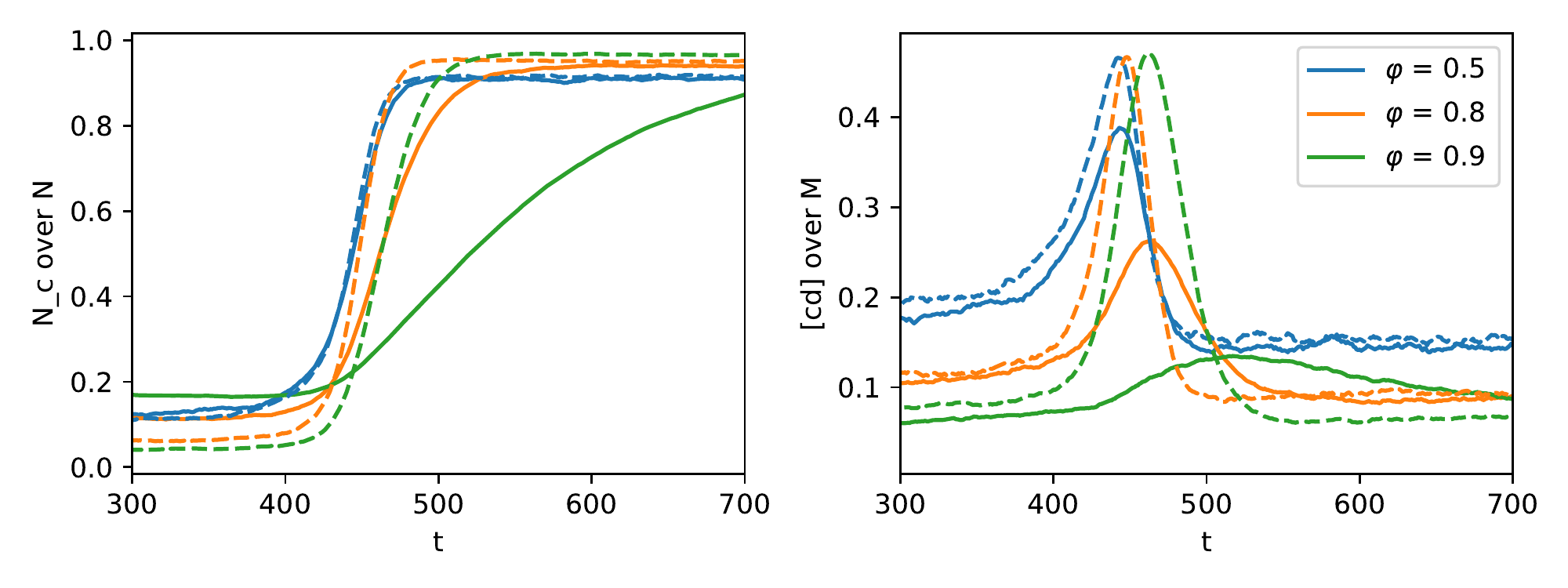}
    \caption{Comparison of microscopic model with adaptive network dynamics with microscopic model with fully connected network for varying rewiring rate $\varphi$. All other parameters are given in Tab.~\ref{tab:Parameter_list}. Solid lines indicate results with network adaptation, dashed lines indicate results with fully connected network. Initial network topology is a Erd\H{o}s-Renyi random graph.}
    \label{fig:compare_fc}
\end{figure*}

We compare the dynamics of the micro model with adaptive network rewiring with the dynamics of micro model with a fully connected acquaintance network. The model with a fully connected acquaintance network is equivalent to a well-mixed model with pairwise interactions between all agents. The results in Fig.~\ref{fig:compare_fc} show, that the well-mixed model approximates the adaptive network model for $\varphi=0.5$ quite well. However, for increasing $\varphi$, the fragmentation increases in the adaptive network model, indicated by the lower fraction of links between agents with different savings decisions (clean and dirty)  $[cd]/M$. This cannot be captured by the fully connected network model. As an economically observable result, this leads to significantly slower tipping in the adaptive network model.

\section{Effects of the rewiring rate $\varphi$ on model dynamics}
\label{app:rewiring}
\begin{figure*}[ht]
    \centering
    \includegraphics[width=.8 \textwidth]{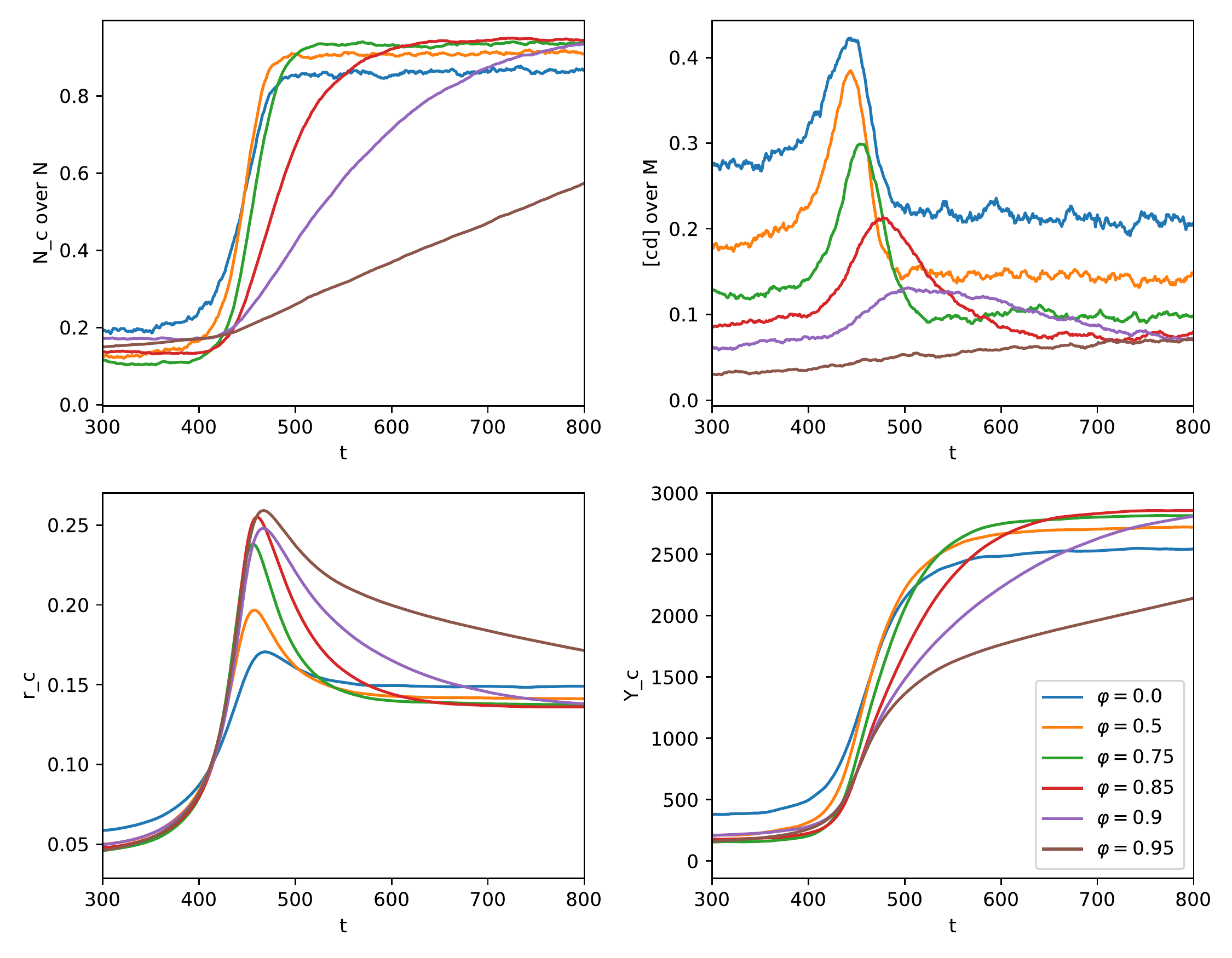}
    \caption{Model trajectories with varying $\varphi$. All other parameters are given in Tab.~\ref{tab:Parameter_list}. Results are ensemble averages from 200 runs. Initial network topology is a Erd\H{o}s-Renyi random graph.}
    \label{fig:varphi_scan}
\end{figure*}

We analyze the effect of changes in the network rewiring rate $\varphi$ on the model dynamics. The results in Fig.~\ref{fig:varphi_scan} indicate that for increasing rewiring rate $\varphi$ the model undergoes a transition from a connected network state with a considerable number of connections between agents investing in different sectors, to a fragmented network state in which such connections are effectively non-existent. This transition is especially apparent in the fraction of $[cd]$ links in the network given in Fig.~\ref{fig:varphi_scan}b. This fragmentation transition is well known for adaptive voter type models \cite{Bohme2011, Demirel2014, Min2017}.

\section{Model dynamics depending on network rewiring}
\label{si:rewiring}
\section{ODEs resulting from approximation}
\label{si:odes}
The following are the full ordinary differential equations resulting from \eqref{eq:lsl_transitions} ,\eqref{eq:lsl_capital}, \eqref{eq:sum_up_resource_depletion} and \eqref{eq:sum_up_learning}
\begin{align}
    \dot{x} =&- \frac{\epsilon x}{\tau} - \frac{p_{cd} z \left(\epsilon - 1\right) \left(\phi - 1\right) \left(x + 1\right)}{\tau \left(y + 1\right)} + \frac{p_{dc} z \left(\epsilon - 1\right) \left(\phi - 1\right) \left(x - 1\right)}{\tau \left(y - 1\right)} \\
    \dot{y} =& - \frac{m \left(p_{cd} z \left(\epsilon - 1\right) \left(\phi - 1\right) - p_{dc} z \left(\epsilon - 1\right) \left(\phi - 1\right) + 0.5 \epsilon \left(y - 1\right) + 0.5 \epsilon \left(y + 1\right)\right)}{\tau} \nonumber \\
    &+ \frac{\left(x - 1\right) \left(0.25 \epsilon z \left(x - 1\right) - 0.25 \epsilon \left(x + 1\right) \left(y + z - 1\right) + 0.5 \phi z \left(\epsilon - 1\right)\right)}{\tau \left(y - 1\right)} \nonumber \\
    &+ \frac{\left(x + 1\right) \left(0.25 \epsilon z \left(x + 1\right) + 0.25 \epsilon \left(x - 1\right) \left(y - z + 1\right) - 0.5 \phi z \left(\epsilon - 1\right)\right)}{\tau \left(y + 1\right)}\\
    \dot{z} =& - \frac{\epsilon m \left(2 z - 1\right)}{\tau} \nonumber \\
    &- \frac{0.5 p_{cd} z \left(\epsilon - 1\right) \left(\phi - 1\right) \left(\left(x + 1\right) \left(y + 1\right) - 2 \left(y - 2 z + 1\right) \left(m y + m - 0.5 x - 0.5\right)\right)}{\tau \left(y + 1\right)^{2}} \nonumber \\
    &- \frac{0.5 p_{dc} z \left(\epsilon - 1\right) \left(\phi - 1\right) \left(\left(x - 1\right) \left(y - 1\right) - 2 \left(y + 2 z - 1\right) \left(m y - m - 0.5 x + 0.5\right)\right)}{\tau \left(y - 1\right)^{2}} \nonumber \\
    &+ \frac{\left(x - 1\right) \left(0.25 \epsilon z \left(x - 1\right) - 0.25 \epsilon \left(x + 1\right) \left(y + z - 1\right) + 0.5 \phi z \left(\epsilon - 1\right)\right)}{\tau \left(y - 1\right)} \nonumber \\
    &- \frac{\left(x + 1\right) \left(0.25 \epsilon z \left(x + 1\right) + 0.25 \epsilon \left(x - 1\right) \left(y - z + 1\right) - 0.5 \phi z \left(\epsilon - 1\right)\right)}{\tau \left(y + 1\right)}\\
    \dot{K}_c^{(c)} =&  K_c^{(c)} \left(- \delta + r_{c} s\right) + K_d^{(c)} r_{d} s + L s w \nonumber \\
    &- \frac{0.5 K_c^{(c)} \left(x + 1\right) \left(p_{cd} z \left(\epsilon - 1\right) \left(\phi - 1\right) + 0.5 \epsilon \left(y + 1\right)\right)}{\tau \left(y + 1\right)} \nonumber \\
    &+ \frac{0.5 K_c^{(d)} \left(x - 1\right) \left(p_{dc} z \left(\epsilon - 1\right) \left(\phi - 1\right) - 0.5 \epsilon \left(y - 1\right)\right)}{\tau \left(y - 1\right)} \\
    \dot{K}_d^{(d)} =& K_d^{(d)} \left(- \delta + r_{d} s\right) + K_c^{(d)} r_{c} s + L s w \nonumber \\
    &+ \frac{0.5 K_d^{(c)} \left(x + 1\right) \left(p_{cd} z \left(\epsilon - 1\right) \left(\phi - 1\right) + 0.5 \epsilon \left(y + 1\right)\right)}{\tau \left(y + 1\right)} \nonumber \\
    &- \frac{0.5 K_d^{(d)} \left(x - 1\right) \left(p_{dc} z \left(\epsilon - 1\right) \left(\phi - 1\right) - 0.5 \epsilon \left(y - 1\right)\right)}{\tau \left(y - 1\right)} \\
    \dot{K}_d^{(c)} =& - K_d^{(c)} \delta - \frac{0.5 K_d^{(c)} \left(x + 1\right) \left(p_{cd} z \left(\epsilon - 1\right) \left(\phi - 1\right) + 0.5 \epsilon \left(y + 1\right)\right)}{\tau \left(y + 1\right)} \nonumber \\
    &+ \frac{0.5 K_d^{(d)} \left(x - 1\right) \left(p_{dc} z \left(\epsilon - 1\right) \left(\phi - 1\right) - 0.5 \epsilon \left(y - 1\right)\right)}{\tau \left(y - 1\right)} \\
    \dot{K}_c^{(d)} =&- K_c^{(d)} \delta + \frac{0.5 K_c^{(c)} \left(x + 1\right) \left(p_{cd} z \left(\epsilon - 1\right) \left(\phi - 1\right) + 0.5 \epsilon \left(y + 1\right)\right)}{\tau \left(y + 1\right)} \nonumber \\
    &- \frac{0.5 K_c^{(d)} \left(x - 1\right) \left(p_{dc} z \left(\epsilon - 1\right) \left(\phi - 1\right) - 0.5 \epsilon \left(y - 1\right)\right)}{\tau \left(y - 1\right)} \\
\end{align}
\begin{align}
    \dot{G} =& - \frac{L^{\pi} b_{d}}{e_{R}} \left(\frac{\left(b_{d} \left(K_d^{(c)} + K_d^{(d)}\right)^{\kappa_{d}}\right)^{\frac{1}{1 - \pi}} \left(1 - \frac{G_{0}^{2} b_{R}}{G^{2} e_{R}}\right)^{\frac{1}{1 - \pi}}}{\left(b_{d} \left(K_d^{(c)} + K_d^{(d)}\right)^{\kappa_{d}}\right)^{\frac{1}{1 - \pi}} \left(1 - \frac{G_{0}^{2} b_{R}}{G^{2} e_{R}}\right)^{\frac{1}{1 - \pi}} + \left(C^{\xi} b_{c} \left(K_c^{(c)} + K_c^{(d)}\right)^{\kappa_{c}}\right)^{\frac{1}{1 - \pi}}}\right)^{\pi} \left(K_d^{(c)} + K_d^{(d)}\right)^{\kappa_{d}}\\
    \dot{C} =& - C \delta + C^{\xi} b_{c} \left(\frac{L \left(C^{\xi} b_{c} \left(K_c^{(c)} + K_c^{(d)}\right)^{\kappa_{c}}\right)^{\frac{1}{1 - \pi}}}{\left(b_{d} \left(K_d^{(c)} + K_d^{(d)}\right)^{\kappa_{d}}\right)^{\frac{1}{1 - \pi}} \left(1 - \frac{G_{0}^{2} b_{R}}{G^{2} e_{R}}\right)^{\frac{1}{1 - \pi}} + \left(C^{\xi} b_{c} \left(K_c^{(c)} + K_c^{(d)}\right)^{\kappa_{c}}\right)^{\frac{1}{1 - \pi}}}\right)^{\pi} \nonumber \\
    & \times\left(K_c^{(c)} \left(\frac{x}{2} + \frac{1}{2}\right) + K_c^{(d)} \left(\frac{1}{2} - \frac{x}{2}\right)\right)^{\kappa_{c}}
\end{align}
 where $p_{cd}$ and $p_{dc}$ are given by Eq.~\eqref{eq:approx_p_cd_final} and \eqref{eq:approx_p_dc_final} and $r_c, r_d$ and $w$ are given by
 \begin{align}
     r_c = & \frac{L^{\pi} \kappa_{c} \left(C^{\xi} b_{c} \left(K_c^{(c)} + K_c^{(d)}\right)^{\kappa_{c}}\right)^{\frac{1}{1 - \pi}} \left(C^{\frac{1 \xi}{1 - \pi}} b_{c}^{\frac{1}{1 - \pi}} \left(K_c^{(c)} + K_c^{(d)}\right)^{\frac{1 \kappa_{c}}{1 - \pi}} + \left(b_{d} \left(K_d^{(c)} + K_d^{(d)}\right)^{\kappa_{d}}\left(1 - \frac{G_{0}^{2} b_{R}}{G^{2} e_{R}}\right)\right)^{\frac{1}{1 - \pi}}\right)^{- \pi}}{K_c^{(c)} + K_c^{(d)}}\\
     r_d = & \frac{L^{\pi} \kappa_{d} \left(b_{d} \left(K_d^{(c)} + K_d^{(d)}\right)^{\kappa_{d}} \left(1 - \frac{G_{0}^{2} b_{R}}{G^{2} e_{R}}\right)\right)^{\frac{1}{1 - \pi}}}{K_d^{(c)} + K_d^{(d)}} \nonumber \\
     & \times \left(C^{\frac{1 \xi}{1 - \pi}} b_{c}^{\frac{1}{1 - \pi}} \left(K_c^{(c)} + K_c^{(d)}\right)^{\frac{1 \kappa_{c}}{1 - \pi}} + \left(b_{d} \left(K_d^{(c)} + K_d^{(d)}\right)^{\kappa_{d}}\left(1 - \frac{G_{0}^{2} b_{R}}{G^{2} e_{R}}\right)\right)^{\frac{1}{1 - \pi}}\right)^{- \pi}\\
     w = & L^{\pi - 1} \pi \left(C^{\frac{1 \xi}{1 - \pi}} b_{c}^{\frac{1}{1 - \pi}} \left(K_c^{(c)} + K_c^{(d)}\right)^{\frac{1 \kappa_{c}}{1 - \pi}} + \left(b_{d} \left(K_d^{(c)} + K_d^{(d)}\right)^{\kappa_{d}}\right)^{\frac{1}{1 - \pi}} \left(1 - \frac{G_{0}^{2} b_{R}}{G^{2} e_{R}}\right)^{\frac{1}{1 - \pi}}\right)^{1 - \pi}
 \end{align}
\end{widetext}

\bibliography{library}

\begin{thebibliography}{109}%
\makeatletter
\providecommand \@ifxundefined [1]{%
 \@ifx{#1\undefined}
}%
\providecommand \@ifnum [1]{%
 \ifnum #1\expandafter \@firstoftwo
 \else \expandafter \@secondoftwo
 \fi
}%
\providecommand \@ifx [1]{%
 \ifx #1\expandafter \@firstoftwo
 \else \expandafter \@secondoftwo
 \fi
}%
\providecommand \natexlab [1]{#1}%
\providecommand \enquote  [1]{``#1''}%
\providecommand \bibnamefont  [1]{#1}%
\providecommand \bibfnamefont [1]{#1}%
\providecommand \citenamefont [1]{#1}%
\providecommand \href@noop [0]{\@secondoftwo}%
\providecommand \href [0]{\begingroup \@sanitize@url \@href}%
\providecommand \@href[1]{\@@startlink{#1}\@@href}%
\providecommand \@@href[1]{\endgroup#1\@@endlink}%
\providecommand \@sanitize@url [0]{\catcode `\\12\catcode `\$12\catcode
  `\&12\catcode `\#12\catcode `\^12\catcode `\_12\catcode `\%12\relax}%
\providecommand \@@startlink[1]{}%
\providecommand \@@endlink[0]{}%
\providecommand \url  [0]{\begingroup\@sanitize@url \@url }%
\providecommand \@url [1]{\endgroup\@href {#1}{\urlprefix }}%
\providecommand \urlprefix  [0]{URL }%
\providecommand \Eprint [0]{\href }%
\providecommand \doibase [0]{https://doi.org/}%
\providecommand \selectlanguage [0]{\@gobble}%
\providecommand \bibinfo  [0]{\@secondoftwo}%
\providecommand \bibfield  [0]{\@secondoftwo}%
\providecommand \translation [1]{[#1]}%
\providecommand \BibitemOpen [0]{}%
\providecommand \bibitemStop [0]{}%
\providecommand \bibitemNoStop [0]{.\EOS\space}%
\providecommand \EOS [0]{\spacefactor3000\relax}%
\providecommand \BibitemShut  [1]{\csname bibitem#1\endcsname}%
\let\auto@bib@innerbib\@empty
\bibitem [{\citenamefont {Grimm}\ and\ \citenamefont
  {Railsback}(2005)}]{Grimm2005}%
  \BibitemOpen
  \bibfield  {author} {\bibinfo {author} {\bibfnamefont {V.}~\bibnamefont
  {Grimm}}\ and\ \bibinfo {author} {\bibfnamefont {S.~F.}\ \bibnamefont
  {Railsback}},\ }\href@noop {} {\emph {\bibinfo {title} {{Individual-based
  modeling and ecology}}}},\ Princeton Series in Theoretical and Computational
  Biology\ (\bibinfo  {publisher} {Princeton University Press},\ \bibinfo
  {address} {Princeton, NJ},\ \bibinfo {year} {2005})\BibitemShut {NoStop}%
\bibitem [{\citenamefont {Bonabeau}(2002)}]{Bonabeau2002}%
  \BibitemOpen
  \bibfield  {author} {\bibinfo {author} {\bibfnamefont {E.}~\bibnamefont
  {Bonabeau}},\ }\bibfield  {title} {\bibinfo {title} {{Agent-based modeling:
  Methods and techniques for simulating human systems}},\ }\href
  {https://doi.org/10.1073/pnas.082080899} {\bibfield  {journal} {\bibinfo
  {journal} {Proceedings of the National Academy of Sciences}\ }\textbf
  {\bibinfo {volume} {99}},\ \bibinfo {pages} {7280} (\bibinfo {year}
  {2002})}\BibitemShut {NoStop}%
\bibitem [{\citenamefont {Macy}\ and\ \citenamefont {Willer}(2002)}]{Macy2002}%
  \BibitemOpen
  \bibfield  {author} {\bibinfo {author} {\bibfnamefont {M.~W.}\ \bibnamefont
  {Macy}}\ and\ \bibinfo {author} {\bibfnamefont {R.}~\bibnamefont {Willer}},\
  }\bibfield  {title} {\bibinfo {title} {{From Factors to Actors: Computational
  Sociology and Agent-Based Modeling}},\ }\href
  {https://doi.org/10.1146/annurev.soc.28.110601.141117} {\bibfield  {journal}
  {\bibinfo  {journal} {Annual Review of Sociology}\ }\textbf {\bibinfo
  {volume} {28}},\ \bibinfo {pages} {143} (\bibinfo {year} {2002})}\BibitemShut
  {NoStop}%
\bibitem [{\citenamefont {Tesfatsion}(2006)}]{Tesfatsion2006}%
  \BibitemOpen
  \bibfield  {author} {\bibinfo {author} {\bibfnamefont {L.}~\bibnamefont
  {Tesfatsion}},\ }\bibfield  {title} {\bibinfo {title} {{Agent-Based
  Computational Economics: A Constructive Approach to Economic Theory}},\ }in\
  \href {https://doi.org/10.1016/S1574-0021(05)02016-2} {\emph {\bibinfo
  {booktitle} {Handbook of Computational Economics Vol. 2}}},\ \bibinfo
  {editor} {edited by\ \bibinfo {editor} {\bibfnamefont {L.}~\bibnamefont
  {Tesfatsion}}\ and\ \bibinfo {editor} {\bibfnamefont {K.~L.}\ \bibnamefont
  {Judd}}}\ (\bibinfo  {publisher} {North Holland},\ \bibinfo {address}
  {Amsterdam},\ \bibinfo {year} {2006})\ Chap.~\bibinfo {chapter} {16}, pp.\
  \bibinfo {pages} {831--880}\BibitemShut {NoStop}%
\bibitem [{\citenamefont {Hamill}\ and\ \citenamefont
  {Gilbert}(2016)}]{Hamill2016}%
  \BibitemOpen
  \bibfield  {author} {\bibinfo {author} {\bibfnamefont {L.}~\bibnamefont
  {Hamill}}\ and\ \bibinfo {author} {\bibfnamefont {N.}~\bibnamefont
  {Gilbert}},\ }\href@noop {} {\emph {\bibinfo {title} {{Agent-Based Modelling
  in Economics}}}}\ (\bibinfo  {publisher} {Wiley},\ \bibinfo {address}
  {Chichester, UK},\ \bibinfo {year} {2016})\BibitemShut {NoStop}%
\bibitem [{\citenamefont {Epstein}(1999)}]{Epstein1999}%
  \BibitemOpen
  \bibfield  {author} {\bibinfo {author} {\bibfnamefont {J.~M.}\ \bibnamefont
  {Epstein}},\ }\bibfield  {title} {\bibinfo {title} {{Agent-based
  computational models and generative social science}},\ }\href
  {https://doi.org/10.1002/(SICI)1099-0526(199905/06)4:5<41::AID-CPLX9>3.0.CO;2-F}
  {\bibfield  {journal} {\bibinfo  {journal} {Complexity}\ }\textbf {\bibinfo
  {volume} {4}},\ \bibinfo {pages} {41} (\bibinfo {year} {1999})}\BibitemShut
  {NoStop}%
\bibitem [{\citenamefont {Gross}\ and\ \citenamefont
  {Blasius}(2008)}]{Gross2008}%
  \BibitemOpen
  \bibfield  {author} {\bibinfo {author} {\bibfnamefont {T.}~\bibnamefont
  {Gross}}\ and\ \bibinfo {author} {\bibfnamefont {B.}~\bibnamefont
  {Blasius}},\ }\bibfield  {title} {\bibinfo {title} {{Adaptive coevolutionary
  networks: a review.}},\ }\href {https://doi.org/10.1098/rsif.2007.1229}
  {\bibfield  {journal} {\bibinfo  {journal} {Journal of the Royal Society,
  Interface / the Royal Society}\ }\textbf {\bibinfo {volume} {5}},\ \bibinfo
  {pages} {259} (\bibinfo {year} {2008})}\BibitemShut {NoStop}%
\bibitem [{\citenamefont {Holme}\ and\ \citenamefont
  {Newman}(2006)}]{Holme2006a}%
  \BibitemOpen
  \bibfield  {author} {\bibinfo {author} {\bibfnamefont {P.}~\bibnamefont
  {Holme}}\ and\ \bibinfo {author} {\bibfnamefont {M.~E.~J.}\ \bibnamefont
  {Newman}},\ }\bibfield  {title} {\bibinfo {title} {{Nonequilibrium phase
  transition in the coevolution of networks and opinions}},\ }\href
  {https://doi.org/10.1103/PhysRevE.74.056108} {\bibfield  {journal} {\bibinfo
  {journal} {Physical Review E}\ }\textbf {\bibinfo {volume} {74}},\ \bibinfo
  {pages} {056108} (\bibinfo {year} {2006})}\BibitemShut {NoStop}%
\bibitem [{\citenamefont {Bargigli}\ and\ \citenamefont
  {Tedeschi}(2014)}]{Bargigli2014}%
  \BibitemOpen
  \bibfield  {author} {\bibinfo {author} {\bibfnamefont {L.}~\bibnamefont
  {Bargigli}}\ and\ \bibinfo {author} {\bibfnamefont {G.}~\bibnamefont
  {Tedeschi}},\ }\bibfield  {title} {\bibinfo {title} {{Interaction in
  agent-based economics: A survey on the network approach}},\ }\href
  {https://doi.org/10.1016/j.physa.2013.12.029} {\bibfield  {journal} {\bibinfo
   {journal} {Physica A: Statistical Mechanics and its Applications}\ }\textbf
  {\bibinfo {volume} {399}},\ \bibinfo {pages} {1} (\bibinfo {year}
  {2014})}\BibitemShut {NoStop}%
\bibitem [{\citenamefont {Granovetter}(2005)}]{Granovetter2005}%
  \BibitemOpen
  \bibfield  {author} {\bibinfo {author} {\bibfnamefont {M.}~\bibnamefont
  {Granovetter}},\ }\bibfield  {title} {\bibinfo {title} {{The Impact of Social
  Structure on Economic Outcomes}},\ }\href@noop {} {\bibfield  {journal}
  {\bibinfo  {journal} {Journal ofEconomic Perspectives}\ }\textbf {\bibinfo
  {volume} {19}},\ \bibinfo {pages} {33} (\bibinfo {year} {2005})}\BibitemShut
  {NoStop}%
\bibitem [{\citenamefont {Delli~Gatti}\ \emph {et~al.}(2008)\citenamefont
  {Delli~Gatti}, \citenamefont {Gaffeo}, \citenamefont {Gallegati},
  \citenamefont {Giulioni},\ and\ \citenamefont {Palestrini}}]{DelliGatti2008}%
  \BibitemOpen
  \bibfield  {author} {\bibinfo {author} {\bibfnamefont {D.}~\bibnamefont
  {Delli~Gatti}}, \bibinfo {author} {\bibfnamefont {E.}~\bibnamefont {Gaffeo}},
  \bibinfo {author} {\bibfnamefont {M.}~\bibnamefont {Gallegati}}, \bibinfo
  {author} {\bibfnamefont {G.}~\bibnamefont {Giulioni}},\ and\ \bibinfo
  {author} {\bibfnamefont {A.}~\bibnamefont {Palestrini}},\ }\href@noop {}
  {\emph {\bibinfo {title} {{Emergent Macroeconomics. An Agent-Based Approach
  to Business Fluctuations}}}},\ New Economic Windows\ (\bibinfo  {publisher}
  {Springer},\ \bibinfo {address} {Milan},\ \bibinfo {year} {2008})\BibitemShut
  {NoStop}%
\bibitem [{Note1()}]{Note1}%
  \BibitemOpen
  \bibinfo {note} {Approaches to represent heterogeneous agents in DSGE models
  have been used to counter this criticism and add more realism regarding the
  distribution of agent attributes \protect \citep [see for example the review
  by][]{Heathcote2009}. Particularly, because the representative agent approach
  cannot account for interactions within a heterogeneous group, models using
  this approach do not allow for the representation of emergent phenomena
  \protect \citet {Kirman1992}. But their solution require complex numerical
  methods and cannot integrate local interactions between agents.}\BibitemShut
  {Stop}%
\bibitem [{\citenamefont {Kirman}(2014)}]{Kirman2014}%
  \BibitemOpen
  \bibfield  {author} {\bibinfo {author} {\bibfnamefont {A.}~\bibnamefont
  {Kirman}},\ }\bibfield  {title} {\bibinfo {title} {{Is it rational to have
  rational expectations?}},\ }\href {https://doi.org/10.1007/s11299-014-0136-x}
  {\bibfield  {journal} {\bibinfo  {journal} {Mind and Society}\ }\textbf
  {\bibinfo {volume} {13}},\ \bibinfo {pages} {29} (\bibinfo {year}
  {2014})}\BibitemShut {NoStop}%
\bibitem [{\citenamefont {Evans}\ and\ \citenamefont
  {Ramey}(2006)}]{Evans2006}%
  \BibitemOpen
  \bibfield  {author} {\bibinfo {author} {\bibfnamefont {G.~W.}\ \bibnamefont
  {Evans}}\ and\ \bibinfo {author} {\bibfnamefont {G.}~\bibnamefont {Ramey}},\
  }\bibfield  {title} {\bibinfo {title} {{Adaptive expectations,
  underparameterization and the Lucas critique}},\ }\href
  {https://doi.org/10.1016/j.jmoneco.2004.12.002} {\bibfield  {journal}
  {\bibinfo  {journal} {Journal of Monetary Economics}\ }\textbf {\bibinfo
  {volume} {53}},\ \bibinfo {pages} {249} (\bibinfo {year} {2006})}\BibitemShut
  {NoStop}%
\bibitem [{Note2()}]{Note2}%
  \BibitemOpen
  \bibinfo {note} {We use here a weak notion of emergence, which allows
  explaining macro-phenomena on the basis of micro-interactions of the systems
  constituents that differ from the explained macro-phenomena. This is opposed
  to strong emergence, that embraces the irreducibility of macro-phenomena to
  lower-level dynamics. For a discussion see \protect \citet
  {Bedau1997}.}\BibitemShut {Stop}%
\bibitem [{\citenamefont {Leombruni}\ and\ \citenamefont
  {Richiardi}(2005)}]{Leombruni2005}%
  \BibitemOpen
  \bibfield  {author} {\bibinfo {author} {\bibfnamefont {R.}~\bibnamefont
  {Leombruni}}\ and\ \bibinfo {author} {\bibfnamefont {M.}~\bibnamefont
  {Richiardi}},\ }\bibfield  {title} {\bibinfo {title} {{Why are economists
  sceptical about agent-based simulations?}},\ }\href
  {https://doi.org/10.1016/j.physa.2005.02.072} {\bibfield  {journal} {\bibinfo
   {journal} {Physica A: Statistical Mechanics and its Applications}\ }\textbf
  {\bibinfo {volume} {355}},\ \bibinfo {pages} {103} (\bibinfo {year}
  {2005})}\BibitemShut {NoStop}%
\bibitem [{\citenamefont {Grimm}\ \emph {et~al.}(2006)\citenamefont {Grimm},
  \citenamefont {Berger}, \citenamefont {Bastiansen}, \citenamefont {Eliassen},
  \citenamefont {Ginot}, \citenamefont {Giske}, \citenamefont {Goss-Custard},
  \citenamefont {Grand}, \citenamefont {Heinz}, \citenamefont {Huse},
  \citenamefont {Huth}, \citenamefont {Jepsen}, \citenamefont {J{\o}rgensen},
  \citenamefont {Mooij}, \citenamefont {M{\"{u}}ller}, \citenamefont {Pe'er},
  \citenamefont {Piou}, \citenamefont {Railsback}, \citenamefont {Robbins},
  \citenamefont {Robbins}, \citenamefont {Rossmanith}, \citenamefont
  {R{\"{u}}ger}, \citenamefont {Strand}, \citenamefont {Souissi}, \citenamefont
  {Stillman}, \citenamefont {Vab{\o}}, \citenamefont {Visser},\ and\
  \citenamefont {DeAngelis}}]{Grimm2006}%
  \BibitemOpen
  \bibfield  {author} {\bibinfo {author} {\bibfnamefont {V.}~\bibnamefont
  {Grimm}}, \bibinfo {author} {\bibfnamefont {U.}~\bibnamefont {Berger}},
  \bibinfo {author} {\bibfnamefont {F.}~\bibnamefont {Bastiansen}}, \bibinfo
  {author} {\bibfnamefont {S.}~\bibnamefont {Eliassen}}, \bibinfo {author}
  {\bibfnamefont {V.}~\bibnamefont {Ginot}}, \bibinfo {author} {\bibfnamefont
  {J.}~\bibnamefont {Giske}}, \bibinfo {author} {\bibfnamefont
  {J.}~\bibnamefont {Goss-Custard}}, \bibinfo {author} {\bibfnamefont
  {T.}~\bibnamefont {Grand}}, \bibinfo {author} {\bibfnamefont {S.~K.}\
  \bibnamefont {Heinz}}, \bibinfo {author} {\bibfnamefont {G.}~\bibnamefont
  {Huse}}, \bibinfo {author} {\bibfnamefont {A.}~\bibnamefont {Huth}}, \bibinfo
  {author} {\bibfnamefont {J.~U.}\ \bibnamefont {Jepsen}}, \bibinfo {author}
  {\bibfnamefont {C.}~\bibnamefont {J{\o}rgensen}}, \bibinfo {author}
  {\bibfnamefont {W.~M.}\ \bibnamefont {Mooij}}, \bibinfo {author}
  {\bibfnamefont {B.}~\bibnamefont {M{\"{u}}ller}}, \bibinfo {author}
  {\bibfnamefont {G.}~\bibnamefont {Pe'er}}, \bibinfo {author} {\bibfnamefont
  {C.}~\bibnamefont {Piou}}, \bibinfo {author} {\bibfnamefont {S.~F.}\
  \bibnamefont {Railsback}}, \bibinfo {author} {\bibfnamefont {A.~M.}\
  \bibnamefont {Robbins}}, \bibinfo {author} {\bibfnamefont {M.~M.}\
  \bibnamefont {Robbins}}, \bibinfo {author} {\bibfnamefont {E.}~\bibnamefont
  {Rossmanith}}, \bibinfo {author} {\bibfnamefont {N.}~\bibnamefont
  {R{\"{u}}ger}}, \bibinfo {author} {\bibfnamefont {E.}~\bibnamefont {Strand}},
  \bibinfo {author} {\bibfnamefont {S.}~\bibnamefont {Souissi}}, \bibinfo
  {author} {\bibfnamefont {R.~A.}\ \bibnamefont {Stillman}}, \bibinfo {author}
  {\bibfnamefont {R.}~\bibnamefont {Vab{\o}}}, \bibinfo {author} {\bibfnamefont
  {U.}~\bibnamefont {Visser}},\ and\ \bibinfo {author} {\bibfnamefont {D.~L.}\
  \bibnamefont {DeAngelis}},\ }\bibfield  {title} {\bibinfo {title} {{A
  standard protocol for describing individual-based and agent-based models}},\
  }\href {https://doi.org/10.1016/j.ecolmodel.2006.04.023} {\bibfield
  {journal} {\bibinfo  {journal} {Ecological Modelling}\ }\textbf {\bibinfo
  {volume} {198}},\ \bibinfo {pages} {115} (\bibinfo {year}
  {2006})}\BibitemShut {NoStop}%
\bibitem [{\citenamefont {Lee}\ \emph {et~al.}(2015)\citenamefont {Lee},
  \citenamefont {Filatova}, \citenamefont {Ligmann-Zielinska}, \citenamefont
  {Hassani-Mahmooei}, \citenamefont {Stonedahl}, \citenamefont {Lorscheid},
  \citenamefont {Voinov}, \citenamefont {Polhill}, \citenamefont {Sun},\ and\
  \citenamefont {Parker}}]{Lee2015}%
  \BibitemOpen
  \bibfield  {author} {\bibinfo {author} {\bibfnamefont {J.-S.}\ \bibnamefont
  {Lee}}, \bibinfo {author} {\bibfnamefont {T.}~\bibnamefont {Filatova}},
  \bibinfo {author} {\bibfnamefont {A.}~\bibnamefont {Ligmann-Zielinska}},
  \bibinfo {author} {\bibfnamefont {B.}~\bibnamefont {Hassani-Mahmooei}},
  \bibinfo {author} {\bibfnamefont {F.}~\bibnamefont {Stonedahl}}, \bibinfo
  {author} {\bibfnamefont {I.}~\bibnamefont {Lorscheid}}, \bibinfo {author}
  {\bibfnamefont {A.}~\bibnamefont {Voinov}}, \bibinfo {author} {\bibfnamefont
  {G.}~\bibnamefont {Polhill}}, \bibinfo {author} {\bibfnamefont
  {Z.}~\bibnamefont {Sun}},\ and\ \bibinfo {author} {\bibfnamefont {D.~C.}\
  \bibnamefont {Parker}},\ }\bibfield  {title} {\bibinfo {title} {{The
  Complexities of Agent-Based Modeling Output Analysis}},\ }\href
  {https://doi.org/10.18564/jasss.2897} {\bibfield  {journal} {\bibinfo
  {journal} {Journal of Artifical Societies and Social Simulations}\ }\textbf
  {\bibinfo {volume} {18}},\ \bibinfo {pages} {4} (\bibinfo {year}
  {2015})}\BibitemShut {NoStop}%
\bibitem [{\citenamefont {Mantegna}\ and\ \citenamefont
  {Stanley}(1999)}]{Mantegna1999}%
  \BibitemOpen
  \bibfield  {author} {\bibinfo {author} {\bibfnamefont {R.~N.}\ \bibnamefont
  {Mantegna}}\ and\ \bibinfo {author} {\bibfnamefont {H.~E.}\ \bibnamefont
  {Stanley}},\ }\href@noop {} {\emph {\bibinfo {title} {{Introduction to
  econophysics: correlations and complexity in finance}}}}\ (\bibinfo
  {publisher} {Cambridge university press},\ \bibinfo {year}
  {1999})\BibitemShut {NoStop}%
\bibitem [{\citenamefont {Castellano}\ \emph {et~al.}(2009)\citenamefont
  {Castellano}, \citenamefont {Fortunato},\ and\ \citenamefont
  {Loreto}}]{castellano2009statistical}%
  \BibitemOpen
  \bibfield  {author} {\bibinfo {author} {\bibfnamefont {C.}~\bibnamefont
  {Castellano}}, \bibinfo {author} {\bibfnamefont {S.}~\bibnamefont
  {Fortunato}},\ and\ \bibinfo {author} {\bibfnamefont {V.}~\bibnamefont
  {Loreto}},\ }\bibfield  {title} {\bibinfo {title} {{Statistical physics of
  social dynamics}},\ }\href@noop {} {\bibfield  {journal} {\bibinfo  {journal}
  {Reviews of Modern Physics}\ }\textbf {\bibinfo {volume} {81}},\ \bibinfo
  {pages} {591} (\bibinfo {year} {2009})}\BibitemShut {NoStop}%
\bibitem [{\citenamefont {Martino}\ and\ \citenamefont
  {Marsili}(2006)}]{Martino2006}%
  \BibitemOpen
  \bibfield  {author} {\bibinfo {author} {\bibfnamefont {A.~D.}\ \bibnamefont
  {Martino}}\ and\ \bibinfo {author} {\bibfnamefont {M.}~\bibnamefont
  {Marsili}},\ }\bibfield  {title} {\bibinfo {title} {{Statistical mechanics of
  socio-economic systems with heterogeneous agents}},\ }\href
  {https://doi.org/10.1088/0305-4470/39/43/R01} {\bibfield  {journal} {\bibinfo
   {journal} {Journal of Physics A: Mathematical and General}\ }\textbf
  {\bibinfo {volume} {39}},\ \bibinfo {pages} {R465} (\bibinfo {year}
  {2006})}\BibitemShut {NoStop}%
\bibitem [{\citenamefont {Acemoglu}\ \emph {et~al.}(2015)\citenamefont
  {Acemoglu}, \citenamefont {Ozdaglar},\ and\ \citenamefont
  {Tahbaz-Salehi}}]{Acemoglu2015}%
  \BibitemOpen
  \bibfield  {author} {\bibinfo {author} {\bibfnamefont {D.}~\bibnamefont
  {Acemoglu}}, \bibinfo {author} {\bibfnamefont {A.}~\bibnamefont {Ozdaglar}},\
  and\ \bibinfo {author} {\bibfnamefont {A.}~\bibnamefont {Tahbaz-Salehi}},\
  }\bibfield  {title} {\bibinfo {title} {{Networks, Shocks, and Systemic
  Risk}}} (\bibinfo {year} {2015})\BibitemShut {NoStop}%
\bibitem [{\citenamefont {Di~Guilmi}\ \emph {et~al.}(2012)\citenamefont
  {Di~Guilmi}, \citenamefont {Gallegati}, \citenamefont {Landini},\ and\
  \citenamefont {Stiglitz}}]{DiGuilmi2012}%
  \BibitemOpen
  \bibfield  {author} {\bibinfo {author} {\bibfnamefont {C.}~\bibnamefont
  {Di~Guilmi}}, \bibinfo {author} {\bibfnamefont {M.}~\bibnamefont
  {Gallegati}}, \bibinfo {author} {\bibfnamefont {S.}~\bibnamefont {Landini}},\
  and\ \bibinfo {author} {\bibfnamefont {J.~E.}\ \bibnamefont {Stiglitz}},\
  }\bibfield  {title} {\bibinfo {title} {{Towards an Analytical Solution for
  Agent Based Models: an Application to a Credit Network Economy}},\ }in\ \href
  {https://doi.org/http://dx.doi.org/10.2139/ssrn.1943280} {\emph {\bibinfo
  {booktitle} {Complexity and Institutions: Market Norms and Corporations}}},\
  \bibinfo {editor} {edited by\ \bibinfo {editor} {\bibfnamefont
  {A.}~\bibnamefont {Masahiko}}, \bibinfo {editor} {\bibfnamefont
  {B.}~\bibnamefont {Kenneth}}, \bibinfo {editor} {\bibfnamefont
  {D.}~\bibnamefont {Simon}},\ and\ \bibinfo {editor} {\bibfnamefont
  {G.}~\bibnamefont {Herbert}}}\ (\bibinfo  {publisher} {Palgrave Macmillan},\
  \bibinfo {address} {New York},\ \bibinfo {year} {2012})\ Chap.~\bibinfo
  {chapter} {3}, pp.\ \bibinfo {pages} {63--80}\BibitemShut {NoStop}%
\bibitem [{\citenamefont {Aoki}(1996)}]{Aoki1998}%
  \BibitemOpen
  \bibfield  {author} {\bibinfo {author} {\bibfnamefont {M.}~\bibnamefont
  {Aoki}},\ }\href@noop {} {\emph {\bibinfo {title} {{New Approaches to
  Macroeconomic Modeling}}}}\ (\bibinfo  {publisher} {Cambridge University
  Press},\ \bibinfo {address} {Cambridge, UK},\ \bibinfo {year}
  {1996})\BibitemShut {NoStop}%
\bibitem [{\citenamefont {Aoki}\ and\ \citenamefont
  {Yoshikawa}(2006)}]{Aoki2007}%
  \BibitemOpen
  \bibfield  {author} {\bibinfo {author} {\bibfnamefont {M.}~\bibnamefont
  {Aoki}}\ and\ \bibinfo {author} {\bibfnamefont {H.}~\bibnamefont
  {Yoshikawa}},\ }\href {https://doi.org/10.1007/s13398-014-0173-7.2} {\emph
  {\bibinfo {title} {{Reconstructing Macroeconomics: A Perspective from
  Statistical Physics and Combinatorial Stochastic Processes}}}}\ (\bibinfo
  {publisher} {Cambridge University Press},\ \bibinfo {address} {Cambridge,
  UK},\ \bibinfo {year} {2006})\BibitemShut {NoStop}%
\bibitem [{\citenamefont {Delli~Gatti}\ \emph {et~al.}(2000)\citenamefont
  {Delli~Gatti}, \citenamefont {Gallegati},\ and\ \citenamefont
  {Kirman}}]{DelliGatti2000}%
  \BibitemOpen
  \bibinfo {editor} {\bibfnamefont {D.}~\bibnamefont {Delli~Gatti}}, \bibinfo
  {editor} {\bibfnamefont {M.}~\bibnamefont {Gallegati}},\ and\ \bibinfo
  {editor} {\bibfnamefont {A.}~\bibnamefont {Kirman}},\ eds.,\ \href@noop {}
  {\emph {\bibinfo {title} {{Interaction and Market Structure}}}},\ Lecture
  Notes in Economics and Mathematical Systems\ (\bibinfo  {publisher}
  {Springer},\ \bibinfo {address} {Berlin},\ \bibinfo {year}
  {2000})\BibitemShut {NoStop}%
\bibitem [{\citenamefont {Gualdi}\ \emph {et~al.}(2015)\citenamefont {Gualdi},
  \citenamefont {Tarzia}, \citenamefont {Zamponi},\ and\ \citenamefont
  {Bouchaud}}]{Gualdi2015}%
  \BibitemOpen
  \bibfield  {author} {\bibinfo {author} {\bibfnamefont {S.}~\bibnamefont
  {Gualdi}}, \bibinfo {author} {\bibfnamefont {M.}~\bibnamefont {Tarzia}},
  \bibinfo {author} {\bibfnamefont {F.}~\bibnamefont {Zamponi}},\ and\ \bibinfo
  {author} {\bibfnamefont {J.~P.}\ \bibnamefont {Bouchaud}},\ }\bibfield
  {title} {\bibinfo {title} {{Tipping points in macroeconomic agent-based
  models}},\ }\href {https://doi.org/10.1016/j.jedc.2014.08.003} {\bibfield
  {journal} {\bibinfo  {journal} {Journal of Economic Dynamics and Control}\
  }\textbf {\bibinfo {volume} {50}},\ \bibinfo {pages} {29} (\bibinfo {year}
  {2015})}\BibitemShut {NoStop}%
\bibitem [{\citenamefont {Gualdi}\ \emph {et~al.}(2017)\citenamefont {Gualdi},
  \citenamefont {Tarzia}, \citenamefont {Zamponi},\ and\ \citenamefont
  {Bouchaud}}]{Gualdi2017}%
  \BibitemOpen
  \bibfield  {author} {\bibinfo {author} {\bibfnamefont {S.}~\bibnamefont
  {Gualdi}}, \bibinfo {author} {\bibfnamefont {M.}~\bibnamefont {Tarzia}},
  \bibinfo {author} {\bibfnamefont {F.}~\bibnamefont {Zamponi}},\ and\ \bibinfo
  {author} {\bibfnamefont {J.-P.}\ \bibnamefont {Bouchaud}},\ }\bibfield
  {title} {\bibinfo {title} {{Monetary Policy and Dark Corners in a stylized
  Agent-Based Model}},\ }\href {https://doi.org/10.1007/s11403-016-0174-z}
  {\bibfield  {journal} {\bibinfo  {journal} {Journal of Economic Interaction
  and Coordination}\ }\textbf {\bibinfo {volume} {12}},\ \bibinfo {pages} {507}
  (\bibinfo {year} {2017})}\BibitemShut {NoStop}%
\bibitem [{\citenamefont {Di~Guilmi}\ \emph {et~al.}(2008)\citenamefont
  {Di~Guilmi}, \citenamefont {Gallegati},\ and\ \citenamefont
  {Landini}}]{DiGuilmi2008}%
  \BibitemOpen
  \bibfield  {author} {\bibinfo {author} {\bibfnamefont {C.}~\bibnamefont
  {Di~Guilmi}}, \bibinfo {author} {\bibfnamefont {M.}~\bibnamefont
  {Gallegati}},\ and\ \bibinfo {author} {\bibfnamefont {S.}~\bibnamefont
  {Landini}},\ }\bibfield  {title} {\bibinfo {title} {{Economic dynamics with
  financial fragility and mean-field interaction: A model}},\ }\href
  {https://doi.org/10.1016/j.physa.2008.01.048} {\bibfield  {journal} {\bibinfo
   {journal} {Physica A: Statistical Mechanics and its Applications}\ }\textbf
  {\bibinfo {volume} {387}},\ \bibinfo {pages} {3852} (\bibinfo {year}
  {2008})}\BibitemShut {NoStop}%
\bibitem [{\citenamefont {Chiarella}\ and\ \citenamefont
  {Di~Guilmi}(2011)}]{Chiarella2011a}%
  \BibitemOpen
  \bibfield  {author} {\bibinfo {author} {\bibfnamefont {C.}~\bibnamefont
  {Chiarella}}\ and\ \bibinfo {author} {\bibfnamefont {C.}~\bibnamefont
  {Di~Guilmi}},\ }\bibfield  {title} {\bibinfo {title} {{The financial
  instability hypothesis: A stochastic microfoundation framework}},\ }\href
  {https://doi.org/10.1016/j.jedc.2011.02.005} {\bibfield  {journal} {\bibinfo
  {journal} {Journal of Economic Dynamics and Control}\ }\textbf {\bibinfo
  {volume} {35}},\ \bibinfo {pages} {1151} (\bibinfo {year}
  {2011})}\BibitemShut {NoStop}%
\bibitem [{\citenamefont {Landini}\ and\ \citenamefont
  {Gallegati}(2014)}]{Landini2014}%
  \BibitemOpen
  \bibfield  {author} {\bibinfo {author} {\bibfnamefont {S.}~\bibnamefont
  {Landini}}\ and\ \bibinfo {author} {\bibfnamefont {M.}~\bibnamefont
  {Gallegati}},\ }\bibfield  {title} {\bibinfo {title} {{Heterogeneity,
  interaction and emergence: effects of composition}},\ }\href
  {https://doi.org/https://doi.org/10.1504/ijcee.2014.064787} {\bibfield
  {journal} {\bibinfo  {journal} {International Journal of Computational
  Economics and Econometrics}\ }\textbf {\bibinfo {volume} {4}},\ \bibinfo
  {pages} {339} (\bibinfo {year} {2014})}\BibitemShut {NoStop}%
\bibitem [{\citenamefont {Bouchaud}(2013)}]{Bouchaud2013}%
  \BibitemOpen
  \bibfield  {author} {\bibinfo {author} {\bibfnamefont {J.-P.}\ \bibnamefont
  {Bouchaud}},\ }\bibfield  {title} {\bibinfo {title} {{Crises and Collective
  Socio-Economic Phenomena: Simple Models and Challenges}},\ }\href
  {https://doi.org/10.1007/s10955-012-0687-3} {\bibfield  {journal} {\bibinfo
  {journal} {Journal of Statistical Physics}\ }\textbf {\bibinfo {volume}
  {151}},\ \bibinfo {pages} {567} (\bibinfo {year} {2013})}\BibitemShut
  {NoStop}%
\bibitem [{\citenamefont {Fiaschi}\ and\ \citenamefont
  {Marsili}(2010)}]{Fiaschi2010}%
  \BibitemOpen
  \bibfield  {author} {\bibinfo {author} {\bibfnamefont {D.}~\bibnamefont
  {Fiaschi}}\ and\ \bibinfo {author} {\bibfnamefont {M.}~\bibnamefont
  {Marsili}},\ }\bibfield  {title} {\bibinfo {title} {{Economic interactions
  and the distribution of wealth}},\ }in\ \href
  {https://doi.org/10.1007/978-88-470-1501-2_8} {\emph {\bibinfo {booktitle}
  {Econophysics and Economics of Games,Social Choices and Quantitative
  Techniques}}},\ \bibinfo {editor} {edited by\ \bibinfo {editor}
  {\bibfnamefont {B.}~\bibnamefont {Basu}}, \bibinfo {editor} {\bibfnamefont
  {S.}~\bibnamefont {Chakravarty}}, \bibinfo {editor} {\bibfnamefont
  {B.}~\bibnamefont {Chakrabarti}},\ and\ \bibinfo {editor} {\bibfnamefont
  {K.}~\bibnamefont {Gangopadhyay}}}\ (\bibinfo  {publisher} {Springer},\
  \bibinfo {address} {Milano},\ \bibinfo {year} {2010})\ pp.\ \bibinfo {pages}
  {61--70}\BibitemShut {NoStop}%
\bibitem [{\citenamefont {Friedkin}\ and\ \citenamefont
  {Johnsen}(2011)}]{Friedkin2011}%
  \BibitemOpen
  \bibfield  {author} {\bibinfo {author} {\bibfnamefont {N.~E.}\ \bibnamefont
  {Friedkin}}\ and\ \bibinfo {author} {\bibfnamefont {E.~C.}\ \bibnamefont
  {Johnsen}},\ }\href@noop {} {\emph {\bibinfo {title} {{Social Influence
  Network Theory}}}}\ (\bibinfo  {publisher} {Cambridge University Press},\
  \bibinfo {address} {New York},\ \bibinfo {year} {2011})\BibitemShut {NoStop}%
\bibitem [{\citenamefont {Alfarano}\ \emph {et~al.}(2008)\citenamefont
  {Alfarano}, \citenamefont {Lux},\ and\ \citenamefont
  {Wagner}}]{Alfarano2008a}%
  \BibitemOpen
  \bibfield  {author} {\bibinfo {author} {\bibfnamefont {S.}~\bibnamefont
  {Alfarano}}, \bibinfo {author} {\bibfnamefont {T.}~\bibnamefont {Lux}},\ and\
  \bibinfo {author} {\bibfnamefont {F.}~\bibnamefont {Wagner}},\ }\bibfield
  {title} {\bibinfo {title} {{Time variation of higher moments in a financial
  market with heterogeneous agents: An analytical approach}},\ }\href
  {https://doi.org/10.1016/j.jedc.2006.12.014} {\bibfield  {journal} {\bibinfo
  {journal} {Journal of Economic Dynamics and Control}\ }\textbf {\bibinfo
  {volume} {32}},\ \bibinfo {pages} {101} (\bibinfo {year} {2008})}\BibitemShut
  {NoStop}%
\bibitem [{\citenamefont {Lux}(2016)}]{Lux2016}%
  \BibitemOpen
  \bibfield  {author} {\bibinfo {author} {\bibfnamefont {T.}~\bibnamefont
  {Lux}},\ }\bibfield  {title} {\bibinfo {title} {{A model of the topology of
  the bank - firm credit network and its role as channel of contagion}},\
  }\href {https://doi.org/10.1016/j.jedc.2016.03.002} {\bibfield  {journal}
  {\bibinfo  {journal} {Journal of Economic Dynamics and Control}\ }\textbf
  {\bibinfo {volume} {66}},\ \bibinfo {pages} {36} (\bibinfo {year}
  {2016})}\BibitemShut {NoStop}%
\bibitem [{\citenamefont {Gross}\ and\ \citenamefont
  {Sayama}(2009)}]{gross2009adaptive}%
  \BibitemOpen
  \bibfield  {author} {\bibinfo {author} {\bibfnamefont {T.}~\bibnamefont
  {Gross}}\ and\ \bibinfo {author} {\bibfnamefont {H.}~\bibnamefont {Sayama}},\
  }\bibfield  {title} {\bibinfo {title} {{Adaptive networks}},\ }in\ \href@noop
  {} {\emph {\bibinfo {booktitle} {Adaptive networks}}}\ (\bibinfo  {publisher}
  {Springer},\ \bibinfo {year} {2009})\ pp.\ \bibinfo {pages}
  {1--8}\BibitemShut {NoStop}%
\bibitem [{\citenamefont {Do}\ and\ \citenamefont {Gross}(2009)}]{Do2009}%
  \BibitemOpen
  \bibfield  {author} {\bibinfo {author} {\bibfnamefont {A.-L.}\ \bibnamefont
  {Do}}\ and\ \bibinfo {author} {\bibfnamefont {T.}~\bibnamefont {Gross}},\
  }\bibfield  {title} {\bibinfo {title} {{Contact processes and moment closure
  on adaptive networks}},\ }in\ \href
  {https://doi.org/10.1007/978-3-642-01284-6{\_}9} {\emph {\bibinfo {booktitle}
  {Adaptive Networks: Theory, Models and Applications}}},\ \bibinfo {editor}
  {edited by\ \bibinfo {editor} {\bibfnamefont {T.}~\bibnamefont {Gross}}\ and\
  \bibinfo {editor} {\bibfnamefont {H.}~\bibnamefont {Sayama}}}\ (\bibinfo
  {publisher} {Springer and NECSI},\ \bibinfo {address} {Cambridge,
  Massachusetts},\ \bibinfo {year} {2009})\ Chap.~\bibinfo {chapter} {9}, pp.\
  \bibinfo {pages} {191--208}\BibitemShut {NoStop}%
\bibitem [{\citenamefont {Demirel}\ \emph {et~al.}(2014)\citenamefont
  {Demirel}, \citenamefont {Vazquez}, \citenamefont {B{\"{o}}hme},\ and\
  \citenamefont {Gross}}]{Demirel2014}%
  \BibitemOpen
  \bibfield  {author} {\bibinfo {author} {\bibfnamefont {G.}~\bibnamefont
  {Demirel}}, \bibinfo {author} {\bibfnamefont {F.}~\bibnamefont {Vazquez}},
  \bibinfo {author} {\bibfnamefont {G.~A.}\ \bibnamefont {B{\"{o}}hme}},\ and\
  \bibinfo {author} {\bibfnamefont {T.}~\bibnamefont {Gross}},\ }\bibfield
  {title} {\bibinfo {title} {{Moment-closure approximations for discrete
  adaptive networks}},\ }\href {https://doi.org/10.1016/j.physd.2013.07.003}
  {\bibfield  {journal} {\bibinfo  {journal} {Physica D: Nonlinear Phenomena}\
  }\textbf {\bibinfo {volume} {267}},\ \bibinfo {pages} {68} (\bibinfo {year}
  {2014})}\BibitemShut {NoStop}%
\bibitem [{\citenamefont {Wiedermann}\ \emph {et~al.}(2015)\citenamefont
  {Wiedermann}, \citenamefont {Donges}, \citenamefont {Heitzig}, \citenamefont
  {Lucht},\ and\ \citenamefont {Kurths}}]{Wiedermann2015}%
  \BibitemOpen
  \bibfield  {author} {\bibinfo {author} {\bibfnamefont {M.}~\bibnamefont
  {Wiedermann}}, \bibinfo {author} {\bibfnamefont {J.~F.}\ \bibnamefont
  {Donges}}, \bibinfo {author} {\bibfnamefont {J.}~\bibnamefont {Heitzig}},
  \bibinfo {author} {\bibfnamefont {W.}~\bibnamefont {Lucht}},\ and\ \bibinfo
  {author} {\bibfnamefont {J.}~\bibnamefont {Kurths}},\ }\bibfield  {title}
  {\bibinfo {title} {{Macroscopic description of complex adaptive networks
  co-evolving with dynamic node states}},\ }\href
  {https://doi.org/10.1103/PhysRevE.91.052801} {\bibfield  {journal} {\bibinfo
  {journal} {Physical Review E}\ }\textbf {\bibinfo {volume} {91}},\ \bibinfo
  {pages} {1} (\bibinfo {year} {2015})}\BibitemShut {NoStop}%
\bibitem [{\citenamefont {Min}\ and\ \citenamefont {Miguel}(2017)}]{Min2017}%
  \BibitemOpen
  \bibfield  {author} {\bibinfo {author} {\bibfnamefont {B.}~\bibnamefont
  {Min}}\ and\ \bibinfo {author} {\bibfnamefont {M.~S.}\ \bibnamefont
  {Miguel}},\ }\bibfield  {title} {\bibinfo {title} {{Fragmentation transitions
  in a coevolving nonlinear voter model}},\ }\href
  {https://doi.org/10.1038/s41598-017-13047-2} {\bibfield  {journal} {\bibinfo
  {journal} {Scientific Reports}\ }\textbf {\bibinfo {volume} {7}},\ \bibinfo
  {pages} {1} (\bibinfo {year} {2017})}\BibitemShut {NoStop}%
\bibitem [{\citenamefont {Kuehn}(2016)}]{Kuehn2016}%
  \BibitemOpen
  \bibfield  {author} {\bibinfo {author} {\bibfnamefont {C.}~\bibnamefont
  {Kuehn}},\ }\bibfield  {title} {\bibinfo {title} {{Moment Closure - A Brief
  Review}},\ }in\ \href {https://doi.org/10.1007/978-3-319-28028-8_13} {\emph
  {\bibinfo {booktitle} {Control of Self-Organizing Nonlinear Systems.
  Understanding Complex Systems}}},\ \bibinfo {editor} {edited by\ \bibinfo
  {editor} {\bibfnamefont {E.}~\bibnamefont {Sch{\"{o}}ll}}, \bibinfo {editor}
  {\bibfnamefont {S.}~\bibnamefont {Klapp}},\ and\ \bibinfo {editor}
  {\bibfnamefont {P.}~\bibnamefont {H{\"{o}}vel}}}\ (\bibinfo  {publisher}
  {Springer},\ \bibinfo {address} {Cham},\ \bibinfo {year} {2016})\
  Chap.~\bibinfo {chapter} {13}, pp.\ \bibinfo {pages} {253--271}\BibitemShut
  {NoStop}%
\bibitem [{\citenamefont {{IPCC}}(2014)}]{IPCC2014}%
  \BibitemOpen
  \bibfield  {author} {\bibinfo {author} {\bibnamefont {{IPCC}}},\ }\href@noop
  {} {\emph {\bibinfo {title} {{Climate Change 2014: Mitigation of Climate
  Change: Contribution of Working Group III to the Fifth Assessment Report of
  the Intergovernmental Panel on}}}}\ (\bibinfo  {publisher} {Cambridge
  University Press},\ \bibinfo {address} {Cambridge},\ \bibinfo {year}
  {2014})\BibitemShut {NoStop}%
\bibitem [{\citenamefont {Ansar}\ \emph {et~al.}(2013)\citenamefont {Ansar},
  \citenamefont {Caldecot},\ and\ \citenamefont {Tibury}}]{Ans2013}%
  \BibitemOpen
  \bibfield  {author} {\bibinfo {author} {\bibfnamefont {A.}~\bibnamefont
  {Ansar}}, \bibinfo {author} {\bibfnamefont {B.}~\bibnamefont {Caldecot}},\
  and\ \bibinfo {author} {\bibfnamefont {J.}~\bibnamefont {Tibury}},\
  }\bibfield  {title} {\bibinfo {title} {{Stranded assets and the fossil fuel
  divestment campaign: what does divestment mean for the valuation of fossil
  fuel assets?}},\ }\href {https://doi.org/10.1177/0149206309337896} {\bibfield
   {journal} {\bibinfo  {journal} {SSEE}\ ,\ \bibinfo {pages} {1}} (\bibinfo
  {year} {2013})}\BibitemShut {NoStop}%
\bibitem [{\citenamefont {Mattauch}\ and\ \citenamefont
  {Hepburn}(2016)}]{Mattauch2016}%
  \BibitemOpen
  \bibfield  {author} {\bibinfo {author} {\bibfnamefont {L.}~\bibnamefont
  {Mattauch}}\ and\ \bibinfo {author} {\bibfnamefont {C.}~\bibnamefont
  {Hepburn}},\ }\bibfield  {title} {\bibinfo {title} {{Climate Policy When
  Preferences Are Endogenous — and Sometimes They Are}},\ }\href
  {https://doi.org/10.1111/misp.12048} {\bibfield  {journal} {\bibinfo
  {journal} {Midwest Studies In Philosophy}\ }\textbf {\bibinfo {volume}
  {40}},\ \bibinfo {pages} {76} (\bibinfo {year} {2016})}\BibitemShut {NoStop}%
\bibitem [{\citenamefont {Mattauch}\ \emph {et~al.}(2018)\citenamefont
  {Mattauch}, \citenamefont {Hepburn},\ and\ \citenamefont
  {Stern}}]{Mattauch2018}%
  \BibitemOpen
  \bibfield  {author} {\bibinfo {author} {\bibfnamefont {L.}~\bibnamefont
  {Mattauch}}, \bibinfo {author} {\bibfnamefont {C.}~\bibnamefont {Hepburn}},\
  and\ \bibinfo {author} {\bibfnamefont {N.}~\bibnamefont {Stern}},\ }\bibfield
   {title} {\bibinfo {title} {{Pigou pushes preferences: decarbonization and
  endogenous values}}} (\bibinfo {year} {2018})\BibitemShut {NoStop}%
\bibitem [{\citenamefont {Gsottbauer}\ and\ \citenamefont
  {Bergh}(2011)}]{Gsottbauer2011}%
  \BibitemOpen
  \bibfield  {author} {\bibinfo {author} {\bibfnamefont {E.}~\bibnamefont
  {Gsottbauer}}\ and\ \bibinfo {author} {\bibfnamefont {J.~C. J. M. V.~D.}\
  \bibnamefont {Bergh}},\ }\bibfield  {title} {\bibinfo {title} {{Environmental
  Policy Theory Given Bounded Rationality and Other-regarding Preferences}},\
  }\href {https://doi.org/10.1007/s10640-010-9433-y} {\bibfield  {journal}
  {\bibinfo  {journal} {Environmental and Resource Economics}\ }\textbf
  {\bibinfo {volume} {49}},\ \bibinfo {pages} {263} (\bibinfo {year}
  {2011})}\BibitemShut {NoStop}%
\bibitem [{\citenamefont {Hong}\ and\ \citenamefont
  {Kacperczyk}(2009)}]{Hong2009}%
  \BibitemOpen
  \bibfield  {author} {\bibinfo {author} {\bibfnamefont {H.}~\bibnamefont
  {Hong}}\ and\ \bibinfo {author} {\bibfnamefont {M.}~\bibnamefont
  {Kacperczyk}},\ }\bibfield  {title} {\bibinfo {title} {{The price of sin: The
  effects of social norms on markets}},\ }\href
  {https://doi.org/10.1016/j.jfineco.2008.09.001} {\bibfield  {journal}
  {\bibinfo  {journal} {Journal of Financial Economics}\ }\textbf {\bibinfo
  {volume} {93}},\ \bibinfo {pages} {15} (\bibinfo {year} {2009})}\BibitemShut
  {NoStop}%
\bibitem [{\citenamefont {Williams}(2007)}]{Williams2007}%
  \BibitemOpen
  \bibfield  {author} {\bibinfo {author} {\bibfnamefont {G.}~\bibnamefont
  {Williams}},\ }\bibfield  {title} {\bibinfo {title} {{Some Determinants of
  the Socially Responsible Investment Decision: A Cross-Country Study}},\
  }\href {https://doi.org/10.1080/15427560709337016} {\bibfield  {journal}
  {\bibinfo  {journal} {Journal of Behavioral Finance}\ }\textbf {\bibinfo
  {volume} {8}},\ \bibinfo {pages} {43} (\bibinfo {year} {2007})}\BibitemShut
  {NoStop}%
\bibitem [{\citenamefont {Griskevicius}\ \emph {et~al.}(2008)\citenamefont
  {Griskevicius}, \citenamefont {Cialdini},\ and\ \citenamefont
  {Goldstein}}]{Griskevicius2008}%
  \BibitemOpen
  \bibfield  {author} {\bibinfo {author} {\bibfnamefont {V.}~\bibnamefont
  {Griskevicius}}, \bibinfo {author} {\bibfnamefont {R.~B.}\ \bibnamefont
  {Cialdini}},\ and\ \bibinfo {author} {\bibfnamefont {N.~J.}\ \bibnamefont
  {Goldstein}},\ }\bibfield  {title} {\bibinfo {title} {{Social norms: an
  underestimated and underemployed lever for managing climate change}},\
  }\href@noop {} {\bibfield  {journal} {\bibinfo  {journal} {International
  Journal of Sustainability Communication}\ }\textbf {\bibinfo {volume} {3}},\
  \bibinfo {pages} {5} (\bibinfo {year} {2008})}\BibitemShut {NoStop}%
\bibitem [{\citenamefont {Masson}\ and\ \citenamefont
  {Fritsche}(2014)}]{Masson2014}%
  \BibitemOpen
  \bibfield  {author} {\bibinfo {author} {\bibfnamefont {T.}~\bibnamefont
  {Masson}}\ and\ \bibinfo {author} {\bibfnamefont {I.}~\bibnamefont
  {Fritsche}},\ }\bibfield  {title} {\bibinfo {title} {{Adherence to climate
  change-related ingroup norms: Do dimensions of group identification
  matter?}},\ }\href {https://doi.org/10.1002/ejsp.2036} {\bibfield  {journal}
  {\bibinfo  {journal} {European Journal of Social Psychology}\ }\textbf
  {\bibinfo {volume} {465}},\ \bibinfo {pages} {455} (\bibinfo {year}
  {2014})}\BibitemShut {NoStop}%
\bibitem [{\citenamefont {Stern}(2011)}]{Stern2011}%
  \BibitemOpen
  \bibfield  {author} {\bibinfo {author} {\bibfnamefont {P.~C.}\ \bibnamefont
  {Stern}},\ }\bibfield  {title} {\bibinfo {title} {{Contributions of
  Psychology to Limiting Climate Change}},\ }\href
  {https://doi.org/10.1037/a0023235} {\bibfield  {journal} {\bibinfo  {journal}
  {American Psychologist}\ }\textbf {\bibinfo {volume} {66}},\ \bibinfo {pages}
  {303} (\bibinfo {year} {2011})}\BibitemShut {NoStop}%
\bibitem [{\citenamefont {Rabinovich}\ \emph {et~al.}(2011)\citenamefont
  {Rabinovich}, \citenamefont {Morton},\ and\ \citenamefont
  {Duke}}]{Rabinovich2011}%
  \BibitemOpen
  \bibfield  {author} {\bibinfo {author} {\bibfnamefont {A.}~\bibnamefont
  {Rabinovich}}, \bibinfo {author} {\bibfnamefont {T.~A.}\ \bibnamefont
  {Morton}},\ and\ \bibinfo {author} {\bibfnamefont {C.~C.}\ \bibnamefont
  {Duke}},\ }\bibfield  {title} {\bibinfo {title} {{Collective Self and
  Individual Choice: The Role of Social Comparisons in Promoting Public
  Engagement with Climate Change}},\ }in\ \href@noop {} {\emph {\bibinfo
  {booktitle} {Engaging the public with climate change: Behaviour change and
  communication}}},\ \bibinfo {editor} {edited by\ \bibinfo {editor}
  {\bibfnamefont {L.}~\bibnamefont {Whitmarsh}}, \bibinfo {editor}
  {\bibfnamefont {S.}~\bibnamefont {O'Neill}},\ and\ \bibinfo {editor}
  {\bibfnamefont {I.}~\bibnamefont {Lorenzoni}}}\ (\bibinfo  {publisher}
  {Earthscan},\ \bibinfo {address} {Oxon, UK, and New York},\ \bibinfo {year}
  {2011})\ Chap.~\bibinfo {chapter} {4}, pp.\ \bibinfo {pages}
  {66--83}\BibitemShut {NoStop}%
\bibitem [{\citenamefont {Nyborg}\ \emph {et~al.}(2016)\citenamefont {Nyborg},
  \citenamefont {Anderies}, \citenamefont {Dannenberg}, \citenamefont
  {Lindahl}, \citenamefont {Schill}, \citenamefont {Schl{\"{u}}ter},
  \citenamefont {Adger}, \citenamefont {Arrow}, \citenamefont {Barrett},
  \citenamefont {Carpenter}, \citenamefont {Stuart}, \citenamefont {Iii},
  \citenamefont {Cr{\'{e}}pin}, \citenamefont {Daily}, \citenamefont {Ehrlich},
  \citenamefont {Folke}, \citenamefont {Jager}, \citenamefont {Kautsky},
  \citenamefont {Levin}, \citenamefont {Madsen}, \citenamefont {Polasky},
  \citenamefont {Scheffer}, \citenamefont {Weber}, \citenamefont {Wilen},
  \citenamefont {Xepapadeas},\ and\ \citenamefont {Zeeuw}}]{Nyborg2016}%
  \BibitemOpen
  \bibfield  {author} {\bibinfo {author} {\bibfnamefont {B.~K.}\ \bibnamefont
  {Nyborg}}, \bibinfo {author} {\bibfnamefont {J.~M.}\ \bibnamefont
  {Anderies}}, \bibinfo {author} {\bibfnamefont {A.}~\bibnamefont
  {Dannenberg}}, \bibinfo {author} {\bibfnamefont {T.}~\bibnamefont {Lindahl}},
  \bibinfo {author} {\bibfnamefont {C.}~\bibnamefont {Schill}}, \bibinfo
  {author} {\bibfnamefont {M.}~\bibnamefont {Schl{\"{u}}ter}}, \bibinfo
  {author} {\bibfnamefont {W.~N.}\ \bibnamefont {Adger}}, \bibinfo {author}
  {\bibfnamefont {K.~J.}\ \bibnamefont {Arrow}}, \bibinfo {author}
  {\bibfnamefont {S.}~\bibnamefont {Barrett}}, \bibinfo {author} {\bibfnamefont
  {S.}~\bibnamefont {Carpenter}}, \bibinfo {author} {\bibfnamefont
  {F.}~\bibnamefont {Stuart}}, \bibinfo {author} {\bibfnamefont
  {C.}~\bibnamefont {Iii}}, \bibinfo {author} {\bibfnamefont {A.-s.}\
  \bibnamefont {Cr{\'{e}}pin}}, \bibinfo {author} {\bibfnamefont
  {G.}~\bibnamefont {Daily}}, \bibinfo {author} {\bibfnamefont
  {P.}~\bibnamefont {Ehrlich}}, \bibinfo {author} {\bibfnamefont
  {C.}~\bibnamefont {Folke}}, \bibinfo {author} {\bibfnamefont
  {W.}~\bibnamefont {Jager}}, \bibinfo {author} {\bibfnamefont
  {N.}~\bibnamefont {Kautsky}}, \bibinfo {author} {\bibfnamefont {S.~A.}\
  \bibnamefont {Levin}}, \bibinfo {author} {\bibfnamefont {O.~J.}\ \bibnamefont
  {Madsen}}, \bibinfo {author} {\bibfnamefont {S.}~\bibnamefont {Polasky}},
  \bibinfo {author} {\bibfnamefont {M.}~\bibnamefont {Scheffer}}, \bibinfo
  {author} {\bibfnamefont {E.~U.}\ \bibnamefont {Weber}}, \bibinfo {author}
  {\bibfnamefont {J.}~\bibnamefont {Wilen}}, \bibinfo {author} {\bibfnamefont
  {A.}~\bibnamefont {Xepapadeas}},\ and\ \bibinfo {author} {\bibfnamefont
  {A.~D.}\ \bibnamefont {Zeeuw}},\ }\bibfield  {title} {\bibinfo {title}
  {{Social norms as solutions}},\ }\href
  {https://doi.org/10.1126/science.aaf8317} {\bibfield  {journal} {\bibinfo
  {journal} {Science}\ }\textbf {\bibinfo {volume} {354}},\ \bibinfo {pages}
  {42} (\bibinfo {year} {2016})}\BibitemShut {NoStop}%
\bibitem [{\citenamefont {Bandura}(1977)}]{bandura1977}%
  \BibitemOpen
  \bibfield  {author} {\bibinfo {author} {\bibfnamefont {A.}~\bibnamefont
  {Bandura}},\ }\bibfield  {title} {\bibinfo {title} {{Social learning
  theory}},\ }in\ \href@noop {} {\emph {\bibinfo {booktitle} {Social learning
  theory}}},\ Vol.~\bibinfo {volume} {1},\ \bibinfo {editor} {edited by\
  \bibinfo {editor} {\bibfnamefont {A.}~\bibnamefont {Bandura}}\ and\ \bibinfo
  {editor} {\bibfnamefont {R.~H.}\ \bibnamefont {Walters}}}\ (\bibinfo
  {publisher} {Prentice-hall},\ \bibinfo {address} {Englewood Cliffs, NJ},\
  \bibinfo {year} {1977})\ pp.\ \bibinfo {pages} {1--46}\BibitemShut {NoStop}%
\bibitem [{\citenamefont {Friedkin}(2001)}]{Friedkin2001}%
  \BibitemOpen
  \bibfield  {author} {\bibinfo {author} {\bibfnamefont {N.~E.}\ \bibnamefont
  {Friedkin}},\ }\bibfield  {title} {\bibinfo {title} {{Norm formation in
  social influence networks}},\ }\href@noop {} {\bibfield  {journal} {\bibinfo
  {journal} {Social Networks}\ }\textbf {\bibinfo {volume} {23}},\ \bibinfo
  {pages} {167} (\bibinfo {year} {2001})}\BibitemShut {NoStop}%
\bibitem [{\citenamefont {Centola}\ \emph {et~al.}(2007)\citenamefont
  {Centola}, \citenamefont {Gonza},\ and\ \citenamefont
  {Miguel}}]{Centola2007HomophilyGroups}%
  \BibitemOpen
  \bibfield  {author} {\bibinfo {author} {\bibfnamefont {D.}~\bibnamefont
  {Centola}}, \bibinfo {author} {\bibfnamefont {J.~C.}\ \bibnamefont {Gonza}},\
  and\ \bibinfo {author} {\bibfnamefont {M.~S.}\ \bibnamefont {Miguel}},\
  }\bibfield  {title} {\bibinfo {title} {{Homophily, Cultural Drift, and the
  Co-Evolution of Cultural Groups}},\ }\href@noop {} {\bibfield  {journal}
  {\bibinfo  {journal} {Journal of Conflict Resolution}\ }\textbf {\bibinfo
  {volume} {2}},\ \bibinfo {pages} {905} (\bibinfo {year} {2007})}\BibitemShut
  {NoStop}%
\bibitem [{\citenamefont {Kimura}\ and\ \citenamefont
  {Hayakawa}(2008)}]{Kimura2008}%
  \BibitemOpen
  \bibfield  {author} {\bibinfo {author} {\bibfnamefont {D.}~\bibnamefont
  {Kimura}}\ and\ \bibinfo {author} {\bibfnamefont {Y.}~\bibnamefont
  {Hayakawa}},\ }\bibfield  {title} {\bibinfo {title} {{Coevolutionary networks
  with homophily and heterophily}},\ }\href
  {https://doi.org/10.1103/PhysRevE.78.016103} {\bibfield  {journal} {\bibinfo
  {journal} {Physical Review E}\ ,\ \bibinfo {pages} {016103}} (\bibinfo {year}
  {2008})}\BibitemShut {NoStop}%
\bibitem [{\citenamefont {{International Monetary Fund}}(2011)}]{IMF2011}%
  \BibitemOpen
  \bibfield  {author} {\bibinfo {author} {\bibnamefont {{International Monetary
  Fund}}},\ }\href@noop {} {\emph {\bibinfo {title} {World economic and
  financial surveys}}},\ \bibinfo {type} {Tech. Rep.}\ (\bibinfo {year}
  {2011})\BibitemShut {NoStop}%
\bibitem [{\citenamefont {H{\"{o}}ssinger}\ \emph {et~al.}(2017)\citenamefont
  {H{\"{o}}ssinger}, \citenamefont {Link}, \citenamefont {Sonntag},
  \citenamefont {Stark}, \citenamefont {H{\"{o}}sslinger}, \citenamefont
  {Link}, \citenamefont {Sonntag},\ and\ \citenamefont
  {Stark}}]{Hosslinger2017}%
  \BibitemOpen
  \bibfield  {author} {\bibinfo {author} {\bibfnamefont {R.}~\bibnamefont
  {H{\"{o}}ssinger}}, \bibinfo {author} {\bibfnamefont {C.}~\bibnamefont
  {Link}}, \bibinfo {author} {\bibfnamefont {A.}~\bibnamefont {Sonntag}},
  \bibinfo {author} {\bibfnamefont {J.}~\bibnamefont {Stark}}, \bibinfo
  {author} {\bibfnamefont {R.}~\bibnamefont {H{\"{o}}sslinger}}, \bibinfo
  {author} {\bibfnamefont {C.}~\bibnamefont {Link}}, \bibinfo {author}
  {\bibfnamefont {A.}~\bibnamefont {Sonntag}},\ and\ \bibinfo {author}
  {\bibfnamefont {J.}~\bibnamefont {Stark}},\ }\bibfield  {title} {\bibinfo
  {title} {{Estimating the price elasticity of fuel demand with stated
  preferences derived from a situational approach}},\ }\href
  {https://doi.org/10.1016/j.tra.2017.06.001} {\bibfield  {journal} {\bibinfo
  {journal} {Transportation Research Part A: Policy and Practice}\ }\textbf
  {\bibinfo {volume} {103}},\ \bibinfo {pages} {154} (\bibinfo {year}
  {2017})}\BibitemShut {NoStop}%
\bibitem [{\citenamefont {Labandeira}\ \emph {et~al.}(2017)\citenamefont
  {Labandeira}, \citenamefont {Labeaga},\ and\ \citenamefont
  {L{\'{o}}pez-otero}}]{Labandeira2017}%
  \BibitemOpen
  \bibfield  {author} {\bibinfo {author} {\bibfnamefont {X.}~\bibnamefont
  {Labandeira}}, \bibinfo {author} {\bibfnamefont {J.~M.}\ \bibnamefont
  {Labeaga}},\ and\ \bibinfo {author} {\bibfnamefont {X.}~\bibnamefont
  {L{\'{o}}pez-otero}},\ }\bibfield  {title} {\bibinfo {title} {{A
  meta-analysis on the price elasticity of energy demand}},\ }\href
  {https://doi.org/10.1016/j.enpol.2017.01.002} {\bibfield  {journal} {\bibinfo
   {journal} {Energy Policy}\ }\textbf {\bibinfo {volume} {102}},\ \bibinfo
  {pages} {549} (\bibinfo {year} {2017})}\BibitemShut {NoStop}%
\bibitem [{\citenamefont {Daly}(1997)}]{Daly1997}%
  \BibitemOpen
  \bibfield  {author} {\bibinfo {author} {\bibfnamefont {H.~E.}\ \bibnamefont
  {Daly}},\ }\bibfield  {title} {\bibinfo {title} {{Georgescu-Roegen versus
  Solow / Stiglitz}},\ }\href {https://doi.org/10.1016/S0921-8009(97)00081-5}
  {\bibfield  {journal} {\bibinfo  {journal} {Ecological Economics}\ }\textbf
  {\bibinfo {volume} {22}},\ \bibinfo {pages} {261} (\bibinfo {year}
  {1997})}\BibitemShut {NoStop}%
\bibitem [{\citenamefont {Georgescu-Roegen}(1975)}]{georgescu1975energy}%
  \BibitemOpen
  \bibfield  {author} {\bibinfo {author} {\bibfnamefont {N.}~\bibnamefont
  {Georgescu-Roegen}},\ }\bibfield  {title} {\bibinfo {title} {{Energy and
  economic myths}},\ }\href {https://doi.org/10.2307/1056148} {\bibfield
  {journal} {\bibinfo  {journal} {Southern Economic Journal}\ }\textbf
  {\bibinfo {volume} {41}},\ \bibinfo {pages} {347} (\bibinfo {year}
  {1975})}\BibitemShut {NoStop}%
\bibitem [{\citenamefont {Georgescu-Roegen}(1979)}]{georgescu1979comments}%
  \BibitemOpen
  \bibfield  {author} {\bibinfo {author} {\bibfnamefont {N.}~\bibnamefont
  {Georgescu-Roegen}},\ }\bibfield  {title} {\bibinfo {title} {{Comments on the
  papers by Daly and Stiglitz}},\ }in\ \href@noop {} {\emph {\bibinfo
  {booktitle} {Scarcity and growth reconsidered}}},\ \bibinfo {editor} {edited
  by\ \bibinfo {editor} {\bibfnamefont {V.~K.}\ \bibnamefont {Smith}}}\
  (\bibinfo  {publisher} {Resources for the Future},\ \bibinfo {address} {New
  York},\ \bibinfo {year} {1979})\ pp.\ \bibinfo {pages} {95--105}\BibitemShut
  {NoStop}%
\bibitem [{\citenamefont {Ayres}\ \emph {et~al.}(2007)\citenamefont {Ayres},
  \citenamefont {Turton},\ and\ \citenamefont {Casten}}]{Ayres2007}%
  \BibitemOpen
  \bibfield  {author} {\bibinfo {author} {\bibfnamefont {R.~U.}\ \bibnamefont
  {Ayres}}, \bibinfo {author} {\bibfnamefont {H.}~\bibnamefont {Turton}},\ and\
  \bibinfo {author} {\bibfnamefont {T.}~\bibnamefont {Casten}},\ }\bibfield
  {title} {\bibinfo {title} {{Energy efficiency, sustainability and economic
  growth}},\ }\href {https://doi.org/10.1016/j.energy.2006.06.005} {\bibfield
  {journal} {\bibinfo  {journal} {Energy}\ }\textbf {\bibinfo {volume} {32}},\
  \bibinfo {pages} {634} (\bibinfo {year} {2007})}\BibitemShut {NoStop}%
\bibitem [{\citenamefont {Ayres}\ \emph {et~al.}(2013)\citenamefont {Ayres},
  \citenamefont {van~den Bergh}, \citenamefont {Lindenberger},\ and\
  \citenamefont {Warr}}]{Ayres2013}%
  \BibitemOpen
  \bibfield  {author} {\bibinfo {author} {\bibfnamefont {R.~U.}\ \bibnamefont
  {Ayres}}, \bibinfo {author} {\bibfnamefont {J.~C. J.~M.}\ \bibnamefont
  {van~den Bergh}}, \bibinfo {author} {\bibfnamefont {D.}~\bibnamefont
  {Lindenberger}},\ and\ \bibinfo {author} {\bibfnamefont {B.~S.}\ \bibnamefont
  {Warr}},\ }\bibfield  {title} {\bibinfo {title} {{The Underestimated
  Contribution of Energy to Economic Growth}},\ }\href
  {https://doi.org/10.1016/j.strueco.2013.07.004} {\bibfield  {journal}
  {\bibinfo  {journal} {Structural Change and Economic Dynamics}\ }\textbf
  {\bibinfo {volume} {27}},\ \bibinfo {pages} {79} (\bibinfo {year}
  {2013})}\BibitemShut {NoStop}%
\bibitem [{\citenamefont {Mulder}\ and\ \citenamefont
  {Groot}(2012)}]{Mulder2012}%
  \BibitemOpen
  \bibfield  {author} {\bibinfo {author} {\bibfnamefont {P.}~\bibnamefont
  {Mulder}}\ and\ \bibinfo {author} {\bibfnamefont {H.~L. F.~D.}\ \bibnamefont
  {Groot}},\ }\bibfield  {title} {\bibinfo {title} {{Structural change and
  convergence of energy intensity across OECD countries, 1970 – 2005}},\
  }\href {https://doi.org/10.1016/j.eneco.2012.07.023} {\bibfield  {journal}
  {\bibinfo  {journal} {Energy Economics}\ }\textbf {\bibinfo {volume} {34}},\
  \bibinfo {pages} {1910} (\bibinfo {year} {2012})}\BibitemShut {NoStop}%
\bibitem [{\citenamefont {Argote}\ and\ \citenamefont
  {Epple}(1990)}]{argote1990learning}%
  \BibitemOpen
  \bibfield  {author} {\bibinfo {author} {\bibfnamefont {L.}~\bibnamefont
  {Argote}}\ and\ \bibinfo {author} {\bibfnamefont {D.~N.}\ \bibnamefont
  {Epple}},\ }\bibfield  {title} {\bibinfo {title} {{Learning curves in
  manufacturing}},\ }\href {https://doi.org/10.1126/science.247.4945.920}
  {\bibfield  {journal} {\bibinfo  {journal} {Science}\ }\textbf {\bibinfo
  {volume} {247}},\ \bibinfo {pages} {920} (\bibinfo {year}
  {1990})}\BibitemShut {NoStop}%
\bibitem [{\citenamefont {Wright}(1936)}]{wright1936factors}%
  \BibitemOpen
  \bibfield  {author} {\bibinfo {author} {\bibfnamefont {T.~P.}\ \bibnamefont
  {Wright}},\ }\bibfield  {title} {\bibinfo {title} {{Factors affecting the
  cost of airplanes}},\ }\href@noop {} {\bibfield  {journal} {\bibinfo
  {journal} {Journal of the aeronautical sciences}\ }\textbf {\bibinfo {volume}
  {3}},\ \bibinfo {pages} {122} (\bibinfo {year} {1936})}\BibitemShut {NoStop}%
\bibitem [{\citenamefont {Nagy}\ \emph {et~al.}(2013)\citenamefont {Nagy},
  \citenamefont {Farmer}, \citenamefont {Bui},\ and\ \citenamefont
  {Trancik}}]{Nagy2013}%
  \BibitemOpen
  \bibfield  {author} {\bibinfo {author} {\bibfnamefont {B.}~\bibnamefont
  {Nagy}}, \bibinfo {author} {\bibfnamefont {J.~D.}\ \bibnamefont {Farmer}},
  \bibinfo {author} {\bibfnamefont {Q.~M.}\ \bibnamefont {Bui}},\ and\ \bibinfo
  {author} {\bibfnamefont {J.~E.}\ \bibnamefont {Trancik}},\ }\bibfield
  {title} {\bibinfo {title} {{Statistical Basis for Predicting Technological
  Progress}},\ }\href {https://doi.org/10.1371/journal.pone.0052669} {\bibfield
   {journal} {\bibinfo  {journal} {PLoS ONE}\ }\textbf {\bibinfo {volume}
  {8}},\ \bibinfo {pages} {1} (\bibinfo {year} {2013})}\BibitemShut {NoStop}%
\bibitem [{\citenamefont {Kahouli-Brahmi}(2008)}]{Kahouli-Brahmi2008}%
  \BibitemOpen
  \bibfield  {author} {\bibinfo {author} {\bibfnamefont {S.}~\bibnamefont
  {Kahouli-Brahmi}},\ }\bibfield  {title} {\bibinfo {title} {{Technological
  learning in energy-environment-economy modelling: A survey}},\ }\href
  {https://doi.org/10.1016/j.enpol.2007.09.001} {\bibfield  {journal} {\bibinfo
   {journal} {Energy Policy}\ }\textbf {\bibinfo {volume} {36}},\ \bibinfo
  {pages} {138} (\bibinfo {year} {2008})}\BibitemShut {NoStop}%
\bibitem [{\citenamefont {Dasgupta}\ and\ \citenamefont
  {Heal}(1974)}]{Dasgupta1974}%
  \BibitemOpen
  \bibfield  {author} {\bibinfo {author} {\bibfnamefont {P.}~\bibnamefont
  {Dasgupta}}\ and\ \bibinfo {author} {\bibfnamefont {G.}~\bibnamefont
  {Heal}},\ }\bibfield  {title} {\bibinfo {title} {{The Optimal Depletion of
  Exhaustible Resources}},\ }\href@noop {} {\bibfield  {journal} {\bibinfo
  {journal} {The Review of Economic Studies, Ltd., Oxford University Press}\
  }\textbf {\bibinfo {volume} {41}},\ \bibinfo {pages} {3} (\bibinfo {year}
  {1974})}\BibitemShut {NoStop}%
\bibitem [{\citenamefont {Perman}\ \emph {et~al.}(2003)\citenamefont {Perman},
  \citenamefont {Ma}, \citenamefont {McGilvray},\ and\ \citenamefont
  {Common}}]{Perman2003}%
  \BibitemOpen
  \bibfield  {author} {\bibinfo {author} {\bibfnamefont {R.}~\bibnamefont
  {Perman}}, \bibinfo {author} {\bibfnamefont {Y.}~\bibnamefont {Ma}}, \bibinfo
  {author} {\bibfnamefont {J.}~\bibnamefont {McGilvray}},\ and\ \bibinfo
  {author} {\bibfnamefont {M.}~\bibnamefont {Common}},\ }\href@noop {} {\emph
  {\bibinfo {title} {{Natural Resource and Environmental Economics}}}},\
  \bibinfo {edition} {3rd}\ ed.\ (\bibinfo  {publisher} {Pearson Education},\
  \bibinfo {year} {2003})\BibitemShut {NoStop}%
\bibitem [{\citenamefont {Simon}(1972)}]{simon1972theories}%
  \BibitemOpen
  \bibfield  {author} {\bibinfo {author} {\bibfnamefont {H.~A.}\ \bibnamefont
  {Simon}},\ }\bibfield  {title} {\bibinfo {title} {{Theories of bounded
  rationality}},\ }in\ \href@noop {} {\emph {\bibinfo {booktitle} {Decision and
  Organization}}},\ \bibinfo {editor} {edited by\ \bibinfo {editor}
  {\bibfnamefont {C.~B.}\ \bibnamefont {McGuire}}\ and\ \bibinfo {editor}
  {\bibfnamefont {R.}~\bibnamefont {Radner}}}\ (\bibinfo  {publisher} {North
  Holland},\ \bibinfo {year} {1972})\ Chap.~\bibinfo {chapter} {8}, pp.\
  \bibinfo {pages} {161--176}\BibitemShut {NoStop}%
\bibitem [{\citenamefont {Simon}(1982)}]{simon1982models}%
  \BibitemOpen
  \bibfield  {author} {\bibinfo {author} {\bibfnamefont {H.~A.}\ \bibnamefont
  {Simon}},\ }\href@noop {} {\emph {\bibinfo {title} {{Models of bounded
  rationality: Empirically grounded economic reason}}}},\ Vol.~\bibinfo
  {volume} {3}\ (\bibinfo  {publisher} {MIT press},\ \bibinfo {year}
  {1982})\BibitemShut {NoStop}%
\bibitem [{\citenamefont {Gigerenzer}\ and\ \citenamefont
  {Selten}(2002)}]{gigerenzer2002bounded}%
  \BibitemOpen
  \bibfield  {author} {\bibinfo {author} {\bibfnamefont {G.}~\bibnamefont
  {Gigerenzer}}\ and\ \bibinfo {author} {\bibfnamefont {R.}~\bibnamefont
  {Selten}},\ }\href@noop {} {\emph {\bibinfo {title} {{Bounded rationality:
  The adaptive toolbox}}}}\ (\bibinfo  {publisher} {MIT Press},\ \bibinfo
  {year} {2002})\BibitemShut {NoStop}%
\bibitem [{\citenamefont {Traulsen}\ \emph {et~al.}(2010)\citenamefont
  {Traulsen}, \citenamefont {Semmann}, \citenamefont {Sommerfeld},
  \citenamefont {Krambeck},\ and\ \citenamefont {Milinski}}]{Traulsen2010}%
  \BibitemOpen
  \bibfield  {author} {\bibinfo {author} {\bibfnamefont {A.}~\bibnamefont
  {Traulsen}}, \bibinfo {author} {\bibfnamefont {D.}~\bibnamefont {Semmann}},
  \bibinfo {author} {\bibfnamefont {R.~D.}\ \bibnamefont {Sommerfeld}},
  \bibinfo {author} {\bibfnamefont {H.-J.}\ \bibnamefont {Krambeck}},\ and\
  \bibinfo {author} {\bibfnamefont {M.}~\bibnamefont {Milinski}},\ }\bibfield
  {title} {\bibinfo {title} {{Human strategy updating in evolutionary
  games.}},\ }\href {https://doi.org/10.1073/pnas.0912515107} {\bibfield
  {journal} {\bibinfo  {journal} {Proceedings of the National Academy of
  Sciences}\ }\textbf {\bibinfo {volume} {107}},\ \bibinfo {pages} {2962}
  (\bibinfo {year} {2010})}\BibitemShut {NoStop}%
\bibitem [{\citenamefont {Barkoczi}\ and\ \citenamefont
  {Galesic}(2016)}]{Barkoczi2016}%
  \BibitemOpen
  \bibfield  {author} {\bibinfo {author} {\bibfnamefont {D.}~\bibnamefont
  {Barkoczi}}\ and\ \bibinfo {author} {\bibfnamefont {M.}~\bibnamefont
  {Galesic}},\ }\bibfield  {title} {\bibinfo {title} {{Social learning
  strategies modify the effect of network structure on group performance}},\
  }\href {https://doi.org/10.1038/ncomms13109} {\bibfield  {journal} {\bibinfo
  {journal} {Nature Communications}\ }\textbf {\bibinfo {volume} {7}},\
  \bibinfo {pages} {1} (\bibinfo {year} {2016})}\BibitemShut {NoStop}%
\bibitem [{\citenamefont {Fisher}\ \emph {et~al.}(2013)\citenamefont {Fisher},
  \citenamefont {Waggle},\ and\ \citenamefont {Leifeld}}]{Fisher2013}%
  \BibitemOpen
  \bibfield  {author} {\bibinfo {author} {\bibfnamefont {D.~R.}\ \bibnamefont
  {Fisher}}, \bibinfo {author} {\bibfnamefont {J.}~\bibnamefont {Waggle}},\
  and\ \bibinfo {author} {\bibfnamefont {P.}~\bibnamefont {Leifeld}},\
  }\bibfield  {title} {\bibinfo {title} {{Where Does Political Polarization
  Come From? Locating Polarization Within the U.S. Climate Change Debate}},\
  }\href {https://doi.org/10.1177/0002764212463360} {\bibfield  {journal}
  {\bibinfo  {journal} {American Behavioral Scientist}\ }\textbf {\bibinfo
  {volume} {57}},\ \bibinfo {pages} {70} (\bibinfo {year} {2013})}\BibitemShut
  {NoStop}%
\bibitem [{\citenamefont {Farrell}(2016)}]{Farrell2016a}%
  \BibitemOpen
  \bibfield  {author} {\bibinfo {author} {\bibfnamefont {J.}~\bibnamefont
  {Farrell}},\ }\bibfield  {title} {\bibinfo {title} {{Corporate funding and
  ideological polarization about climate change}},\ }\href
  {https://doi.org/10.1073/pnas.1509433112} {\bibfield  {journal} {\bibinfo
  {journal} {Proceedings of the National Academy of Sciences}\ }\textbf
  {\bibinfo {volume} {113}},\ \bibinfo {pages} {92} (\bibinfo {year}
  {2016})}\BibitemShut {NoStop}%
\bibitem [{\citenamefont {Dunlap}\ \emph {et~al.}(2016)\citenamefont {Dunlap},
  \citenamefont {McCright},\ and\ \citenamefont {Yarosh}}]{Dunlap2016}%
  \BibitemOpen
  \bibfield  {author} {\bibinfo {author} {\bibfnamefont {R.~E.}\ \bibnamefont
  {Dunlap}}, \bibinfo {author} {\bibfnamefont {A.~M.}\ \bibnamefont
  {McCright}},\ and\ \bibinfo {author} {\bibfnamefont {J.~H.}\ \bibnamefont
  {Yarosh}},\ }\bibfield  {title} {\bibinfo {title} {{The Political Divide on
  Climate Change: Partisan Polarization Widens in the U.S.}},\ }\href
  {https://doi.org/10.1080/00139157.2016.1208995} {\bibfield  {journal}
  {\bibinfo  {journal} {Environment: Science and Policy for Sustainable
  Development}\ }\textbf {\bibinfo {volume} {58}},\ \bibinfo {pages} {4}
  (\bibinfo {year} {2016})}\BibitemShut {NoStop}%
\bibitem [{\citenamefont {McCright}\ and\ \citenamefont
  {Dunlap}(2011)}]{McCright2011}%
  \BibitemOpen
  \bibfield  {author} {\bibinfo {author} {\bibfnamefont {A.~M.}\ \bibnamefont
  {McCright}}\ and\ \bibinfo {author} {\bibfnamefont {R.~E.}\ \bibnamefont
  {Dunlap}},\ }\bibfield  {title} {\bibinfo {title} {{THE POLITICIZATION OF
  CLIMATE CHANGE AND POLARIZATION IN THE AMERICAN PUBLIC'S VIEWS OF GLOBAL
  WARMING, 2001 – 2010}},\ }\href@noop {} {\bibfield  {journal} {\bibinfo
  {journal} {The Sociological Quarterly}\ }\textbf {\bibinfo {volume} {52}},\
  \bibinfo {pages} {155} (\bibinfo {year} {2011})}\BibitemShut {NoStop}%
\bibitem [{\citenamefont {Hart}\ and\ \citenamefont {Nisbet}(2012)}]{Hart2012}%
  \BibitemOpen
  \bibfield  {author} {\bibinfo {author} {\bibfnamefont {P.~S.}\ \bibnamefont
  {Hart}}\ and\ \bibinfo {author} {\bibfnamefont {E.~C.}\ \bibnamefont
  {Nisbet}},\ }\bibfield  {title} {\bibinfo {title} {{Boomerang Effects in
  Science Communication: How Motivated Reasoning and Identity Cues Amplify
  Opinion Polarization About Climate Mitigation Policies}},\ }\href
  {https://doi.org/10.1177/0093650211416646} {\bibfield  {journal} {\bibinfo
  {journal} {Communication Research}\ }\textbf {\bibinfo {volume} {39}},\
  \bibinfo {pages} {701 } (\bibinfo {year} {2012})}\BibitemShut {NoStop}%
\bibitem [{\citenamefont {Williams}\ \emph {et~al.}(2015)\citenamefont
  {Williams}, \citenamefont {McMurray}, \citenamefont {Kurz},\ and\
  \citenamefont {{Hugo Lambert}}}]{Williams2015}%
  \BibitemOpen
  \bibfield  {author} {\bibinfo {author} {\bibfnamefont {H.~T.}\ \bibnamefont
  {Williams}}, \bibinfo {author} {\bibfnamefont {J.~R.}\ \bibnamefont
  {McMurray}}, \bibinfo {author} {\bibfnamefont {T.}~\bibnamefont {Kurz}},\
  and\ \bibinfo {author} {\bibfnamefont {F.}~\bibnamefont {{Hugo Lambert}}},\
  }\bibfield  {title} {\bibinfo {title} {{Network analysis reveals open forums
  and echo chambers in social media discussions of climate change}},\ }\href
  {https://doi.org/10.1016/j.gloenvcha.2015.03.006} {\bibfield  {journal}
  {\bibinfo  {journal} {Glob. Environ. Chang.}\ }\textbf {\bibinfo {volume}
  {32}},\ \bibinfo {pages} {126} (\bibinfo {year} {2015})}\BibitemShut
  {NoStop}%
\bibitem [{\citenamefont {McPherson}\ \emph {et~al.}(2001)\citenamefont
  {McPherson}, \citenamefont {Smith-Lovin},\ and\ \citenamefont
  {Cook}}]{McPherson2007}%
  \BibitemOpen
  \bibfield  {author} {\bibinfo {author} {\bibfnamefont {M.}~\bibnamefont
  {McPherson}}, \bibinfo {author} {\bibfnamefont {L.}~\bibnamefont
  {Smith-Lovin}},\ and\ \bibinfo {author} {\bibfnamefont {J.~M.}\ \bibnamefont
  {Cook}},\ }\bibfield  {title} {\bibinfo {title} {{Birds of a Feather:
  Homophily in Social Networks}},\ }\href@noop {} {\bibfield  {journal}
  {\bibinfo  {journal} {Annual Review of Sociology}\ }\textbf {\bibinfo
  {volume} {27}},\ \bibinfo {pages} {415} (\bibinfo {year} {2001})}\BibitemShut
  {NoStop}%
\bibitem [{\citenamefont {Centola}(2011)}]{Centola2011}%
  \BibitemOpen
  \bibfield  {author} {\bibinfo {author} {\bibfnamefont {D.}~\bibnamefont
  {Centola}},\ }\bibfield  {title} {\bibinfo {title} {{An experimental study of
  homophily in the adoption of health behavior}},\ }\href
  {https://doi.org/10.1126/science.1207055} {\bibfield  {journal} {\bibinfo
  {journal} {Science}\ }\textbf {\bibinfo {volume} {334}},\ \bibinfo {pages}
  {1269} (\bibinfo {year} {2011})}\BibitemShut {NoStop}%
\bibitem [{\citenamefont {Asano}\ \emph {et~al.}(2019)\citenamefont {Asano},
  \citenamefont {Kolb}, \citenamefont {Heitzig},\ and\ \citenamefont
  {Farmer}}]{Asano2019}%
  \BibitemOpen
  \bibfield  {author} {\bibinfo {author} {\bibfnamefont {Y.~M.}\ \bibnamefont
  {Asano}}, \bibinfo {author} {\bibfnamefont {J.~J.}\ \bibnamefont {Kolb}},
  \bibinfo {author} {\bibfnamefont {J.}~\bibnamefont {Heitzig}},\ and\ \bibinfo
  {author} {\bibfnamefont {J.~D.}\ \bibnamefont {Farmer}},\ }\bibfield  {title}
  {\bibinfo {title} {{Emergent inequality and endogenous dynamics in a simple
  behavioral macroeconomic model}},\ }\href@noop {} {\bibfield  {journal}
  {\bibinfo  {journal} {PNAS - In Review}\ } (\bibinfo {year}
  {2019})}\BibitemShut {NoStop}%
\bibitem [{\citenamefont {Kolb}(2018)}]{kolb2018}%
  \BibitemOpen
  \bibfield  {author} {\bibinfo {author} {\bibfnamefont {J.~J.}\ \bibnamefont
  {Kolb}},\ }\href {https://github.com/jakobkolb/pydivest} {\bibinfo {title}
  {{github.com/jakobkolb/pydivest}}} (\bibinfo {year} {2018})\BibitemShut
  {NoStop}%
\bibitem [{\citenamefont {Gross}\ \emph {et~al.}(2006)\citenamefont {Gross},
  \citenamefont {D'Lima},\ and\ \citenamefont {Blasius}}]{Gross2006}%
  \BibitemOpen
  \bibfield  {author} {\bibinfo {author} {\bibfnamefont {T.}~\bibnamefont
  {Gross}}, \bibinfo {author} {\bibfnamefont {C.}~\bibnamefont {D'Lima}},\ and\
  \bibinfo {author} {\bibfnamefont {B.}~\bibnamefont {Blasius}},\ }\bibfield
  {title} {\bibinfo {title} {{Epidemic Dynamics on an Adaptive Network}},\
  }\href {https://doi.org/10.1103/PhysRevLett.96.208701} {\bibfield  {journal}
  {\bibinfo  {journal} {Phys. Rev. Lett.}\ }\textbf {\bibinfo {volume} {96}},\
  \bibinfo {pages} {208701} (\bibinfo {year} {2006})}\BibitemShut {NoStop}%
\bibitem [{\citenamefont {B{\"{o}}hme}\ and\ \citenamefont
  {Gross}(2011)}]{Bohme2011}%
  \BibitemOpen
  \bibfield  {author} {\bibinfo {author} {\bibfnamefont {G.~A.}\ \bibnamefont
  {B{\"{o}}hme}}\ and\ \bibinfo {author} {\bibfnamefont {T.}~\bibnamefont
  {Gross}},\ }\bibfield  {title} {\bibinfo {title} {{Analytical calculation of
  fragmentation transitions in adaptive networks}},\ }\href
  {https://doi.org/10.1103/PhysRevE.83.035101} {\bibfield  {journal} {\bibinfo
  {journal} {Phys. Rev. E - Stat. Nonlinear, Soft Matter Phys.}\ }\textbf
  {\bibinfo {volume} {83}},\ \bibinfo {pages} {4} (\bibinfo {year} {2011})},\
  \Eprint {https://arxiv.org/abs/1012.1213} {arXiv:1012.1213} \BibitemShut
  {NoStop}%
\bibitem [{\citenamefont {Rogers}\ \emph {et~al.}(2012)\citenamefont {Rogers},
  \citenamefont {Clifford-Brown}, \citenamefont {Mills},\ and\ \citenamefont
  {Galla}}]{Rogers2012}%
  \BibitemOpen
  \bibfield  {author} {\bibinfo {author} {\bibfnamefont {T.}~\bibnamefont
  {Rogers}}, \bibinfo {author} {\bibfnamefont {W.}~\bibnamefont
  {Clifford-Brown}}, \bibinfo {author} {\bibfnamefont {C.}~\bibnamefont
  {Mills}},\ and\ \bibinfo {author} {\bibfnamefont {T.}~\bibnamefont {Galla}},\
  }\bibfield  {title} {\bibinfo {title} {{Stochastic oscillations of adaptive
  networks: application to epidemic modelling}},\ }\href
  {https://doi.org/10.1088/1742-5468/2012/08/P08018} {\bibfield  {journal}
  {\bibinfo  {journal} {Journal of Statistical Mechanics: Theory and
  Experiment}\ }\textbf {\bibinfo {volume} {2012}},\ \bibinfo {pages} {P08018}
  (\bibinfo {year} {2012})}\BibitemShut {NoStop}%
\bibitem [{\citenamefont {Rogers}\ and\ \citenamefont
  {Gross}(2013)}]{Rogers2013}%
  \BibitemOpen
  \bibfield  {author} {\bibinfo {author} {\bibfnamefont {T.}~\bibnamefont
  {Rogers}}\ and\ \bibinfo {author} {\bibfnamefont {T.}~\bibnamefont {Gross}},\
  }\bibfield  {title} {\bibinfo {title} {{Consensus time and conformity in the
  adaptive voter model}},\ }\href {https://doi.org/10.1103/PhysRevE.88.030102}
  {\bibfield  {journal} {\bibinfo  {journal} {Phys. Rev. E - Stat. Nonlinear,
  Soft Matter Phys.}\ }\textbf {\bibinfo {volume} {88}},\ \bibinfo {pages} {1}
  (\bibinfo {year} {2013})},\ \Eprint {https://arxiv.org/abs/1304.4742}
  {arXiv:1304.4742} \BibitemShut {NoStop}%
\bibitem [{\citenamefont {Nitzbon}\ \emph {et~al.}(2017)\citenamefont
  {Nitzbon}, \citenamefont {Heitzig},\ and\ \citenamefont
  {Parlitz}}]{Nitzbon2017}%
  \BibitemOpen
  \bibfield  {author} {\bibinfo {author} {\bibfnamefont {J.}~\bibnamefont
  {Nitzbon}}, \bibinfo {author} {\bibfnamefont {J.}~\bibnamefont {Heitzig}},\
  and\ \bibinfo {author} {\bibfnamefont {U.}~\bibnamefont {Parlitz}},\
  }\bibfield  {title} {\bibinfo {title} {{Sustainability, collapse and
  oscillations of global climate, population and economy in a simple
  World-Earth model}},\ }\bibfield  {journal} {\bibinfo  {journal}
  {Environmental Research Letters}\ }\textbf {\bibinfo {volume} {12}},\ \href
  {https://doi.org/10.1088/1748-9326/aa7581} {10.1088/1748-9326/aa7581}
  (\bibinfo {year} {2017})\BibitemShut {NoStop}%
\bibitem [{\citenamefont {Van~Kampen}(1992)}]{VanKampen1992}%
  \BibitemOpen
  \bibfield  {author} {\bibinfo {author} {\bibfnamefont {N.~G.}\ \bibnamefont
  {Van~Kampen}},\ }\href@noop {} {\emph {\bibinfo {title} {{Stochastic
  Processes in Physics and Chemistry}}}},\ \bibinfo {edition} {2nd}\ ed.\
  (\bibinfo  {publisher} {North-Holland Personal Library},\ \bibinfo {year}
  {1992})\BibitemShut {NoStop}%
\bibitem [{\citenamefont {M{\"{u}}ller-Hansen}\ \emph
  {et~al.}(2017)\citenamefont {M{\"{u}}ller-Hansen}, \citenamefont
  {Schl{\"{u}}ter}, \citenamefont {M{\"{a}}s}, \citenamefont {Donges},
  \citenamefont {Kolb}, \citenamefont {Thonicke},\ and\ \citenamefont
  {Heitzig}}]{Mueller-Hansen2017}%
  \BibitemOpen
  \bibfield  {author} {\bibinfo {author} {\bibfnamefont {F.}~\bibnamefont
  {M{\"{u}}ller-Hansen}}, \bibinfo {author} {\bibfnamefont {M.}~\bibnamefont
  {Schl{\"{u}}ter}}, \bibinfo {author} {\bibfnamefont {M.}~\bibnamefont
  {M{\"{a}}s}}, \bibinfo {author} {\bibfnamefont {J.~F.}\ \bibnamefont
  {Donges}}, \bibinfo {author} {\bibfnamefont {J.~J.}\ \bibnamefont {Kolb}},
  \bibinfo {author} {\bibfnamefont {K.}~\bibnamefont {Thonicke}},\ and\
  \bibinfo {author} {\bibfnamefont {J.}~\bibnamefont {Heitzig}},\ }\bibfield
  {title} {\bibinfo {title} {{Towards representing human behavior and decision
  making in Earth system models - an overview of techniques and approaches}},\
  }\href {https://doi.org/10.5194/esd-8-977-2017} {\bibfield  {journal}
  {\bibinfo  {journal} {Earth System Dynamics}\ }\textbf {\bibinfo {volume}
  {8}},\ \bibinfo {pages} {977} (\bibinfo {year} {2017})}\BibitemShut {NoStop}%
\bibitem [{\citenamefont {Strogatz}(1994)}]{Strogatz1994}%
  \BibitemOpen
  \bibfield  {author} {\bibinfo {author} {\bibfnamefont {S.~H.}\ \bibnamefont
  {Strogatz}},\ }\href@noop {} {\emph {\bibinfo {title} {Book}}}\ (\bibinfo
  {publisher} {Perseus Books},\ \bibinfo {address} {Reading, MA},\ \bibinfo
  {year} {1994})\BibitemShut {NoStop}%
\bibitem [{\citenamefont {Kuznetsov}(2013)}]{Kuznetsov1998}%
  \BibitemOpen
  \bibfield  {author} {\bibinfo {author} {\bibfnamefont {Y.~A.}\ \bibnamefont
  {Kuznetsov}},\ }\href@noop {} {\emph {\bibinfo {title} {{Elements of Applied
  Bifurcation Theory}}}},\ \bibinfo {edition} {2nd}\ ed.,\ Vol.\ \bibinfo
  {volume} {112}\ (\bibinfo  {publisher} {Springer Science {\&} Business
  Media},\ \bibinfo {address} {New York, Berlin, Heidelberg},\ \bibinfo {year}
  {2013})\BibitemShut {NoStop}%
\bibitem [{\citenamefont {Allgower}\ and\ \citenamefont
  {Georg}(2003)}]{Allgower2003}%
  \BibitemOpen
  \bibfield  {author} {\bibinfo {author} {\bibfnamefont {E.~L.}\ \bibnamefont
  {Allgower}}\ and\ \bibinfo {author} {\bibfnamefont {K.}~\bibnamefont
  {Georg}},\ }\href@noop {} {\emph {\bibinfo {title} {{Introduction to
  Numerical Continuation Methods}}}}\ (\bibinfo  {publisher} {SIAM},\ \bibinfo
  {address} {Philadelphia, PA},\ \bibinfo {year} {2003})\BibitemShut {NoStop}%
\bibitem [{\citenamefont {Clewley~RH}(2007)}]{pydstool}%
  \BibitemOpen
  \bibfield  {author} {\bibinfo {author} {\bibfnamefont {L.~M. G.~J.}\
  \bibnamefont {Clewley~RH}, \bibfnamefont {Sherwood~WE}},\ }\href
  {http://pydstool.sourceforge.net} {\bibinfo {title} {{PyDSTool, a software
  environment for dynamical systems modeling}}} (\bibinfo {year}
  {2007})\BibitemShut {NoStop}%
\bibitem [{\citenamefont {Clewley}(2012)}]{10.1371/journal.pcbi.1002628}%
  \BibitemOpen
  \bibfield  {author} {\bibinfo {author} {\bibfnamefont {R.}~\bibnamefont
  {Clewley}},\ }\bibfield  {title} {\bibinfo {title} {{Hybrid Models and
  Biological Model Reduction with PyDSTool}},\ }\href
  {https://doi.org/10.1371/journal.pcbi.1002628} {\bibfield  {journal}
  {\bibinfo  {journal} {PLOS Computational Biology}\ }\textbf {\bibinfo
  {volume} {8}},\ \bibinfo {pages} {1} (\bibinfo {year} {2012})}\BibitemShut
  {NoStop}%
\bibitem [{Note3()}]{Note3}%
  \BibitemOpen
  \bibinfo {note} {PyDSTool is building on the AUTO-07p continuation library
  \protect \citep {Doedel07auto-07p:continuation}.}\BibitemShut {Stop}%
\bibitem [{\citenamefont {Heitzig}\ \emph {et~al.}(2016)\citenamefont
  {Heitzig}, \citenamefont {Kittel}, \citenamefont {Donges},\ and\
  \citenamefont {Molkenthin}}]{Heitzig2016}%
  \BibitemOpen
  \bibfield  {author} {\bibinfo {author} {\bibfnamefont {J.}~\bibnamefont
  {Heitzig}}, \bibinfo {author} {\bibfnamefont {T.}~\bibnamefont {Kittel}},
  \bibinfo {author} {\bibfnamefont {J.~F.}\ \bibnamefont {Donges}},\ and\
  \bibinfo {author} {\bibfnamefont {N.}~\bibnamefont {Molkenthin}},\ }\bibfield
   {title} {\bibinfo {title} {{Topology of sustainable management of dynamical
  systems with desirable states: from defining planetary boundaries to safe
  operating spaces in the Earth system}},\ }\href
  {https://doi.org/10.5194/esd-7-21-2016} {\bibfield  {journal} {\bibinfo
  {journal} {Earth System Dynamics}\ }\textbf {\bibinfo {volume} {7}},\
  \bibinfo {pages} {21} (\bibinfo {year} {2016})}\BibitemShut {NoStop}%
\bibitem [{\citenamefont {Zhang}\ \emph {et~al.}(2016)\citenamefont {Zhang},
  \citenamefont {Brackbill}, \citenamefont {Yang}, \citenamefont {Becker},
  \citenamefont {Herbert},\ and\ \citenamefont {Centola}}]{Zhang2016}%
  \BibitemOpen
  \bibfield  {author} {\bibinfo {author} {\bibfnamefont {J.}~\bibnamefont
  {Zhang}}, \bibinfo {author} {\bibfnamefont {D.}~\bibnamefont {Brackbill}},
  \bibinfo {author} {\bibfnamefont {S.}~\bibnamefont {Yang}}, \bibinfo {author}
  {\bibfnamefont {J.}~\bibnamefont {Becker}}, \bibinfo {author} {\bibfnamefont
  {N.}~\bibnamefont {Herbert}},\ and\ \bibinfo {author} {\bibfnamefont
  {D.}~\bibnamefont {Centola}},\ }\bibfield  {title} {\bibinfo {title}
  {{Support or competition? How online social networks increase physical
  activity: A randomized controlled trial}},\ }\href
  {https://doi.org/10.1016/j.pmedr.2016.08.008} {\bibfield  {journal} {\bibinfo
   {journal} {Preventive Medicine Reports}\ }\textbf {\bibinfo {volume} {4}},\
  \bibinfo {pages} {453} (\bibinfo {year} {2016})}\BibitemShut {NoStop}%
\bibitem [{\citenamefont {Zhang}\ \emph {et~al.}(2015)\citenamefont {Zhang},
  \citenamefont {Brackbill}, \citenamefont {Yang},\ and\ \citenamefont
  {Centola}}]{Zhang2015}%
  \BibitemOpen
  \bibfield  {author} {\bibinfo {author} {\bibfnamefont {J.}~\bibnamefont
  {Zhang}}, \bibinfo {author} {\bibfnamefont {D.}~\bibnamefont {Brackbill}},
  \bibinfo {author} {\bibfnamefont {S.}~\bibnamefont {Yang}},\ and\ \bibinfo
  {author} {\bibfnamefont {D.}~\bibnamefont {Centola}},\ }\bibfield  {title}
  {\bibinfo {title} {{Efficacy and causal mechanism of an online social media
  intervention to increase physical activity: Results of a randomized
  controlled trial}},\ }\href {https://doi.org/10.1016/j.pmedr.2015.08.005}
  {\bibfield  {journal} {\bibinfo  {journal} {Preventive Medicine Reports}\
  }\textbf {\bibinfo {volume} {2}},\ \bibinfo {pages} {651} (\bibinfo {year}
  {2015})}\BibitemShut {NoStop}%
\bibitem [{\citenamefont {Peattie}(2010)}]{Peattie2010}%
  \BibitemOpen
  \bibfield  {author} {\bibinfo {author} {\bibfnamefont {K.}~\bibnamefont
  {Peattie}},\ }\bibfield  {title} {\bibinfo {title} {{Green Consumption:
  Behavior and Norms}},\ }\href
  {https://doi.org/10.1146/annurev-environ-032609-094328} {\bibfield  {journal}
  {\bibinfo  {journal} {Annual Review of Environment and Resources}\ }\textbf
  {\bibinfo {volume} {35}},\ \bibinfo {pages} {195} (\bibinfo {year}
  {2010})}\BibitemShut {NoStop}%
\bibitem [{\citenamefont {Heathcote}\ \emph {et~al.}(2009)\citenamefont
  {Heathcote}, \citenamefont {Storesletten},\ and\ \citenamefont
  {Violante}}]{Heathcote2009}%
  \BibitemOpen
  \bibfield  {author} {\bibinfo {author} {\bibfnamefont {J.}~\bibnamefont
  {Heathcote}}, \bibinfo {author} {\bibfnamefont {K.}~\bibnamefont
  {Storesletten}},\ and\ \bibinfo {author} {\bibfnamefont {G.~L.}\ \bibnamefont
  {Violante}},\ }\bibfield  {title} {\bibinfo {title} {{Quantitative
  Macroeconomics with Heterogeneous Households}},\ }\href
  {https://doi.org/10.1146/annurev.economics.050708.142922} {\bibfield
  {journal} {\bibinfo  {journal} {Annual Review of Economics}\ }\textbf
  {\bibinfo {volume} {1}},\ \bibinfo {pages} {319} (\bibinfo {year}
  {2009})}\BibitemShut {NoStop}%
\bibitem [{\citenamefont {Kirman}(1992)}]{Kirman1992}%
  \BibitemOpen
  \bibfield  {author} {\bibinfo {author} {\bibfnamefont {A.~P.}\ \bibnamefont
  {Kirman}},\ }\bibfield  {title} {\bibinfo {title} {{Whom or What Does the
  Representative Individual Represent?}},\ }\href
  {https://doi.org/https://doi.org/10.1257/jep.6.2.117} {\bibfield  {journal}
  {\bibinfo  {journal} {Journal of Economic Perspectives}\ }\textbf {\bibinfo
  {volume} {6}},\ \bibinfo {pages} {117} (\bibinfo {year} {1992})}\BibitemShut
  {NoStop}%
\bibitem [{\citenamefont {Bedau}(1997)}]{Bedau1997}%
  \BibitemOpen
  \bibfield  {author} {\bibinfo {author} {\bibfnamefont {M.~A.}\ \bibnamefont
  {Bedau}},\ }\bibfield  {title} {\bibinfo {title} {{Weak Emergence}},\ }\href
  {https://doi.org/10.1111/0029-4624.31.s11.17} {\bibfield  {journal} {\bibinfo
   {journal} {No{\^{u}}s}\ }\textbf {\bibinfo {volume} {31}},\ \bibinfo {pages}
  {375} (\bibinfo {year} {1997})}\BibitemShut {NoStop}%
\bibitem [{\citenamefont {Doedel}\ \emph {et~al.}(2007)\citenamefont {Doedel},
  \citenamefont {Fairgrieve}, \citenamefont {Sandstede}, \citenamefont
  {Champneys}, \citenamefont {Kuznetsov},\ and\ \citenamefont
  {Wang}}]{Doedel07auto-07p:continuation}%
  \BibitemOpen
  \bibfield  {author} {\bibinfo {author} {\bibfnamefont {E.~J.}\ \bibnamefont
  {Doedel}}, \bibinfo {author} {\bibfnamefont {T.~F.}\ \bibnamefont
  {Fairgrieve}}, \bibinfo {author} {\bibfnamefont {B.}~\bibnamefont
  {Sandstede}}, \bibinfo {author} {\bibfnamefont {A.~R.}\ \bibnamefont
  {Champneys}}, \bibinfo {author} {\bibfnamefont {Y.~A.}\ \bibnamefont
  {Kuznetsov}},\ and\ \bibinfo {author} {\bibfnamefont {X.}~\bibnamefont
  {Wang}},\ }\href@noop {} {\emph {\bibinfo {title} {{AUTO-07P: Continuation
  and bifurcation software for ordinary differential equations}}}},\ \bibinfo
  {type} {Tech. Rep.}\ (\bibinfo {year} {2007})\BibitemShut {NoStop}%
\end{thebibliography}%

\end{document}